\documentclass[final,3p,times]{elsarticle}
\usepackage{amssymb}
\usepackage[utf8]{inputenc}
\usepackage{subcaption}
\usepackage{caption}
\usepackage{xcolor}
\usepackage{amsmath}
\usepackage{float}
\usepackage{xcolor}
\usepackage{tabularx}

\newcommand{\intL}{\int_{\cal L}}

\newcommand{\beq}{\begin{equation}}
\newcommand{\beql}[1]{\begin{equation}\label{#1}}
\newcommand{\eeq}{\end{equation}}
\newcommand{\bea}{\begin{eqnarray}}
\newcommand{\beal}[1]{\begin{eqnarray}\label{#1}}
\newcommand{\eea}{\end{eqnarray}}
\newcommand{\bean}{\begin{eqnarray*}}
\newcommand{\eean}{\end{eqnarray*}}

\newcommand{\ignore}[1]{}

\newcommand{\cE}{{\cal E}}
\newcommand{\cH}{{\cal H}}
\newcommand{\cA}{{\cal A}}

\newcommand{\half}{\mbox{$\frac 1 2$}}

\newcommand{\dx}{\,\mbox{d}x}
\newcommand{\dxi}{\,\mbox{d}\xi}

\newcommand{\dt}{\,\mbox{d}t}

\newcommand{\new}[1]{#1}

\journal{International Journal of Engineering Science 205 (2024) 104147
}

\begin{document}
\begin{frontmatter}



\title{Integral Micromorphic Model Reproducing Dispersion in 1D Continuum}

\author[1]{Michal \v{S}mejkal\corref{cor1}}
\ead{michal.smejkal@fsv.cvut.cz}

\author[1]{Milan Jir\'{a}sek}
\ead{milan.jirasek@cvut.cz}

\author[1,2]{Martin Hor\'{a}k}
\ead{martin.horak@cvut.cz}

\cortext[cor1]{Corresponding author}
% \fntext[fn1]{This is the first author footnote.}
% \fntext[fn2]{Another author footnote, this is a very long
% footnote and it should be a really long footnote. But this
% footnote is not yet sufficiently long enough to make two
% lines of footnote text.}
% \fntext[fn3]{Yet another author footnote.}
\address[1]{Czech Technical University in Prague, Faculty of Civil Engineering, Department of Mechanics, Thákurova 2077/7, 166 29 Prague 6, Czechia}
\address[2]{Czech Academy of Sciences, Institute of Information Theory and Automation, Pod vod\'{a}renskou
v\v{e}\v{z}\'{\i}~4, 182~00~Prague~8, Czechia}
%\affiliation[1,2,3]{organization={huhu},%Department and Organization
%             addressline={1}, 
%             city={ag},
%             postcode={10}, 
%%             state={vgy},
%             country={huhu}}

\begin{abstract}
The paper develops a new integral micromorphic elastic continuum model, which can describe dispersion properties of band-gap metamaterials, i.e., metamaterials that inhibit propagation of waves in a certain frequency range. The enrichment consists in nonlocal treatment of three terms in the 
expression for the potential energy density of the standard micromorphic continuum. 
After proper calibration, such a formulation can \textbf{exactly} reproduce two given branches of the dispersion curve (acoustic and optical), even in cases with a band gap. The calibration process exploits Fourier images of the unknown weight functions, which are analytically deduced from the dispersion relation of the material of interest. The weight functions are then reconstructed in the spatial domain by numerical evaluation of the inverse Fourier transform. The presented approach is validated on several examples, including discrete mass-spring chains with alternating masses, for which the dispersion relation has an explicit analytical form and the optical and acoustic branches are separated by a band gap.     

% Moreover, it is explained how the unknown weight functions are determined for a given dispersion relation of a certain metamaterial. This procedure is demonstrated on the particular case of a discrete mass-spring chain with alternating masses, for which the dispersion relation can be analytically derived and a band gap in the dispersion curve is observed.       
 
\end{abstract}


% \begin{highlights}
% \end{highlights}

\begin{keyword}
Micromorphic model \sep nonlocal continuum \sep dispersion \sep band gap



\end{keyword}

\end{frontmatter}


\section{Introduction}
\subsection{Background}
Over the past few years, there has been growing interest in mechanical metamaterials, stimulated by their ability to exhibit unique and customized properties. These materials are typically constructed using repeating unit cells arranged in a specific pattern. The advancement of 3D printing techniques has played a crucial role in developing and producing metamaterials, offering an efficient and reproducible manufacturing method \cite{nevzerka2018jigsaw,lei20193d}.
Among the various types of metamaterials, the Locally Resonant Acoustic Metamaterial (LRAM) has received significant attention \cite{craster2012acoustic, ma2016acoustic,sugino2016mechanism}. LRAM exhibits a dispersion curve with multiple branches, known as the acoustic and optical branches, which are separated by a band gap. As a result, LRAM can effectively attenuate acoustic waves within a specific frequency range. This unique property opens up numerous possibilities for engineering applications, including acoustic cloaking, acoustic filters, and vibration isolation control \cite{norris2015acoustic, miniaci2016large}.

The development of metamaterials heavily relies on the availability of efficient and reliable computational tools that enable numerical simulation of their behavior and properties. One commonly used technique is direct numerical simulation, typically based on the finite element method, which takes into account the complete microstructure of the material. However, when dealing with large metamaterial structures, the computational costs associated with representing each unit cell's geometric details can quickly become prohibitive, making direct numerical simulations impractical for engineering applications.
To address this challenge, homogenization techniques tailored to capture the behavior of Locally Resonant Acoustic Metamaterials have been developed, see, e.g., \cite{roca2019computational,liu2021computational,van2019transient}.

In addition to homogenization methods, various phenomenological theories have been proposed to incorporate the influence of material microstructure on the macroscopic behavior, e.g., on wave propagation. Extended continuum theories, such as higher-gradient elasticity \cite{askes2008four, fish2002non} and integral-type nonlocal elasticity \cite{eringen1972linear, lim2015higher}, have been introduced as approaches to account for the effects of microstructure. A comprehensive investigation of dispersion properties for various integral- and gradient-type elasticity theories was presented in \cite{jirasek2004nonlocal}. Even though gradient and integral continuum enrichments can model dispersive behavior, their dispersion diagrams have only a single branch.
To capture the acoustic and optical branches in the dispersion diagram, an alternative approach is to utilize the micromorphic theory \cite{eringen1964nonlinear,mindlin}. Micromorphic continua introduce additional degrees of freedom that describe the deformation of the microstructure and offer a more comprehensive representation. However, classical micromorphic continua, including well-known models such as the Cosserat continuum \cite{cosserat1909theorie,herrmann1968applications}, microstretch continuum \cite{eringen1990theory, tomar2006propagation}, and full micromorphic continuum \cite{eringen1964nonlinear, berezovski2013dispersive}, are unable to capture the band-gap phenomenon, as demonstrated in \cite{madeo2016complete}. Furthermore, a detailed analysis of wave propagation in one-dimensional micromorphic continua, presented in \cite{berezovski2013dispersive}, did not report any band gap. \new{
A one-dimensional micromorphic model with two internal variables was studied in \cite{berezovski2020wave}, and it was demonstrated that this formulation can capture the band-gap phenomenon. In the variational derivation of the model, the kinetic energy density was constructed using a non-positive-definite mass matrix supplemented by a gyroscopic term, which enables the presence of multiple branches in the dispersion curve. In the conclusions, the authors pointed out that the mass matrix must be positive definite, and they left the removal of this deficiency for future investigations. 
% The researchers provide two alternative derivations of the model, one based on the balance of linear momentum and Clausius-Duhem inequality and the second one based on the variational principle. In the latter approach, the kinetic energy density is constructed using a nonpositive-definite mass matrix supplemented by a gyroscopic term, which enables the presence of multiple branches in the dispersion curve. In the conclusions, the authors point out that the mass matrix must be positive definite, leaving this deficiency for future investigations. 
}

Recently, a novel relaxed micromorphic continuum has been introduced to effectively capture the band-gap phenomenon \cite{neff2014unifying}. Extensive research on this continuum model has been conducted in a series of papers, see, e.g., \cite{ghiba2015relaxed, madeo2015band, neff2017real, madeo2017role, d2020effective, ghiba2021existence}. However, it is important to note that in one-dimensional scenarios, the micromorphic effects of the relaxed micromorphic continuum vanish since they describe relative rotations between the microstructure and the macroscopic matrix. In \cite{madeo2016complete}, the authors claimed that the relaxed micromorphic model was the only nonlocal continuum model capable of describing band gaps. Nevertheless, in a subsequent study \cite{madeo2017role}, the same authors demonstrated that by incorporating a term with mixed temporal and spatial derivatives of displacement, classical micromorphic models could also account for the band-gap phenomenon. 

A micromorphic model with a mixed temporal-spatial derivative of the micromorphic variable, developed based on microstructural arguments, was proposed for modeling band gaps in \cite{nejadsadeghi2020role}.
In addition, the so-called Maxwell-Rayleigh model \cite{sherbiny2018bandgaps} was shown to simulate band gaps, with experimental verification in \cite{placidi2021experbandgap}. This model possesses an appealing physical background. However, it only introduces an enhancement in the inertia term and does not consider the effects of higher-order gradients on the potential energy. Consequently, it cannot predict size effects typically observed in materials with microstructure.

Recently, a new micromorphic integral model was proposed in \cite{jirasek2022integral}, which combines the micromorphic continuum theory with nonlocal integral averaging. This model can represent two branches of the dispersion curve separated by a band gap. Furthermore, it was shown in \cite{jirasek2022integral} that, for the standard local micromorphic continuum in 1D, a band gap appears only when the micromorphic modulus vanishes, i.e., in the particular case when the kinetic energy is enriched by a term dependent on the rate of the micromorphic variable and the potential energy by a penalty term that forces the micromorphic variable to remain close to the local strain but {\bf not} by a term dependent on the spatial gradient of the micromorphic variable. Moreover, the authors showed that, in this special case, the micromorphic model leads to the same one-dimensional dispersion relation as the Maxwell-Rayleigh model in the form presented in \cite{sherbiny2018bandgaps}. Both models also share another feature: under static conditions, they reduce to the standard elastic continuum. 

\new{ Furthermore, the authors of \cite{placidi2024variational} obtained a different version of the Maxwell-Rayleigh-type model by heuristic homogenization of a one-dimensional mass-spring chain with internal resonators. Such a model also leads to a band gap in the dispersion curve, tunable by two parameters. 
% The paper discusses also solution of the standing waves for the chain with finite length and application of the findings for design of tensegrity structures.
This approach can be classified as a subclass of continualization methods \cite{askes2005higher}. In the standard formulation, either the equations of motion or the Lagrange functional describing a specific discrete structure are transformed into their continuum counterparts by approximating the displacement difference of neighboring particles by a truncated Taylor expansion of a continuous macro displacement field, leading to variants of gradient continuum models. However, due to the truncation of the Taylor expansion, the resulting continuum potential energy density functional might lose positive definiteness, leading to a mathematically ill-posed problem. Therefore, enhanced continualization schemes were developed to tackle such pathological effects. For instance, the authors of \cite{bacigalupo2019generalized} introduced a regularized continualization technique preserving positive definiteness of the potential energy density functional for one-dimensional beam lattices and further generalized the procedure to multidimensional beam-like lattices \cite{bacigalupo2021dynamic} and to block-lattice materials \cite{diana2023thermodinamically}. By considering an increasing number of terms in the expansion, a gradient continuum of increasing order can be obtained, leading to a better match of the dispersion properties of the discrete lattice and the corresponding enriched continuum model. In the limit case, when infinitely many terms are considered in the expansion series, the resulting equations of motions can alternatively be obtained from an integral-type continuum with a specific averaging kernel, and the dispersion properties of such an enriched continuum model would be identical to those of the analyzed discrete structure. }

\new{
In the present paper, we propose an enriched continuum model that combines the nonlocal integral continuum theory with the micromorphic approach. This model is able to \textbf{exactly} reproduce the dispersive behavior of an arbitrary metamaterial with two branches in the dispersion diagram, including cases in which the branches are separated by a band gap. For calibration of the averaging kernels, only the knowledge of the standard macroscopic material parameters and of the dispersion relation of the analyzed material is needed. This is a novelty compared to the attempts to capture the band gap found in the literature, which either provide only an approximation of the dispersion relation, or require the knowledge of the microstructure (e.g., in the continualization methods).  
The whole paper is dedicated exclusively to infinite domains; therefore, the question of suitable boundary conditions for finite domains is left open for future research. Similarly, wave attenuation caused by damping is not considered in this work. In \cite{roubivcek2024some} the author studied wave dispersion and attenuation for various visco-elastic models enriched by gradient extensions. A similar analysis could also be performed for the micromorphic and integral extensions presented in this paper, but this remains to be explored in the future.   
%
% In the previously mentioned procedure, the form of the enriched continuum model is taylored to a specific discrete model. Opposed to that, the aim of our paper is to introduce generalized continuum model with fixed form of the potential energy density functional, which can exactly reproduce dispersive behavior of arbitrary material based only on the information of standard macroscopic material parameters and of the dispersion relation of the analysed material.  
}

% \new{
% % There are various open topics for a possible future research which are not considered in the present work. 
% Whole paper is dedicated purely to infinite domains; therefore, the question of admissible boundary conditions for finite domains is left open and will be treated in the upcoming publication. Similarly, the topic of wave attenuation and damping is not considered in this paper. In \cite{roubivcek2024some} the author studies wave dispersion and attenuation for various visco-elastic models enhanced by .    
% }

%{\bf ???}
%the classical micromorphic continuum can the 1D micromorphic model is generalized by incorporation of nonlocal averaging in two terms in the free energy density function. 

\begin{table}
\begin{center}
\begin{tabular}{|ll|}
\hline
% \multicolumn{2}{|l|}{\textbf{Nomenclature}}    \\ 
\multicolumn{1}{|l|}{\textbf{symbol}}  & \textbf{meaning}\\
\hline
\multicolumn{1}{|l|}{$x,\ \xi$}  & space coordinates\\
\multicolumn{1}{|l|}{$r=|x-\xi|$}  & spatial distance of points $x$ and $\xi$\\
\multicolumn{1}{|l|}{$t$}  & time coordinate\\
\multicolumn{1}{|l|}{$\bar{E}$} & elastic modulus  \\
\multicolumn{1}{|l|}{$\rho$} & mass density \\
\multicolumn{1}{|l|}{$\eta$} & micromorphic density \\
\multicolumn{1}{|l|}{$\tau = \sqrt{\eta / \bar{E}}$}  & characteristic time\\
\multicolumn{1}{|l|}{$l = \sqrt{\eta / \rho}$}  & characteristic length\\
\multicolumn{1}{|l|}{$\tilde r = r/l$}  & dimensionless distance\\
\multicolumn{1}{|l|}{$i$}  & imaginary unit\\
\multicolumn{1}{|l|}{$k$}  & wave number\\
\multicolumn{1}{|l|}{$\tilde k = l k$}  & dimensionless wave number\\
\multicolumn{1}{|l|}{$\omega$}  & circular frequency\\
\multicolumn{1}{|l|}{$\omega_1$, $\omega_2$}  & circular frequency of acoustic and optical branch\\
\multicolumn{1}{|l|}{$\tilde\omega = \tau \omega$}  & dimensionless circular frequency\\
\multicolumn{1}{|l|}{$u$}  & displacement field\\
\multicolumn{1}{|l|}{$\hat{u}$}  & amplitude of harmonic displacement field\\
\multicolumn{1}{|l|}{$\chi$} & micromorphic variable field \\
\multicolumn{1}{|l|}{$\hat{\chi}$}  & amplitude of harmonic micromorphic strain field\\
\multicolumn{1}{|l|}{$\bar{H}$} & coupling stiffness \\ 
\multicolumn{1}{|l|}{$\bar{A}$} & micromorphic stiffness \\ 
\multicolumn{1}{|l|}{$l_s = \sqrt{\bar{A} / \bar{E}}$}  & characteristic length related to stiffness\\
\multicolumn{1}{|l|}{$\kappa = {\bar{H} / \bar{E}}$}  & ratio of coupling micromorphic stiffness and the standard stiffness\\
\multicolumn{1}{|l|}{$B,\ C$} & additional material parameters \\ 
\multicolumn{1}{|l|}{$\Psi$} & free energy \\ 
\multicolumn{1}{|l|}{$\bar{\Psi}$} & spatially averaged free energy \\
\multicolumn{1}{|l|}{$\bar{\Psi}_E$,\ $\bar{\Psi}_A$,\ $\bar{\Psi}_H$} & contributions of the standard, micromorphic, and coupling terms to  spatially averaged free energy\\ 
% \multicolumn{1}{|l|}{} & spatially averaged free energy\\
\multicolumn{1}{|l|}{$T$} & kinetic energy \\ 
\multicolumn{1}{|l|}{$\bar{T}$} & spatially averaged kinetic energy \\ 
\multicolumn{1}{|l|}{$\bar{T}_{\rho}$, $\bar{T}_{\eta}$} & standard and micromorphic contributions to spatially averaged kinetic energy \\ 
\multicolumn{1}{|l|}{$\mathcal{S}$} & action functional \\ 
\multicolumn{1}{|l|}{$E,\ A,\ H_1,\ H_2$} & weight functions \\ 
\multicolumn{1}{|l|}{$E_s,\ A_s,\ H_{1,s},\ H_{2,s}$} & symmetrized weight functions \\ 
\multicolumn{1}{|l|}{$E_0,\ A_0,\ H_{1,0},\ H_{2,0}$} & weight functions dependent only on a single variable (distance) \\ 
\multicolumn{1}{|l|}{$\cE$,\ $\cA$,\ $\cH_1$,\ $\cH_2$} & Fourier images of weight functions $E_0$,\ $A_0$,\ $H_{1,0}$,\ $H_{2,0}$ \\ 
\multicolumn{1}{|l|}{$\cH$} & Fourier image of weight function $H_{1,0}=H_{2,0}$ \\ 
\multicolumn{1}{|l|}{$\tilde E,\ \tilde A,\ \tilde H$} & dimensionless weight functions \\ 
\multicolumn{1}{|l|}{$\tilde \cE$,\ $\tilde \cA$,\ $\tilde \cH$} & dimensionless version of Fourier images $\cE$,\ $\cA$,\ $\cH$ \\ 
\multicolumn{1}{|l|}{$\gamma=\tilde \cA/\tilde \cE$} & ratio of dimensionless Fourier images $\tilde \cA$ and $\tilde \cE$\\
\multicolumn{1}{|l|}{$m,\ M$} & masses of the mass-spring chain\\ 
\multicolumn{1}{|l|}{$\beta = m/M$} & ratio of the masses $m$ and $M$\\ 
\multicolumn{1}{|l|}{$K,\ K_1,\ K_2$} & stiffnesses of the mass-spring chains\\ 
\multicolumn{1}{|l|}{$\xi = K_2/K_1$} & stiffness ratio of the mass-spring chain with second-nearest-neighbour interactions\\ 
\multicolumn{1}{|l|}{$a$} & mass-spring chain particle distance\\ 
\multicolumn{1}{|l|}{$\lambda=l/a$} & ratio of characteristic length and particle distance\\ 
\multicolumn{1}{|l|}{$\tilde k_0$} & length of the sampling interval of the discrete Fourier transform\\
\multicolumn{1}{|l|}{$N$} & number of sampling points of the discrete Fourier transform\\ 
\multicolumn{1}{|l|}{$h$} & spacing in the space domain of the discrete Fourier transform\\ 
\multicolumn{1}{|l|}{$\phi_j$, $\Phi_j$} & $j$-th hat function and its Fourier transform\\ 
\hline
\end{tabular}
\end{center}
\caption{\new{List of symbols}}
\end{table}

\subsection{Motivation}
In the present work, we build upon the results obtained with the micromorphic integral model introduced in \cite{jirasek2022integral} and propose its further generalization. Nonlocal averaging is now applied to all three free energy density function terms, i.e., the strain-related term, the micromorphic gradient term, and the coupling term. 
Each of the three nonlocal weight functions is then determined based on the given dispersion relation of the material to be reproduced. It turns out that by properly adjusting the weight functions, the present enriched continuum model can be tuned up to reproduce an arbitrary one-dimensional dispersion diagram consisting of two branches.

The developed identification procedure will be validated on three simple test cases for which the dispersion relation is available in an explicit analytical form. These examples will start from a discrete mass-spring chain with two alternating masses---perhaps one of the simplest discrete models showing a band gap in the dispersion curve. The second example will look at the micromorphic continuum with a vanishing micromorphic modulus, which is the only case of the standard 1D micromorphic continuum model that leads to a band gap. Finally, a mass-spring chain with two alternating masses and second-nearest-neighbor interactions will be considered. Based on the provided dispersion relations, the appropriate weight functions of the proposed micromorphic integral continuum will be constructed. The error will be assessed by comparison of the originally given dispersion relations with those obtained based on the present model. 
% The paper is structured as follows. Firstly, the equations of motion for the proposed integral micromorphic continuum are derived using the Hamilton variational principle. Subsequently, assuming  harmonic wave propagation, the dispersion equation 
% is obtained, and explicit formulas for the Fourier images of the unknown weight functions are then derived. Afterwards in the results section the dispersion relation of the double mass-spring chain is exactly reproduced by the proposed enriched continuum model.
% Afterwards in the results section 
For simplicity, and to allow for analytical derivations as much as possible, the present study will be performed in the one-dimensional setting.

Before diving into the derivation of the model, we outline the line of reasoning and motivate the proposed form of the potential and kinetic energies. 
In \cite{nejadsadeghi2020role}, using micro-mechanical arguments, the authors developed a micromorphic model with a {\em mixed temporal-spatial derivative of the micromorphic variable}, governed by  equations of motion
\bea\label{eq_mix1mod}
\rho \ddot{u}(x,t) &=& (\bar E +\bar H)\, u''(x,t) -\bar H\chi'(x,t) \\ \label{eq_mix2mod}
\eta\ddot{\chi}(x,t)-C\ddot{\chi}''(x,t) &=& \bar A\chi''(x,t)-\bar H\chi(x,t)+\bar Hu'(x,t)
\eea
Here, $x$ is the spatial coordinate and $t$ is the time, 
primes denote spatial derivatives and overdots time derivatives,
$u(x,t)$ is the macroscopic displacement field and $\chi(x,t)$ is the micromorphic variable field, which has the physical meaning of micro-level deformation, $\rho$ is the macroscopic mass density, $\eta$ and $C$ stand for nonstandard micromorphic density measures, which can be computed based on the given microscopic density field, $\bar E$ denotes the elastic modulus, and $\bar A$ and $\bar H$ are additional micromorphic stiffness parameters. 

In a subsequent study \cite{madeo2017role}, another micromorphic model incorporating a  {\em mixed temporal and spatial derivative of the displacement field} was proposed and analyzed. The governing equations of that model read
\bea\label{eq_mix1}
\rho \ddot{u}(x,t) - B\ddot{u}''(x,t) &=& (\bar E +\bar H)\, u''(x,t) -\bar H\chi'(x,t) \\\label{eq_mix2}
\eta\ddot{\chi}(x,t) &=& \bar A\chi''(x,t)-\bar H\chi(x,t)+\bar Hu'(x,t)
\eea
where $B$ is an additional material parameter.
Interestingly, both enhanced micromorphic models, i.e., model (\ref{eq_mix1mod})--(\ref{eq_mix2mod}) with a mixed derivative of the micromorphic variable $\chi$ and model (\ref{eq_mix1})--(\ref{eq_mix2}) with a mixed derivative of displacement $u$, lead to a band gap in the dispersion diagram.  Note that the introduction of mixed derivatives can be traced back to \cite{fish2002non}, where a higher-order mathematical homogenization theory was used to derive a nonlocal dispersive model with mixed temporal and spatial derivatives of the displacement for wave propagation in heterogeneous materials.
  %Symbol ${\cal L}$ is the one-dimensional domain representing the analyzed body (here we consider ${\cal L}=(-\infty,\infty)$). The superimposed dot denotes a derivative with respect to time. The same equations of motions can also be derived from the postulated free and kinetic energy densities of the form
%\bea \label{free_en_mic} 
%\Psi(x,t) &=& \intL \frac{1}{2}\bar Eu'(x,t)^2 \dx  +\intL \frac{1}{2}\bar A \chi'(x,t)^2 \dx+\intL \frac{1}{2} \bar H (u'(x,t)-\chi(x,t))^2 \dx\\
%{T}(x,t) &=& \intL \frac{1}{2}\rho \dot{u}(x,t)^2 \dx +\intL \frac{1}{2}\eta \dot{\chi}(x,t)^2 \dx  + \intL \frac{1}{2} C \dot{\chi}'(x,t)^2 \dx
% + \intL \frac{1}{2} B \dot{u}'(x,t)^2 \dx
%\eea  
%Except for the last term in the kinetic energy function, containing the mixed spatial-temporal derivative of the micromorphic variable field, the previous equations have the standard form describing the micromorphic continuum theory. Motivated by the work \cite{fish2002non}, also the term with a mixed spatial-temporal derivative  of the displacement field can be added to the kinetic energy function, which then reads  
%\beq\label{kinEnFun}
%{T}(x,t) = \intL \frac{1}{2}\rho \dot{u}(x,t)^2 \dx +\intL \frac{1}{2}\eta \dot{\chi}(x,t)^2 \dx  + \intL \frac{1}{2} C \dot{\chi}'(x,t)^2 \dx
%+ \intL \frac{1}{2} B \dot{u}'(x,t)^2 \dx
%\eeq  
%where $B$ is an additional material parameter. Subsequently, the derived equations of motion yield

It is now straightforward to combine both above-mentioned models into a generalized micromorphic model with mixed temporal and spatial derivatives of the displacement and micromorphic fields,
described by equations of motion
\bea\label{eq_mix1x}
\rho \ddot{u}(x,t) - B\ddot{u}''(x,t) &=& (\bar E +\bar H)\, u''(x,t) -\bar H\chi'(x,t) \\\label{eq_mix2x}
\eta\ddot{\chi}(x,t)-C\ddot{\chi}''(x,t) &=& \bar A\chi''(x,t)-\bar H\chi(x,t)+\bar Hu'(x,t)
\eea
However, for this kind of model, the dispersion diagram has certain features encoded in the formulation, which cannot be changed even if the material parameters are adjusted. For example, this model cannot capture the dispersion diagram of the double-mass-spring model, which is one of the simplest one-dimensional cases that exhibit a band gap. 

Based on ideas from \cite{jirasek2004nonlocal}, differential equations (\ref{eq_mix1x})--(\ref{eq_mix2x}) may alternatively be reformulated in the integral form
\bea
\rho \ddot{u}(x,t) &=& (\bar E+\bar H)\intL G_1(x,\xi)u''(\xi,t)\dxi  - \bar H \intL G_1(x,\xi)\chi'(\xi,t)\dxi \label{iem1}
\\
\eta\ddot{\chi}(x,t) &=&\bar A \intL G_2(x,\xi)\chi''(\xi,t)\dxi +\bar H\intL G_2(x,\xi)u'(\xi,t)\dxi -\bar H \intL G_{2}(x,\xi)\chi(\xi,t) \dxi \label{iem2}
\eea 
in which ${\cal L}$ is the one-dimensional spatial domain (interval) on which the problem is solved, and $G_1(x,\xi)$ and $G_2(x,\xi)$ denote the Green functions of the spatial differential operators that convert the 
left-hand sides of \eqref{iem1} and \eqref{iem2} into the left-hand sides of 
\eqref{eq_mix1x} and \eqref{eq_mix2x},
respectively,
i.e., of operators ${\cal I}-(B/\rho)\,\partial^2/\partial^2x$ and  ${\cal I}-(C/\rho)\,\partial^2/\partial^2x$ where ${\cal I}$ is the identity operator.
Equations of motion \eqref{iem1}--\eqref{iem2} are taken as the starting point and further generalized by replacing the Green functions $G_1$ and $G_2$ by arbitrary averaging kernels, which enhances the flexibility of the model.
%In \cite{jirasek2004nonlocal}, it is shown that model \cite{fish2002non}, which is derived from standard potential energy and kinetic energy enriched by the term with mixed spatial-temporal derivative, can alternatively also be obtained by introducing nonlocal averaging in the free energy function while keeping the kinetic energy in the classical form. Therefore, the integral averaging by a special weight function in the free energy function is, in a sense, an alternative to the local formulation with enriched kinetic energy. This statement can also be generalized for the case of a micromorphic continuum derived from free and kinetic energies \eqref{free_en_mic} and \eqref{kinEnFun}. This motivates the proposed model for which the averaging is introduced in all the three terms of the micromorphic free energy  \eqref{free_en_mic}. As will be shown later, the equations of motion \eqref{iem1} and \eqref{iem2} can be derived from the proposed general integral micromorphic model with a special choice of weight functions. 

\section{General One-Dimensional Integral Micromorphic Model}\label{sec:2}

\subsection{Model formulation}
% The proposed model is a straightforward generalization of the previously mentioned one; its free energy function reads  
% with free and kinetic energies \eqref{psi1} and \eqref{T1}.  
Instead of directly postulating a generalized form of equations of motion \eqref{iem1}--\eqref{iem2}, we will use
an energy-based approach.
For a standard (local) micromorphic model, the free energy is expressed as
\beq \label{eq9}
\Psi = \intL \left(\half\bar Eu'^2(x)+\half\bar A\chi'^2(x)+\half\bar H(u'(x)-\chi(x))^2\right)\dx
\eeq 
where the first term in the integrand represents the energy stored in strain, the second term is the energy stored in the gradient of micromorphic strain, and the third term penalizes the difference between the strain evaluated from the displacement field and the micromorphic strain.

The newly proposed nonlocal version of the micromorphic model defines the free energy as 
%\beq\label{free_en}
%\Psi(x,t) = \intL \frac{1}{2}\sigma(x,t) u'(x,t)\dx  + \intL \frac{1}{2}m(x,t) \chi'(x,t)\dx  + \intL \frac{1}{2}s(x,t) \left(u'(x,t)-\chi(x,t)\right)\dx
%\eeq
%where
%\bea
%\sigma(x,t) &=& \intL E(x,\xi)u'(\xi,t) \;{\rm d}\xi \\
%m(x,t) &=& \intL A(x,\xi)\chi'(\xi,t) \;{\rm d}\xi
%\\
%s(x,t) &=& \intL \left(H_1(x,\xi)u'(\xi,t)-H_2(x,\xi)\chi(\xi,t)%\right) \;{\rm d}\xi
%\eea
\bea\nonumber
\Psi(t) &=& \intL \half\left(\intL E(x,\xi)u'(\xi,t) \;{\rm d}\xi\right) u'(x,t)\dx  + \intL \half\left(\intL A(x,\xi)\chi'(\xi,t) \;{\rm d}\xi\right) \chi'(x,t)\dx  +
\\
&&+ \intL \half\left(\intL \left(H_1(x,\xi)u'(\xi,t)-H_2(x,\xi)\chi(\xi,t)\right) \;{\rm d}\xi\right) \left(u'(x,t)-\chi(x,t)\right)\dx
\label{free_en}
\eea
%\bea
%\sigma(x,t) &=&  \\
%m(x,t) &=& 
%\\
%s(x,t) &=& 
%\eea
%are the generalized stress-like nonlocal quantities. 
where $E(x,\xi)$, $A(x,\xi)$, $H_1(x,\xi)$ and $H_2(x,\xi)$ are suitable weight functions, and it is marked explicitly that 
the displacement, micromorphic strain and free energy vary in time. 
% Note that compared to Equation\ \eqref{psi1} the parameters $\bar E$, $\bar A$ are included in the
% Function $u(x,t)$ denotes the displacement field and $\chi(x,t)$ stands for the micromorphic variable field, which introduces the additional degree of freedom into the model. 
% Symbol ${\cal L}$ is the one-dimensional domain representing the analyzed body (here we consider ${\cal L}=(-\infty,\infty)$).    
The kinetic energy 
%is generalized only in the micromorphic (local) sense and is given by
\begin{equation}\label{eq3}
{T}(t) = \intL \left(\half\rho{\dot{u}(x,t)}^2 + \half\eta{\dot{\chi}(x,t)}^2 \right)\; \dx
\end{equation}
consists of a standard term related to the displacement rate and
a micromorphic enhancement related to the rate of the micromorphic strain.
% where  $\rho$ is the standard mass density and $\eta$ is a nonstandard, micromorphic density.
% The superimposed dot denotes a derivative with respect to time.
We are interested in the structure of the governing equations
and, for simplicity, we assume that the weight functions as well as the kinematic fields are sufficiently regular, so that all integrals involved in the subsequent derivations remain finite.
Also, ${\cal L}$ should be understood as a finite interval in all
intermediate steps, but the resulting equations can be extended to the whole real axis by a limit process. 

In the spirit of the Hamilton principle, let us define the action functional
\beq 
{\cal S}=\int_{t_1}^{t_2} \left(T(t)-\Psi(t) \right)\dt
\eeq 
and set its first variation to zero, assuming that the state of the system at times $t_1$ and $t_2$ is fixed. Substituting from (\ref{free_en})--(\ref{eq3}),
evaluating the variation $\delta{\cal S}$, integrating by parts with respect to time (to convert the rates of variations $\delta u$ and $\delta\chi$ into the variations), and taking into account that the variations $\delta u$ and $\delta\chi$ at $t_1$ and $t_2$ vanish, we obtain the stationarity condition
\beq \label{eq5}
-\intL\left[\rho\ddot{u}(x,t)\delta u(x,t)+\eta\ddot{\chi}(x,t)\delta\chi(x,t)+  \Sigma(x,t) \delta u'(x,t)\dx + \Theta_0(x,t)  \delta\chi(x,t)\dx+  \Theta_1(x,t)  \delta\chi'(x,t)\right]\dx = 0
\eeq 
in which
\bea
\nonumber
\Sigma(x,t) &=& \intL (E_s(x,\xi)+H_{1,s}(x,\xi))u'(\xi,t)\dxi  -  \half\intL \left(H_2(x,\xi)+ H_1(\xi,x)\right) \chi(\xi,t)\dxi
\\
\nonumber
\Theta_0(x,t) &=&  \intL H_{2,s}(x,\xi)\chi(\xi,t) \dxi - \half\intL \left(H_1(x,\xi)+H_2(\xi,x)\right)u'(\xi,t)\dxi 
\\
\nonumber
\Theta_1(x,t) &=& \intL A_s(x,\xi)\chi'(\xi,t)\dxi
\eea
The added subscript $s$ denotes symmetrization with respect to coordinates $x$ and $\xi$, i.e., for a general function $f(x,\xi)$ we define $f_s(x,\xi)=\half \left(f(x,\xi)+f(\xi,x)\right)$.

In the next step, let us integrate \eqref{eq5} by parts with respect to the spatial coordinate $x$ (to convert the spatial derivatives of variations into the variations), and take into account the independence of variations $\delta u$ and $\delta\chi$.
The resulting strong form of the equations of motion reads
\bea
\rho \ddot{u}(x,t) &=& \left(\intL (E_s(x,\xi)+H_{1,s}(x,\xi))u'(\xi,t)\dxi  -  \half\intL \left(H_2(x,\xi) + H_1(\xi,x)\right)\chi(\xi,t)\dxi\right)' \label{eq.1}
\\
\eta\ddot{\chi}(x,t) &=& \left(\intL A_s(x,\xi)\chi'(\xi,t)\dxi\right)' +\half\intL \left(H_1(x,\xi)+H_2(\xi,x)\right)u'(\xi,t)\dxi -\intL H_{2,s}(x,\xi)\chi(\xi,t) \dxi \label{eq.2}
\eea

If ${\cal L}=(-\infty,\infty)$ and the weight functions depend only on the distance, i.e., if $E(x,\xi)=E_0(x-\xi)$, $A(x,\xi)=A_0(x-\xi)$, $H_{1}(x,\xi)=H_{1,0}(x-\xi)$ and $H_{2}(x,\xi)=H_{2,0}(x-\xi)$ where all functions with subscript 0 are even,
then the ``outer derivative'' on the right-hand sides of (\ref{eq.1}) and (\ref{eq.2}) can be ``shifted'' 
into the nonlocal integrals. The procedure is demonstrated for the first term on the right-hand side of (\ref{eq.1}):
\bea \nonumber
\left(\intL E_0(x-\xi)u'(\xi,t)  \dxi \right)'&=&\intL E_0'(x-\xi)u'(\xi,t)  \dxi
= -\intL E_0'(\xi-x)u'(\xi,t)  \dxi =
\\  &=&
-\left[E_0(\xi-x)u'(\xi,t)\right]_{\xi=-\infty}^\infty + \intL E_0(\xi-x)u''(\xi,t)  \dxi = \intL E_0(x-\xi)u''(\xi,t)  \dxi 
\eea 
Here we have assumed that the weight functions tend to zero as the interaction distance increases, e.g., $E_0(r)\to 0$ as $|r|\to\infty$.   
The remaining terms can be simplified in an analogous way,
and equations \eqref{eq.1} and \eqref{eq.2} can be rewritten as
\bea 
\rho \ddot{u}(x,t) &=& \intL (E_s(x,\xi)+H_{1,s}(x,\xi))u''(\xi,t)\dxi  -  \half\intL \left(H_{2}(x,\xi) + H_{1}(\xi,x)\right)\chi'(\xi,t)\dxi 
\\
\eta\ddot{\chi}(x,t) &=& \intL A_s(x,\xi)\chi''(\xi,t)\dxi +\half\intL \left(H_{1}(x,\xi)+H_{2}(\xi,x)\right)u'(\xi,t)\dxi -\intL H_{2,s}(x,\xi)\chi(\xi,t) \dxi 
\eea 

\bea \label{eq.3}
\rho \ddot{u}(x,t) &=& \intL (E_0(x-\xi)+H_{1,0}(x-\xi))u''(\xi,t)\dxi  -  \half\intL \left(H_{2,0}(x-\xi) + H_{1,0}(\xi-x)\right)\chi'(\xi,t)\dxi 
\\
\eta\ddot{\chi}(x,t) &=& \intL A_0(x-\xi)\chi''(\xi,t)\dxi +\half\intL \left(H_{1,0}(x-\xi)+H_{2,0}(\xi-x)\right)u'(\xi,t)\dxi -\intL H_{2,0}(x-\xi)\chi(\xi,t) \dxi 
\label{eq.4}
\eea 

Comparing the resulting equations of motion (\ref{eq.3})--(\ref{eq.4}) with eqs.~\eqref{iem1}--\eqref{iem2} derived from the micromorphic continuum model with mixed derivatives, it is apparent that 
their structure is very similar. Both formulations 
can be made equivalent (on an infinite domain) if 
the parameters  in (\ref{eq_mix1x})--(\ref{eq_mix2x})
satisfy condition $B/\rho=C/\eta$, which leads to identical Green
functions that can be presented
in the form $G_i(x,\xi)=G_{0}(x-\xi)$, $i=1,2$.
It is then sufficient to set $H_{1,0}(r)=H_{2,0}(r)=\bar{H}\,G_{0}(r)$,
$A(r)=\bar{A}\,G_{0}(r)$ and $E(r)=\bar{E}\,G_0(r)$. In general, the weight functions can have different forms,
which makes the proposed integral formulation more flexible,
leading to a rich family of dispersion diagrams.

\subsection{Dispersion relation}

Let us consider solutions that have the form of a harmonic wave, described in the complex representation by
\bea
u(x,t) &=& \hat{u}\,{\rm e}^{i(kx-\omega t)} \\
\chi(x,t) &=& \hat{\chi}\,{\rm e}^{i(kx-\omega t)} 
\eea
where $\hat{u}$ and $\hat{\chi}$ are the amplitudes of displacement and micromorphic strain, $k$ is the wave number, $\omega$ is the circular frequency, and $i=\sqrt{-1}$ is the imaginary unit. 
Substituting this ansatz into the equations of motion \eqref{eq.3}--\eqref{eq.4}, we obtain, after some manipulations, a set of two homogeneous
linear algebraic equations 
\beq\label{eqHom}
\begin{pmatrix}
\left[\cE(k)+\cH_1(k)\right]k^2-\rho\omega^2 & \half\left[\cH_1(k)+\cH_2(k)\right]ik 
\\
-\half\left[\cH_1(k)+\cH_2(k)\right]ik  &   k^2\cA(k) +\cH_2(k)-\eta\omega^2
\end{pmatrix} \begin{pmatrix}
\hat u 
\\
\hat \chi
\end{pmatrix} = \begin{pmatrix}
0
\\
0
\end{pmatrix}
\eeq
in which $\cE(k)$, $\cA(k)$, $\cH_1(k)$ and $\cH_2(k)$ are Fourier images of weight functions $E_0$, $A_0$, $H_{1,0}$ and $H_{2,0}$, resp.,
considered as dependent only on the distance $r=\vert x-\xi\vert$, which is a reasonable assumption in an infinite
homogeneous body. 
For the present purpose, the Fourier transform ${\cal F}(E_0)={\cal E}$ is defined by
\beq \label{four1}
{\cal E}(k) = \int_{-\infty}^\infty E_0(r)\, {\rm e}^{-ikr}{\rm d}r
\eeq 
and its inverse, $E_0={\cal F}^{-1}({\cal E})$ is evaluated as
\beq \label{four2}
E_0(r) =\frac{1}{2\pi} \int_{-\infty}^\infty {\cal E}(k)\, {\rm e}^{ikr}{\rm d}k 
\eeq 
Since the weight functions are real and even,
their Fourier images are also real and even, and relations (\ref{four1})--(\ref{four2}) can be written in their 
equivalent form
\bea 
{\cal E}(k) &=& \int_{-\infty}^\infty E_0(r)\, \cos kr\, {\rm d}r = 2 \int_0^\infty E_0(r)\, \cos kr\, {\rm d}r \\
E_0(r) &=&\frac{1}{2\pi} \int_{-\infty}^\infty {\cal E}(k)\, \cos kr\,{\rm d}k =  \frac{1}{\pi} \int_{0}^\infty {\cal E}(k)\, \cos kr\,{\rm d}k 
\eea 
The dependence of the Fourier images on the wave number $k$ will not be marked in the subsequent derivation anymore.

Equations \eqref{eqHom} have a nontrivial solution only if
the determinant of the $2\times 2$ matrix on the left-hand side vanishes.
To simplify the analysis, let us assume that $H_1=H_2$ and denote the corresponding Fourier image as $\cH$. The condition of zero determinant then leads to
\beq\label{mm4}
\rho\eta\omega^4 - \left( \rho(\cA k^2+\cH)+\eta(\cE+\cH)k^2\right)\omega^2+\cA(\cE+\cH)k^4+\cE\cH k^2 =0
\eeq
This dispersion equation is quadratic in terms of $\omega^2$ and may provide two positive values of circular frequency $\omega$
for each given positive wave number $k$. Note that $\cE$, $\cA$, and $\cH$ are functions of $k$. The problem can
now be inverted: Suppose that two branches of the dispersion diagram are known and are described by
functions $\omega_1(k)$ and $\omega_2(k)$, with $\omega_1(0)=0$ (acoustic branch) and $\omega_2(0)>0$ (optical branch). We are looking for appropriate weight functions, represented by their Fourier images, such that
the dispersion equation be satisfied for each wave number $k$ combined with the corresponding circular frequency
$\omega_1(k)$, and also with $\omega_2(k)$. Functions  $\cE$, $\cA$, and $\cH$ are now treated as unknowns,
for which we can set up two equations (to be satisfied for all values of $k$):
\bea\label{mm5}
\rho\eta\omega_1^4 - \left( \rho(\cA k^2+\cH)+\eta(\cE+\cH)k^2\right)\omega_1^2+\cA(\cE+\cH)k^4+\cE\cH k^2 &=&0 \\
\label{mm6}
\rho\eta\omega_2^4 - \left(\rho(\cA k^2+\cH)+\eta(\cE+\cH)k^2\right)\omega_2^2+\cA(\cE+\cH)k^4+\cE\cH k^2 &=&0 
\eea
In terms of $\cE$, $\cA$, and $\cH$, these equations are nonlinear, but nonlinearity resides only in the
terms independent of $\omega$. When (\ref{mm5}) is subtracted from (\ref{mm6}), the resulting
equation 
\beq
\rho\eta(\omega_2^4-\omega_1^4) - \left( \rho(\cA k^2+\cH)+\eta(\cE+\cH)k^2\right)(\omega_2^2-\omega_1^2) =0 
\eeq 
is linear (in terms of $\cE$, $\cA$, and $\cH$), and it can be rewritten as
\beq
 \rho(\cA k^2+\cH)+\eta(\cE+\cH)k^2 = \rho\eta(\omega_1^2+\omega_2^2)
\eeq 
It is now easy to express
\beq\label{mm9}
\cH = \frac{\rho\eta(\omega_1^2+\omega_2^2)-(\rho\cA+\eta\cE)k^2}{\rho+\eta k^2}
\eeq
and substitute back into (\ref{mm5}), which yields a seemingly complicated equation
\beq
\rho\eta\omega_1^4 - (\rho\cA+\eta\cE)k^2\omega_1^2
- (\rho+\eta k^2)\omega_1^2\frac{\rho\eta(\omega_1^2+\omega_2^2)-(\rho\cA+\eta\cE)k^2}{\rho+\eta k^2}
+\cA\cE k^4+(\cA k^2+\cE)k^2\frac{\rho\eta(\omega_1^2+\omega_2^2)-(\rho\cA+\eta\cE)k^2}{\rho+\eta k^2}=0
\eeq
However, many terms cancel out and the equation can be simplified to 
\beq\label{eqAE}
- \rho\eta(\rho+\eta k^2)\omega_1^2\omega_2^2
+\rho\eta(\omega_1^2+\omega_2^2)(\cA k^2+\cE)k^2
-\rho\cA^2 k^6-\eta\cE^2 k^4=0
\eeq
which is a quadratic equation with two unknowns, $\cE$ and $\cA$. The mixed term with the product
$\cE\cA$ is missing and the terms with $\cE^2$ and $\cA^2$ have coefficients of the same sign.
Therefore, the set of all solutions is in the $\cE-\cA$ plane graphically represented by an ellipse with
axes parallel to the coordinate axes.

A special case arises for $k=0$, because many coefficients in 
\eqref{eqAE} vanish and the
equation reduces to the identity
\beq
- \rho^2\eta\omega_1^2(0)\omega_2^2(0) = 0
\eeq
which is always satisfied, since $\omega_1(0)=0$. This means that the values of $\cE(0)$ and $\cA(0)$
are arbitrary, but the value of 
\beq
\cH(0) = \eta(\omega_2^2(0)+\omega_1^2(0)) = \eta\omega_2^2(0)
\eeq
can be determined from (\ref{mm9}) with $k$ set to zero. This result indicates that the circular frequency on the optical branch at $k=0$ is linked to the model parameters by
\beq
\omega_2(0) = \sqrt{\frac{\cH(0)}{\eta}}
\eeq
where
\beq 
\cH(0) = \int_{-\infty}^\infty H_0(r)\;{\rm d}r
\eeq 
is the nonlocal counterpart of the coupling modulus $\bar H$ from (\ref{eq9}).

For very small wave numbers $k$, the terms with $k^4$ and $k^6$ can be neglected and \eqref{eqAE}
reduces to 
\beq\label{eq32}
-  \rho\eta(\rho+\eta k^2)\omega_1^2\omega_2^2
+\rho\eta(\omega_1^2+\omega_2^2)\cE k^2
=0
\eeq
Since $\omega_1$ is expected to be of the order of $k$ as $k\to 0^+$, we can neglect terms with $\omega_1^2 k^2$ and also approximate
${\cal E}(k)\,k^2$ by ${\cal E}(0)\,k^2$, neglecting terms
with the third and higher powers of $k$.
Equation \eqref{eq32} is then approximated by
\beq
- \rho^2\eta\omega_1^2\omega_2^2
+\rho\eta\omega_2^2\cE(0) k^2
=0
\eeq
from which
\beq\label{mm17}
\omega_1(k) = \sqrt{\frac{\cE(0)}{\rho}}\,k
\eeq
Here,
\beq\label{eq35}
\cE(0) = \int_{-\infty}^\infty E_0(r)\,{\rm d}r = \bar E
\eeq
plays the role of the macroscopic elastic modulus, and $\sqrt{\cE(0)/\rho}$ is the standard wave speed
of long elastic waves, i.e., the theoretical limit approached by the wave speed as the wave length tends to infinity. Of course, (\ref{mm17}) is only a first-order approximation of function $\omega_1(k)$
valid near $k=0$.

\subsection{Dimensionless form}\label{dimFor}

To simplify further derivations, the relevant equations are cast into a dimensionless form. For easy reference, the dimensions of individual model parameters (including the weight functions) are shown in Table \ref{tab:param}.
\begin{table}[ht]
    \centering
    \begin{tabular}{|c|cccccccc|}
    \hline
        parameter/variable & $\rho$   & $\eta$ & $E_0(r)$ & $H_0(r)$ & $A_0(r)$ & $\cE(k)$, $\bar{E}$ &$\cH(k)$, $\bar{H}$& $\cA(k)$, $\bar{A}$   \\
        \hline
        unit & kg/m$^3$   & kg/m & kg/(m$^2\cdot$s$^2$) &kg/(m$^2\cdot$s$^2$) &kg$/s^2$ & kg/(m$\cdot$s$^2$) &kg/(m$\cdot$s$^2$) &kg$\cdot$ m$/s^2$\\
        \hline
      \end{tabular}
    \caption{Parameters and their units }
    \label{tab:param}
\end{table}
By combining the basic parameters, we can define a characteristic time, e.g., $\tau=\sqrt{\eta/\bar{E}}$, and a characteristic length, e.g., $l=\sqrt{\eta/\rho}$, and then transform the 
circular frequency $\omega$ into a dimensionless
variable $\tilde\omega=\tau\omega$ and the wave number $k$ into a dimensionless variable $\tilde{k}=lk$.
The role of characteristic stress/modulus can be played by the macroscopic elastic modulus $\bar E$.
The dimensionless counterparts of weight functions $E_0(r)$, $A_0(r)$, and $H_0(r)$ are defined as
\bea\label{eq:scaleE}
\tilde E(\tilde r) &=& \dfrac{l}{\bar{E}} {E_0\left(l\tilde r\right)} 
 \\ \label{eq:scaleA}
\tilde A(\tilde r) &=&  \dfrac{1}{l\bar{E}} {A_0\left(l\tilde r\right)} \\ 
\tilde H(\tilde r) &=& \dfrac{l}{\bar{E}}H_0\left(l\tilde r\right) \label{eq:scaleH}
\eea
where $\tilde r = r/l$ is the dimensionless distance,
%For simplicity, we drop subscript 0 at the dimensionless weight functions, because confusion with the weight functions dependent on two coordinates cannot arise.
and their Fourier images are
\bea
\tilde\cE(\tilde k) &=& 
 \dfrac{1}{\bar E}\cE\left(\tilde k /l\right)
\\
\tilde\cA(\tilde k) &=& \dfrac{1}{l^2\bar E } \cA\left(\tilde k/l\right)\\ 
\tilde\cH(\tilde k) &=&\dfrac{1}{\bar E} \cH\left(\tilde k /l\right)
\eea
It is worth noting that, due to the choice of the normalizing factor $\bar E$, the value of $\tilde\cE(0)$ is not arbitrary---it is always equal to 1. In fact, $\tilde\cE(\tilde k)$ is the Fourier
image of the normalized weight function, which is then multiplied
by the physical modulus to construct the actual weight function
that represents the nonlocal stiffness.

After the replacement of all variables by their dimensionless counterparts, Equation\ \eqref{eqAE} is reformulated in the dimensionless form
\beq\label{constEq}
- (1+\tilde k^2)\tilde\omega_1^2\tilde\omega_2^2
+(\tilde\omega_1^2+\tilde\omega_2^2)(\tilde\cA \tilde k^2+\tilde\cE)\tilde k^2
-\tilde\cA^2 \tilde k^6-\tilde\cE^2 \tilde k^4=0
\eeq
Recall that $\tilde\cE$ and $\tilde\cA$ are so far unknown functions of the dimensionless wave number $\tilde k$.
For a given $\tilde k$,
 the set of pairs $[\tilde\cE,\tilde\cA]$ that satisfy \eqref{constEq} can be graphically represented by an ellipse in the $\tilde\cE-\tilde\cA$ space.  
This is best seen if equation \eqref{constEq} is rearranged into the standard form 
\beq\label{ellGen}
\dfrac{\left(\tilde\cA-\dfrac{\tilde\omega_1^2+\tilde\omega_2^2}{2\tilde k^2}\right)^2}{\dfrac{(\tilde\omega_1^2-\tilde\omega_2^2)^2(\tilde k^2+1)}{4\tilde k^6} }+\dfrac{\left(\tilde\cE-\dfrac{\tilde\omega_1^2+\tilde\omega_2^2}{2\tilde k^2}\right)^2}{\dfrac{(\tilde\omega_1^2-\tilde\omega_2^2)^2(\tilde k^2+1)}{4\tilde k^4} } = 1
\eeq
from which one can identify that the center of the ellipse is located at $[\tilde\cE,\tilde\cA]=\left[(\tilde\omega_1^2+\tilde\omega_2^2)/2\tilde k^2,(\tilde\omega_1^2+\tilde\omega_2^2)/2\tilde k^2\right]$ and the length of its axes can be obtained by doubling the square roots of the denominators in \eqref{ellGen}. 
Moreover,  the ratio of the lengths of the axes in the $\tilde\cE$ and the $\tilde\cA$ directions turns out to be equal to $\tilde k$.
Since both denominators in \eqref{ellGen} are always nonnegative, the solution exists for any positive wave number. 
In the special case when $\omega_1(k)=\omega_2(k)$ for some $k$, i.e., when the branches of the dispersion diagram intersect, the solution is unique (the ellipse degenerates into a single point), but in a general case there
exist infinitely many pairs $[\tilde\cE,\tilde\cA]$ satisfying the equation. For each of them, it is possible to evaluate the
corresponding dimensionless function $\tilde\cH$
using the dimensionless version of \eqref{mm9}:
\beq\label{mm9x}
\tilde\cH=\frac{1}{\bar E}
\cH = \frac{\tilde\omega_2^2+\tilde\omega_1^2-(\tilde\cA+\tilde\cE)\,\tilde k^2}{1+\tilde k^2}
\eeq

\subsection{Solution} \label{sol}
In the foregoing derivation it has been shown that, for a given dispersion diagram with two distinct branches and any given wave number, there exist infinitely many solutions $\tilde \cE(\tilde k)$ and $\tilde \cA(\tilde k)$ which satisfy equation \eqref{ellGen}, derived from \eqref{eqAE}. The task is now to choose, for each $\tilde k$, one of the admissible solutions such that the resulting Fourier images of the weight functions, and the weight functions themselves, have meaningful properties. 

To simplify the model and make the solution unique, let us enforce a constant ratio
$\gamma=\tilde\cA(\tilde k)/\tilde\cE(\tilde k)$,
which leads to $\cA(k)=\gamma l^2 \cE(k)$ and eventually to
$A_0(r)=\gamma l^2 E_0(r)$, i.e., to weight functions of the same shape.
If the relation  $\tilde\cA(\tilde k)=\gamma\tilde\cE(\tilde k)$ 
is substituted into equation\ \eqref{constEq}, we obtain
\beq\label{eq45}
-(\gamma^2 \tilde k^6+ \tilde k^4)\,\tilde\cE^2
+(\tilde\omega_1^2+\tilde\omega_2^2)(\gamma \tilde k^2+1)\, \tilde k^2\tilde\cE- (1+\tilde k^2)\,\tilde\omega_1^2\tilde\omega_2^2
=0
\eeq
This is a quadratic equation for the unknown weight function $\tilde\cE$, which has two roots
\beq\label{Efsol0}
\tilde\cE_{1,2} = \dfrac{(\tilde\omega_1^2+\tilde\omega_2^2)(\gamma \tilde k^2+1)  \pm \sqrt{{(\tilde\omega_1^2+\tilde\omega_2^2)}^2{(\gamma \tilde k^2+1)}^2 - 4(\gamma^2 \tilde k^2+ 1)(1+\tilde k^2)\tilde\omega_1^2\tilde\omega_2^2 }}{2(\gamma^2 \tilde k^4+ \tilde k^2)} 
\eeq
The proper root can be selected based on the condition $\tilde\cE(0)=1$,
which originates from the scaling used in the definition of the dimensionless 
weight function $\tilde E(\tilde r)$ and its Fourier image $\tilde\cE(\tilde k)$. For $\tilde k=0$, the denominator of the fraction in 
\eqref{Efsol0} is zero, and the fraction can have a finite limit at $\tilde k\to 0$ only if the numerator vanishes as well. 
The numerator evaluated for $\tilde k=0$ reduces to $\tilde\omega_1^2+\tilde\omega_2^2\pm\vert\tilde\omega_2^2-\tilde\omega_1^2\vert$. If $\tilde \omega_1$
corresponds to the acoustic branch and $\tilde\omega_2$ to the 
optical branch, i.e., if $\tilde\omega_1(0)=0$ and $\tilde\omega_2(0)>0$,
the numerator equals $\tilde\omega_2^2\pm\tilde\omega_2^2$, which vanishes
if the negative sign is used. Therefore, to ensure continuity,
we select the smaller root (even if $\tilde k$ is arbitrary) and write the unique solution as
\beq\label{Efsol}
\tilde\cE = \dfrac{(\tilde\omega_1^2+\tilde\omega_2^2)(\gamma \tilde k^2+1)  - \sqrt{{(\tilde\omega_1^2+\tilde\omega_2^2)}^2{(\gamma \tilde k^2+1)}^2 - 4(\gamma^2 \tilde k^2+ 1)(1+\tilde k^2)\tilde\omega_1^2\tilde\omega_2^2 }}{2(\gamma^2 \tilde k^4+ \tilde k^2)} 
\eeq
However, the selection of the smaller root only makes sure that the numerator
vanishes at $\tilde k=0$ and the fraction becomes of the $0/0$ type,
but it is not yet guaranteed that the fraction tends to 1 as $\tilde k\to 0$. Detailed analysis based on the L'Hospital rule reveals
that the limit value of the fraction is $({\rm d}\tilde\omega_1(0)/{\rm d}\tilde k)^2$ and the normalizing condition is satisfied 
only if ${\rm d}\tilde\omega_1(0)/{\rm d}\tilde k=1$.
This is closely related to the fact that the present choice of
characteristic time and characteristic length leads to a unit 
speed of long elastic waves in the dimensionless representation.

Once function $\tilde\cE$ has been determined, it is straightforward to evaluate the corresponding $\tilde\cA = \gamma \tilde\cE$ and, based on \eqref{mm9x}, also
\beq\label{Hfun}
\tilde\cH = \frac{\tilde\omega_1^2+\tilde\omega_2^2} {1+\tilde k^2}-(1+\gamma)\frac{\tilde k^2 }{1+\tilde k^2} \tilde\cE
\eeq 
Subsequently, the dimensionless wave number $\tilde k$ is replaced by the
product $lk$, and simple scaling leads to the ``true'' Fourier images
$\cE=\bar E\tilde\cE$, $\cA=l^2\bar E\tilde\cE$ and $\cH=\bar E\tilde\cH$. Finally, the inverse Fourier transform yields the
weight functions $E_0$, $A_0$, and $H_0$ in the physical space.

As the weight functions are supposed to be real functions of the distance $|x-\xi|$, and thus even functions of the difference $x-\xi$, their Fourier images should also be real functions.
The outlined procedure may not always give physically meaningful results since the discriminant of quadratic equation \eqref{eq45} may become
negative. Graphically, this means that the ellipse in the $(\cE,\cA)$  space may have
no intersection with the straight line of selected slope $\gamma$.
Interestingly, for the particular choice of $\gamma=1$,
the discriminant is always nonnegative and, moreover, the solution
greatly simplifies and formulae \eqref{Efsol}--\eqref{Hfun} reduce to
\bea 
\tilde\cE &=& \frac{\tilde\omega_1^2}{\tilde k^2} \\
\tilde\cH &=& \frac{\tilde\omega_2^2-\tilde\omega_1^2}{1+\tilde k^2}
\eea 
Here we consider that the subscripts are assigned such that
$\tilde\omega_1\le\tilde\omega_2$.  

In general, a fixed straight line of slope $\gamma$  has
intersections with  all the ellipses in the $(\cE,\cA)$  space that correspond
to all possible values of $\tilde k\ge 0$  only if $\gamma$
is selected in a certain range.
The limiting slopes can be found from the condition of 
zero discriminant of \eqref{eq45}, which is in terms of $\gamma$
again a quadratic equation
\beq\label{eq:53}
\left(\tilde k^4{(\tilde\omega_1^2+\tilde\omega_2^2)}^2-4\tilde k^2(1+\tilde k^2)\tilde\omega_1^2\tilde\omega_2^2\right)\gamma^2+2{(\tilde\omega_1^2+\tilde\omega_2^2)}^2\tilde k^2\gamma+ {(\tilde\omega_1^2+\tilde\omega_2^2)}^2-4(1+\tilde k^2)\tilde\omega_1^2\tilde\omega_2^2=0
\eeq
and its roots,
\bea\label{gamm12}
\gamma_{1,2}&=& \dfrac{-{(\tilde\omega_1^2+\tilde\omega_2^2)}^2\tilde k^2 \pm \sqrt{ {(\tilde\omega_1^2+\tilde\omega_2^2)}^4\tilde k^4- \left(\tilde k^4{(\tilde\omega_1^2+\tilde\omega_2^2)}^2-4\tilde k^2(1+\tilde k^2)\tilde\omega_1^2\tilde\omega_2^2 \right) \left({(\tilde\omega_1^2+\tilde\omega_2^2)}^2-4(1+\tilde k^2)\tilde\omega_1^2\tilde\omega_2^2\right)   }}{\tilde k^4{(\tilde\omega_1^2+\tilde\omega_2^2)}^2-4\tilde k^2(1+\tilde k^2)\tilde\omega_1^2\tilde\omega_2^2} =  \nonumber \\
&=&
\dfrac{-{(\tilde\omega_1^2+\tilde\omega_2^2)}^2\tilde k \pm  
2 (1+\tilde k^2) \tilde\omega_1\tilde\omega_2
{(\tilde\omega_2^2-\tilde\omega_1^2)} }{\tilde k\left[\tilde k^2{(\tilde\omega_1^2+\tilde\omega_2^2)}^2-4(1+\tilde k^2)\tilde\omega_1^2\tilde\omega_2^2\right]}
\eea
are always real. 
Let us order the roots such that $\gamma_1\le \gamma_2$.
Exceptionally, it can happen that the coefficient at $\gamma^2$ in (\ref{eq:53}) vanishes,
and then there is only one root,
\beq \label{eq:55}
 \gamma_1=\frac{ {4(1+\tilde k^2)\tilde\omega_1^2\tilde\omega_2^2-(\tilde\omega_1^2+\tilde\omega_2^2)}^2}{2{(\tilde\omega_1^2+\tilde\omega_2^2)}^2\tilde k^2}
\eeq 
For further discussion, it turns out to be useful to formally set
in this case $\gamma_2=\infty$.

The admissible range of  slopes $\gamma$ for which the expression in \eqref{Efsol} is real is   $[\gamma_1,\gamma_2]$ 
if the coefficient at $\gamma^2$ in (\ref{eq:53}) is negative,
and $[-\infty,\gamma_1] \cup [\gamma_2,\infty]$ if that
coefficient is positive. 
In the exceptional case when the coefficient is zero,
the admissible range is $[\gamma_,\infty)$, where
$\gamma_1$ is given by (\ref{eq:55}). In this case, we formally set
$\gamma_2=\infty$, so that we can use the same description
$[\gamma_1,\gamma_2]$ as for a negative coefficient.
The coefficient at $\gamma^2$ in (\ref{eq:53}) is positive if
\beq \label{eq55}
\vert\tilde\omega_2^2-\tilde\omega_1^2\vert\;\tilde k\;
>\; 2\tilde\omega_1\tilde\omega_2 
\eeq 
Let $\mathcal{K}_a$ be the set of wave numbers $\tilde k\in\mathbb{R}^+$ for which  condition (\ref{eq55}) is satisfied,  and $\mathcal{K}_b=\mathbb{R}^+\setminus \mathcal{K}_a$. 
The set of slopes $\gamma$ that lead to a real 
value of the expression in \eqref{Efsol} for all wave numbers $\tilde k$
can formally be described as
\beq \label{eq:57}
\Gamma = \mathcal{I}_a \cap \mathcal{I}_b
\eeq 
where
\bea
\mathcal{I}_{a} &=& (-\infty,\inf_{\tilde k\in\mathcal{K}_{a}}\gamma_1]\cup[\sup_{\tilde k\in\mathcal{K}_{a}}\gamma_2,\infty)
\\
\mathcal{I}_{b} &=& [\sup_{\tilde k\in\mathcal{K}_{b}}\gamma_1,\inf_{\tilde k\in\mathcal{K}_{b}}\gamma_2]
\eea
In most cases, the set $\Gamma$ needs to be determined numerically. However, we can be sure that it is nonempty,
because it always contains $\gamma=1$.

\section{Results and Discussion}
\subsection{Discrete chain of particles with alternating masses}
\label{sec:doublemasschain}

\subsubsection{Dispersion diagram}

\begin{figure}[H]
\centering%
\includegraphics[width=0.45\textwidth, keepaspectratio=true]{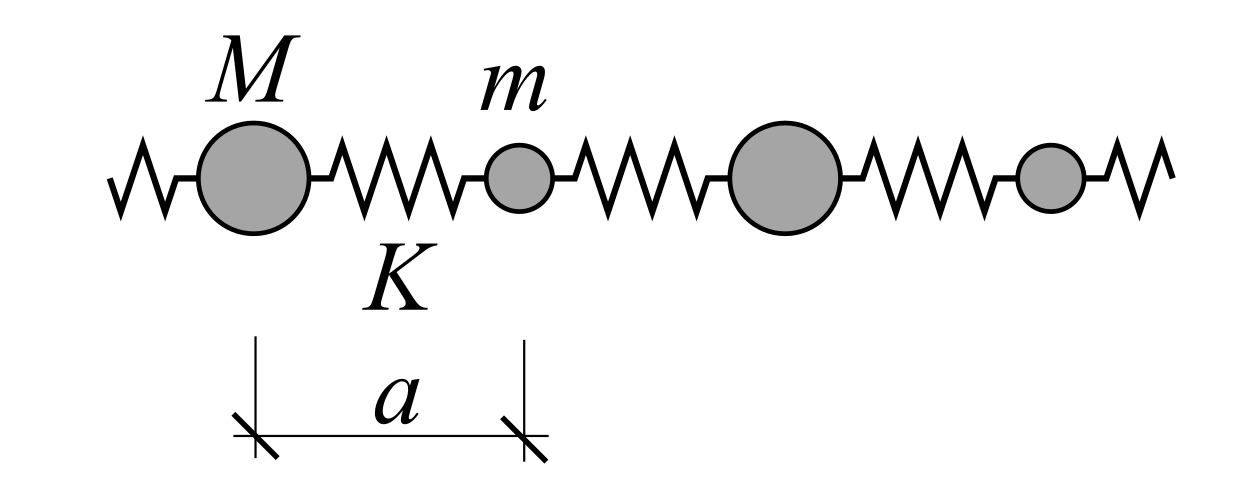}
                \caption{Schematic picture of a mass-spring chain with two alternating masses        and only nearest-neighbor interactions         
                }
                \label{chain_scheme}
\end{figure}

The enriched continuum model proposed in the previous section is now tuned up to reproduce the dispersion curves of the discrete chain of two alternating masses $m$ and $M$ connected by springs with stiffness $K$, shown in Figure \ref{chain_scheme}. The particle spacing is denoted as $a$. 

The dispersion relation of the considered discrete model is given by \cite{ashcroft1976solid}
\begin{equation}\label{eq53}
\omega_{1,2}^2 = \dfrac{K}{m} \left(1+\beta\pm \sqrt{(1+ \beta)^2  - 4\beta\sin^2{ k a}}\right)= \dfrac{K}{m} \left(1+\beta\pm \sqrt{1+\beta^2  +2\beta\cos 2ka}\right)
\end{equation}
where $\beta = m/M$ is the ratio of the two masses. 
To fully specify the integral micromorphic model that has the same dispersion properties, we need to determine parameters
$\rho$ and $\eta$, and also weight functions $E_0$, $H_0$ and
$A_0$. Recall that the integral of
weight function $E_0$ over the whole real axis is the macroscopic modulus $\bar E$ defined in \eqref{eq35}. 

Before we apply the developed procedure that provides 
Fourier images of the weight functions, it is necessary 
to convert the given dispersion diagram into the dimensionless
form. The transformation is based on the choice of a characteristic length $l$ and characteristic time $\tau$, 
which should be related to the properties of the considered discrete model. The choice is not completely arbitrary, because the
acoustic branch of the dimensionless dispersion diagram
should have a unit slope at the origin. The acoustic branch 
$\omega_1$ is obtained by using the negative sign in \eqref{eq53}.
For small wave numbers $k$, the formula can be approximated by
\beq 
\omega_1^2 \approx \dfrac{K}{m} \left(1+\beta- \sqrt{(1+ \beta)^2  - 4\beta k^2 a^2}\right)\approx 
\dfrac{K}{m} \left(1+\beta- (1+ \beta)\left(1-  \frac{1}{2}\frac{4\beta}{(1+\beta)^2} k^2 a^2\right)\right) = \frac{K}{m} \frac{2\beta}{1+\beta}k^2a^2
\eeq 
This indicates that
\beq 
\frac{{\rm d}\omega_1(0)}{{\rm d}k} = \sqrt{\frac{2\beta K}{(1+\beta)m}}\;a = \sqrt{\frac{2K}{M+m}}\;a 
\eeq 
Since the scaling of the circular frequency by characteristic time $\tau$ and of the wave number by characteristic length $l$
leads to ${\rm d}\tilde\omega/{\rm d}\tilde k=(\tau/l)\,{\rm d}\omega/{\rm d}k$, condition ${\rm d}\tilde\omega(0)/{\rm d}\tilde k=1$ is satisfied if
\beq \label{eq64}
\frac{l}{\tau} = \sqrt{\frac{2K}{M+m}}\,a  
\eeq 
The ratio between scaling parameters $l$ and $\tau$ is thus fixed, but one of these parameters can be freely selected and the other is then deduced from (\ref{eq64}). 

Since the length scale of the
discrete model is controlled by the particle spacing $a$, one natural
choice could be $l=a$, leading to $\tau=\sqrt{(M+m)/(2K)}$. 
However, nothing prevents us from setting $l=\lambda a$ where
$\lambda$ is a positive number. The corresponding characteristic time is then $\tau=\lambda\sqrt{(M+m)/(2K)}$
and the dimensionless dispersion relation is given by
\beq \label{eq57}
\tilde\omega_{1,2}^2 = \lambda^2\dfrac{1+\beta}{2\beta} \left(1+\beta\pm \sqrt{(1 +\beta)^2  - 4\beta\sin^2\frac{\tilde k}{\lambda}}\right)
\eeq 

% {\bf Old text, to be deleted:}
% Normalizing the circular frequency and the wave number in the same way as in the continuum model, one obtains 
% % \bea\label{}
% % \tilde\omega_{1,2}^2 &=& \dfrac{K \eta}{E m} \left(1+\beta\pm \sqrt{1+2 \beta +\beta^2  - 4\beta\sin^2{\left( a \sqrt{\frac{\rho}{\eta}}\tilde k  \right)}}\right) \\ 
% % \tilde\omega_{1,2}^2 &=& \alpha \left(1+\beta\pm \sqrt{1+2 \beta +\beta^2  - 4\beta\sin^2{\delta \tilde k }}\right)
% % \eea
% \beq\label{doubmass}
%  \tilde\omega_{1,2}^2 = \alpha \left(1+\beta\pm \sqrt{1+2 \beta +\beta^2  - 4\beta\sin^2{\delta \tilde k }}\right)
% \eeq
% where $\alpha= K \eta/(\bar E m)$ and $\delta = a \sqrt{\rho/\eta}$ are additional dimensionless model parameters. Subsequently, one can show that relations
% \bea
% \tilde\omega_2^2+\tilde\omega_1^2 &=& 2\alpha (1+\beta) \label{omSum} \\
% \tilde\omega_2^2\tilde\omega_1^2 &=& 
% 4\alpha^2\beta \sin^2{\delta \tilde k} \label{omMult}\\ 
% \tilde\omega_2^2-\tilde\omega_1^2 &=& 
% 2\alpha \sqrt{\beta^2  + 2\beta\cos{2\delta \tilde k }+ 1} \label{omDif}
% \eea
% hold. 

\begin{figure}[H]
\centering%
\includegraphics[width=0.55\textwidth, keepaspectratio=true]{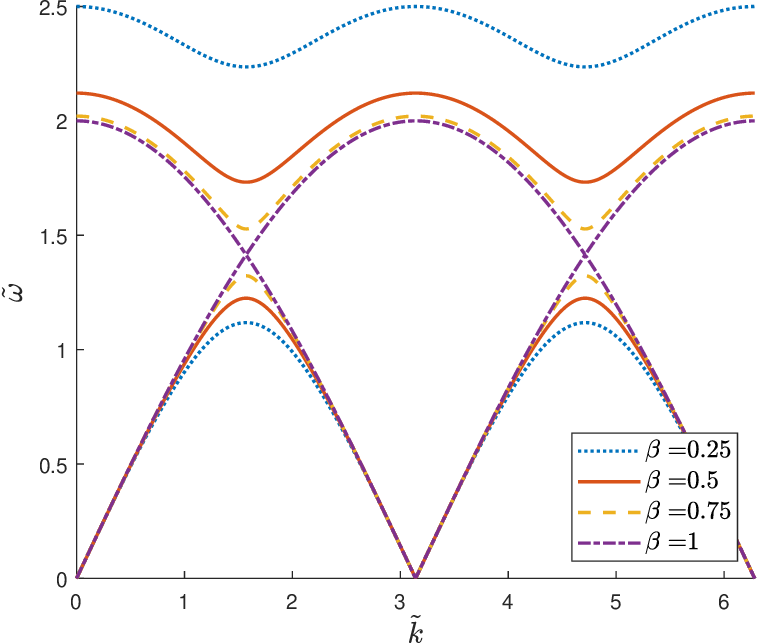}
                \caption{
                % Dispersion diagram of double mass-spring chain, $\alpha=\delta=1$ {\bf Please replot. I suggest to consider $\beta=0.25,0.5,0.75$ and 1. All diagrams should have a unit slope at the origin. Set $\lambda=1$.}
                Dispersion diagram of the mass-spring chain with alternating masses, $\lambda=1$ 
                }
                \label{doubleMassDispRef}
\end{figure}

The dispersion diagrams are visualized in Figure \ref{doubleMassDispRef} for various coefficients $\beta$ and for $\lambda=1$. It is sufficient to consider $\beta\in(0,1]$ because the curves do not change if $\beta$
is replaced by its reciprocal value, $1/\beta$. If $\lambda$ is changed, the curves keep the same shape, 
only the scale on both axes is adjusted by the same factor. 

\begin{figure}[H]
\centering%
\includegraphics[width=0.55\textwidth, keepaspectratio=true]{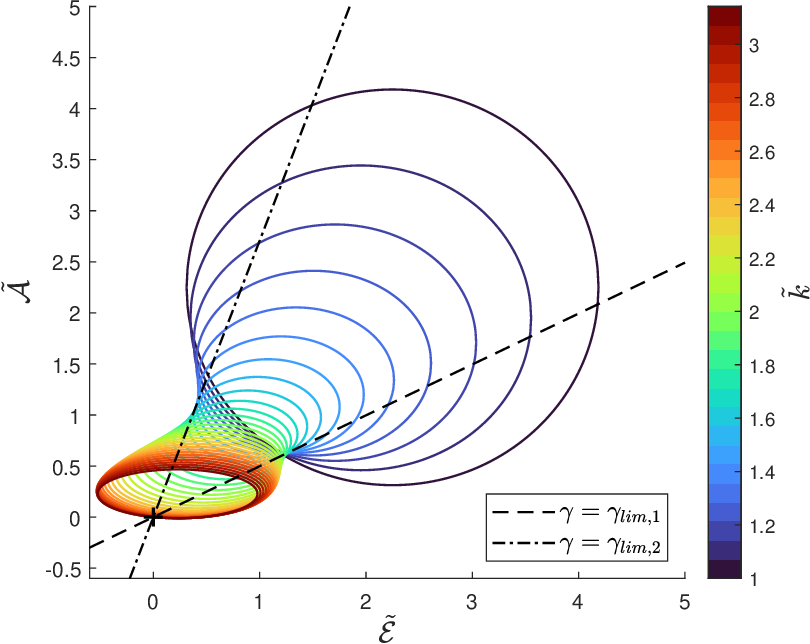}
                \caption{Solutions $\tilde \cA$ and $\tilde \cE$ visualized in the $\tilde\cE-\tilde\cA$ space for various wave numbers, with parameters $\beta=0.5$ and $\lambda=1$ }
                \label{ellipse}
\end{figure}

In the subsequent calculations, we focus on the case of $\beta=0.5$, which means that $m=M/2$. Parameter $\lambda$ is set to~1.
The ellipses that are obtained by substituting \eqref{eq57} into \eqref{ellGen} are plotted in Figure \ref{ellipse}
for dimensionless wave numbers ranging between $1$ and $\pi$.
% {\bf (why not between 0 and $\pi/2$??)}.

% {\bf Old text, to be deleted:} Substituting relations \eqref{omSum} and \eqref{omDif} into Equation \eqref{ellGen} yields 
% \beq\label{EqEll}
% \dfrac{\left(\tilde\cA-\dfrac{\alpha (1+\beta) }{\tilde k^2}\right)^2}{\alpha^2\dfrac{(\tilde k^2+1) (\beta^2  + 2\beta\cos{2\delta \tilde k }+1)}{\tilde k^6} }+
% \dfrac{\left(\tilde\cE-\dfrac{\alpha (1+\beta) }{\tilde k^2}\right)^2}{\alpha^2\dfrac{(\tilde k^2+1) (\beta^2  + 2\beta\cos{2\delta \tilde k } +1) }{\tilde k^4} } = 1
% \eeq
% As was outlined in section \ref{dimFor}, the previous relation is an equation of an ellipse with coordinates of the center and lengths of the half axes dependent on the wave number $\tilde k$. 
% In Figure \ref{ellipse}, these ellipses are visualized for wave numbers ranging between $1$ to $2\pi$ and for $\beta=0.5$.
% \begin{figure}[H]
% \centering%
% \includegraphics[width=0.55\textwidth, keepaspectratio=true]{ellipses.eps}
%                 \caption{{\bf OLD} Solutions $\tilde \cA$ and $\tilde \cE$ visualized in the $\tilde\cE-\tilde\cA$ space for various wave numbers, $\alpha=\delta=1$, $\beta=0.5$}
%                 \label{ellipse}
% \end{figure}
% \begin{figure}[H]
% \centering%
% \includegraphics[width=0.55\textwidth, keepaspectratio=true]{ellipsesCorDet.eps}
%                 \caption{{\bf JUST FOR ILLUSTRATION}, Solutions $\tilde \cA$ and $\tilde \cE$ visualized in the $\tilde\cE-\tilde\cA$ space for various wave numbers, $\beta=0.5$, $\lambda=1$ }
% \end{figure} 

\subsubsection{Evaluation of Fourier images}

\begin{figure}[H]
\centering%
\begin{subfigure}[b]{0.49\textwidth}    
               
                \includegraphics[width=\textwidth, keepaspectratio=true]{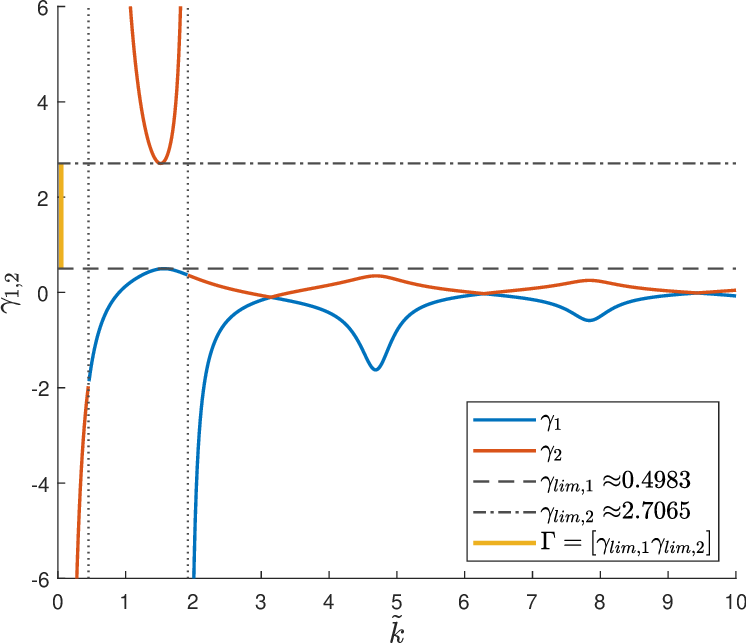}
                \subcaption{Dependence of the limiting value of $\gamma$ on the wave number} \label{gammLim}
\end{subfigure}
\begin{subfigure}[b]{0.49\textwidth}     
                     \includegraphics[width=\textwidth, keepaspectratio=true]{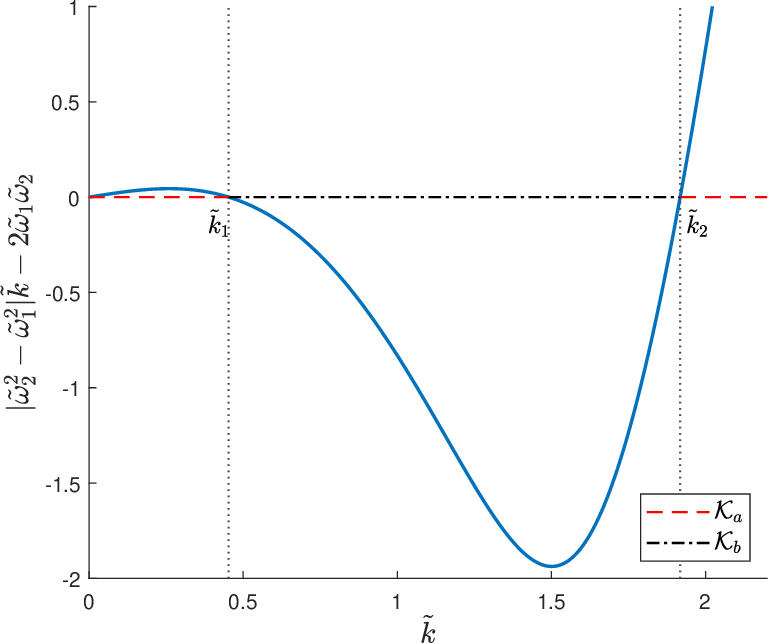}
                     \subcaption{Dependence of the difference of the two sides of \eqref{eq55} on the wave number} \label{LHSRHS} 
 \end{subfigure}               
                \caption{Graphical illustration of the admissible set $\Gamma$ for parameters $\beta=0.5$ and $\lambda=1$}
                \label{}
\end{figure}

Following the approach described in section \ref{sol} and assuming $\tilde \cA = \gamma \tilde \cE$, we express the solution according to formula \eqref{Efsol} as
\beq\label{ellPar}
\tilde\cE =
\lambda^2\frac{(1+\beta)^2}{2\beta(\gamma^2 \tilde k^4+ \tilde k^2)}\left(\gamma \tilde k^2+1
-
\sqrt{(\gamma \tilde k^2+1)^2- 4(\gamma^2 \tilde k^2+ 1)(1+\tilde k^2)\frac{\beta}{(1+\beta)^2} \sin^2\frac{\tilde k}{\lambda}} \right)
\eeq
Thus, from a geometric point of view, the solution to Equation\ \eqref{ellGen} is chosen as the intersection point of the ellipse and a straight line with slope $\gamma$ passing through the origin (the intersection point closer to the origin is selected). To ensure that this line intersects the ellipse, and thus the solution is real, parameter $\gamma$ needs to be chosen from the set $\Gamma$ specified in (\ref{eq:57}). Substituting the given dispersion relations, formula \eqref{gamm12} is rewritten as 
\beq
\gamma_{1,2}
=
\dfrac{-(1+\beta)^2\tilde k \pm  
2\sqrt{\beta}\sin\dfrac{\tilde k}{\lambda}(1+\tilde k^2)\,
\sqrt{\beta^2 +2\beta \cos\dfrac{2\tilde k}{\lambda}+1 }}{\tilde k^3(1+\beta)^2-4\tilde k(1+\tilde k^2)\beta\sin^2{\dfrac{\tilde k}{\lambda}}}
\eeq
For a given wave number $\tilde k$, lines with the slope $\gamma_1$ or $\gamma_2$ are tangent to the ellipse that depicts the set of possible solutions. 
The dependence of two roots on the wave number is visualized in Figure \ref{gammLim}, with the smaller root $\gamma_1$ plotted in blue and the larger root $\gamma_2$ plotted in red.  The difference between the two sides of inequality \eqref{eq55} as a function of the wave number is shown in Figure  \ref{LHSRHS}. As was discussed in Section~\ref{sol}, the sign of this expression determines the sign of the coefficient at $\gamma^2$ in Eq.~(\ref{eq:53}) and hence the type of the interval for admissible $\gamma$ coefficients. 
As seen in the graph, the sign is positive for  $\tilde{k}\in\mathcal{K}_a=(0,\tilde{k}_1) \cup (\tilde{k}_2,\infty)$ and nonpositive for $\tilde{k}\in\mathcal{K}_b=[\tilde{k}_1,\tilde{k}_2]$, 
where $\tilde{k}_1\approx 0.45$ and $\tilde{k}_2\approx 1.92$. By applying the rules described in Section~\ref{sol}, it can be deduced from Figure  \ref{gammLim} that
 the set of admissible $\gamma$ parameters is $\Gamma=[\gamma_{lim,1} ,\gamma_{lim,2}]\approx [0.4983 , 2.7065]$, i.e., the interval between the dashed-dotted and dashed lines. From Figure \ref{ellipse}, where the limiting slopes are also plotted by dashed lines, one can visually check that for $\gamma_{lim,1}<\gamma<\gamma_{lim,2}$ the straight lines with slope $\gamma$ passing through the origin indeed intersect all the ellipses.       
% \begin{figure}[H]
% \centering%       
%              \includegraphics[width=0.55\textwidth, keepaspectratio=true]{gammaLim.eps}

%                 \caption{Dependence of the limiting $\gamma$ parameter on the wave number, $\beta=0.5$, $\lambda=1$}
%                 \label{gammLim}
% \end{figure} 

\begin{figure}[H]
\centering%
\begin{subfigure}[b]{0.49\textwidth}    
               
                \includegraphics[width=\textwidth, keepaspectratio=true]{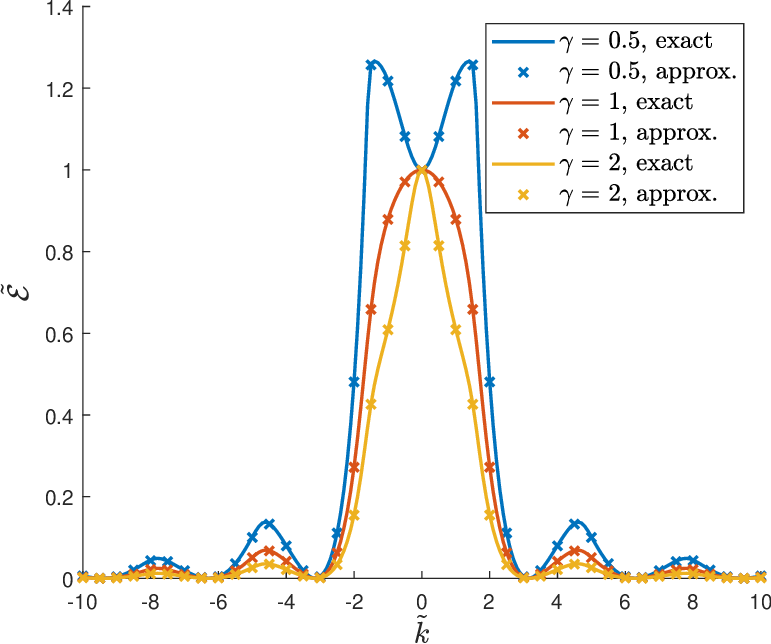}
                \subcaption{Function $\tilde\cE$} \label{Efall}
\end{subfigure}
\begin{subfigure}[b]{0.49\textwidth}     
                     \includegraphics[width=\textwidth, keepaspectratio=true]{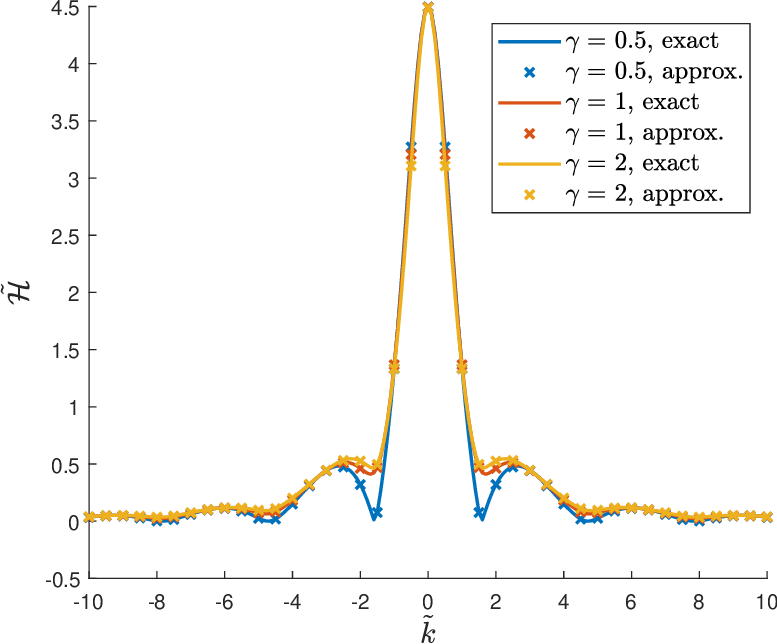}
                     \subcaption{Function $\tilde \cH$} \label{Hfall} 
 \end{subfigure}               
                \caption{Fourier images of weight functions and their approximations  evaluated for a double-mass-spring chain with $\beta = m/M=0.5$ and $\lambda=1$ (the cross symbols correspond to numerical evaluation according to (\ref{Efapprox}) and their meaning will be described later)}
                \label{EfHf}
\end{figure}

 In Figure \ref{Efall}, the Fourier images $\tilde \cE(\tilde k)$ calculated from (\ref{ellPar}) are plotted by solid lines for various choices of parameter $\gamma$.
Function $\tilde\cH$ is then obtained by substituting \eqref{eq57}--(\ref{ellPar}) into   \eqref{Hfun}, which yields 
% \bea\nonumber
% \tilde\cH&=& 2\alpha\frac{1+\beta}{1+\tilde k^2}
%  -
% \alpha\dfrac{(\gamma+1) (1+\beta)(\gamma \tilde k^2+1)  }{(\gamma^2 \tilde k^2+  1)({1+\tilde k^2} )}
% +
% \alpha(\gamma+1) \dfrac{\sqrt{{(1+\beta)}^2{(\gamma \tilde k^2+1)}^2- 4(\gamma^2 \tilde k^2+ 1)(1+\tilde k^2)\beta \sin^2{\delta\tilde k} }}{(\gamma^2 \tilde k^2+  1)({1+\tilde k^2} )} = \\
% &=&
% \tilde\cH_a+\tilde\cH_b+\tilde\cH_c 
% \label{Hfun2}
% \eea
\beq\label{Hfun2a}
\tilde\cH = \tilde\cH_a+\tilde\cH_b+\tilde\cH_c 
\eeq
where
\bea 
 \tilde\cH_a&=& \lambda^2\frac{\left(1+\beta\right)^2}{\beta\left(1+\tilde k^2\right)}
 \\
  \tilde\cH_b &=&
 -
 \lambda^2\dfrac{(\gamma+1) (1+\beta)^2(\gamma \tilde k^2+1)  }{2\beta(\gamma^2 \tilde k^2+  1)({1+\tilde k^2} )}
\\
 \tilde\cH_c &=&
 \lambda^2\dfrac{\left(1+\beta\right)(\gamma+1)}{2\beta} \dfrac{\sqrt{{(1+\beta)}^2{(\gamma \tilde k^2+1)}^2- 4(\gamma^2 \tilde k^2+ 1)(1+\tilde k^2)\beta \sin^2(\tilde k/\lambda) }}{(\gamma^2 \tilde k^2+  1)({1+\tilde k^2} )} 
\label{Hfun2b}
\eea
are three terms to be processed separately, as will be shown later. 
The graphs of $\tilde\cH$ obtained for selected values of parameter $\gamma$ are plotted by solid lines in Figure \ref{Hfall}. The value of $\tilde\cH$ at $\tilde k =0$ is $\tilde\cH(0)=\lambda^2{\left(1+\beta\right)^2}/{\beta}$, which indicates that the integral $\intL \tilde H(\tilde r) \,{\rm d }\tilde r$ is finite.  
% \begin{figure}[H]
% \centering%
% \begin{subfigure}[b]{0.49\textwidth}    
               
%                 \includegraphics[width=\textwidth, keepaspectratio=true]{GM_EfappAll.eps}
%                 \subcaption{Function $\tilde\cE$} \label{Efall}
% \end{subfigure}
% \begin{subfigure}[b]{0.49\textwidth}     
%                      \includegraphics[width=\textwidth, keepaspectratio=true]{GM_HfappAll.eps}
%                      \subcaption{Function $\tilde \cH$} \label{Hfall} 
%  \end{subfigure}               
%                 \caption{{\bf OLD} Fourier images of weight functions and their approximations evaluated for a double mass-spring chain with $\beta = m/M=0.5$}
%                 \label{EfHf}
% \end{figure} 
% 
The last weight function to be determined is $\tilde\cA=\gamma\tilde\cE$. Its graph is only a scaled graph of the already known function $\tilde\cE$. 

\subsubsection{Reconstruction of the weight functions}
So far, we have described the procedure leading to the Fourier images of the weight functions. To reconstruct the original weight functions $\tilde E$, $\tilde A$, and $\tilde H$ in the space domain, the inverse Fourier transform (IFT) is applied to the already determined functions $\tilde\cE$, $\tilde\cA$, and $\tilde\cH$. It is not always possible to perform the inverse transformation analytically; therefore, the inverse discrete Fourier transform (IDFT) is adopted in order to approximate the exact result. According to \eqref{Hfun2a},
the Fourier image $\tilde\cH$ is composed of three parts, from which the first two,  $\tilde\cH_a$ and $\tilde\cH_b$, can be transformed analytically, and only the third part $\tilde\cH_c$ needs to be transformed numerically. The analytical inverse transforms read
\bea
\mathcal{F}^{-1} \left(\tilde\cH_a \right)&=&\mathcal{F}^{-1} \left(\lambda^2\frac{\left(1+\beta\right)^2}{\beta\left(1+\tilde k^2\right)}\right) = \lambda^2\dfrac{(1+\beta)^2}{2\beta}  {\rm e}^{-|\tilde r|} \\
\mathcal{F}^{-1} \left(\tilde\cH_b \right)&=&\mathcal{F}^{-1} \left( -
 \lambda^2\dfrac{(\gamma+1) (1+\beta)^2(\gamma \tilde k^2+1)  }{2\beta(\gamma^2 \tilde k^2+  1)({1+\tilde k^2} )}\right) = - \lambda^2\dfrac{(1+\beta)^2}{4\beta}\left(  {\rm e}^{-|\tilde r|} +  {\rm e}^{-|\tilde r|/\gamma}\right)
\eea
and the complete weight function is given by  
% \beq
% \tilde H = \alpha (1+\beta) \exp (-|\tilde r|) -\half \alpha (1+\beta) \left(  \exp (-|\tilde r|) +  \exp \left(-\dfrac{|\tilde r|}{\gamma} \right) \right) + \mathcal{F}^{-1} \left(\tilde\cH_c
% \right)
% \eeq
\beq
\tilde H =  \lambda^2\dfrac{(1+\beta)^2}{4\beta} \left(  {\rm e}^{-|\tilde r|} -  {\rm e}^{-|\tilde r|/\gamma} \right) + \mathcal{F}^{-1} \left(\tilde\cH_c
\right)
\eeq
Function $\tilde \cE$ given by \eqref{ellPar} can also
be additively decomposed into two parts,  the first of which could be transformed analytically. However, both parts have an infinite limit for $\tilde k \to 0$, which would result in numerical problems when solely the second part is transformed using the IDFT method. Therefore, the complete function $\tilde \cE$ is transformed numerically.  

\begin{figure}[H]
\centering%
\begin{subfigure}[b]{0.49\textwidth}

                \includegraphics[width=\textwidth, keepaspectratio=true]{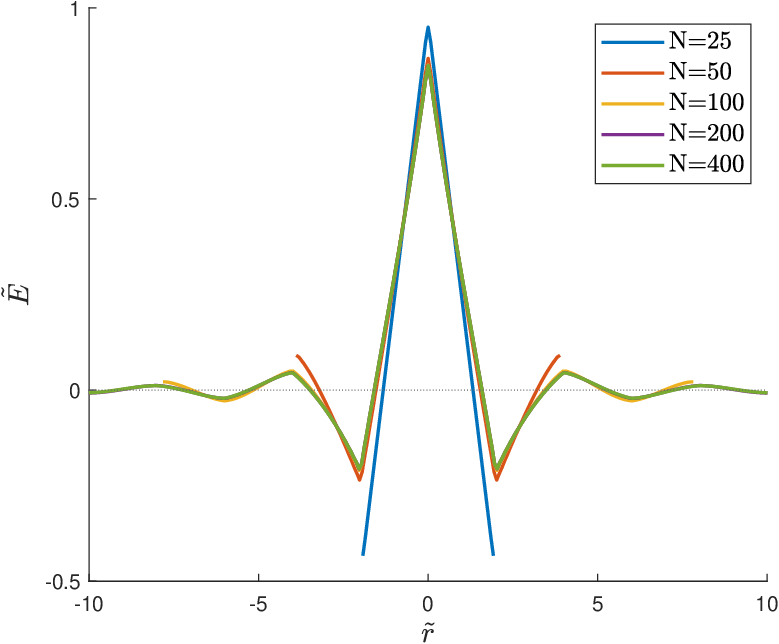}
                 \subcaption{Weight function $\tilde E$}
\end{subfigure}
\begin{subfigure}[b]{0.49\textwidth}            
                
                \includegraphics[width=\textwidth, keepaspectratio=true]{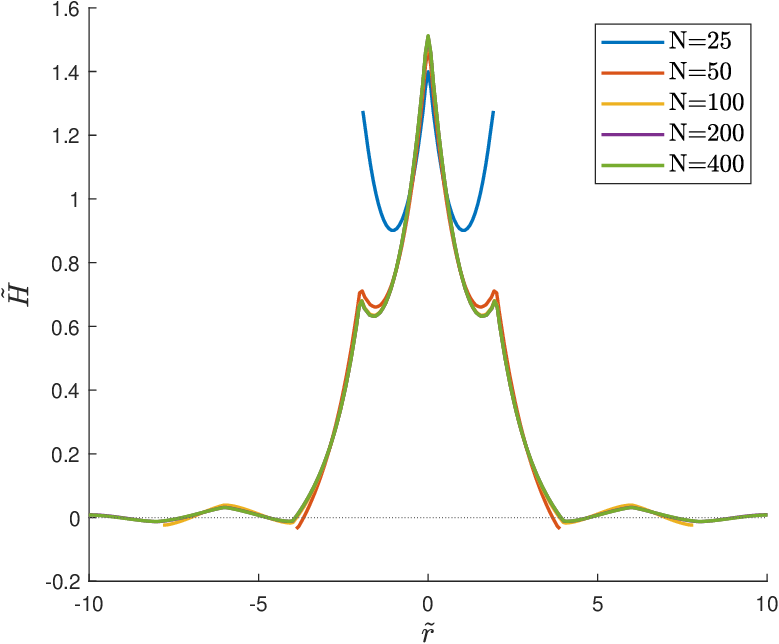}
                \subcaption{Weight function $\tilde H$}
\end{subfigure}

                \caption{Weight functions computed by IDFT for the double-mass-spring chain for various numbers of positive sampling points $N$; model parameters $\beta =0.5$, $\gamma= 0.5$, $\lambda=1$ and numerical parameter $\tilde k_0=40$}
                \label{FH_Nconv}
\end{figure}

The accuracy of the IDFT depends on the number of (positive) sampling points $N$ in the wave number domain and on the length of the finite sampling interval, denoted by $\tilde k_0$. The influence of both parameters on the resulting accuracy is now studied for the case when $\gamma=0.5$.  
Figure \ref{FH_Nconv} contains graphs of the approximated weight functions $\tilde E(\tilde r)$ and $\tilde H(\tilde r)$ computed for fixed $\tilde k_0=40$ and a varying number of sampling points. The sampling points are located 
at $\tilde k_j=j\,\Delta\tilde{k}$, $j=0,1,2,\ldots N$,
in which $\Delta\tilde{k}=\tilde{k}_0/N$ is the spacing (in the wave number domain).
% \begin{figure}[H]
% \centering%
% \begin{subfigure}[b]{0.49\textwidth}

%                 \includegraphics[width=\textwidth, keepaspectratio=true]{GM_Ere_N_gam05.eps}
%                  \subcaption{Weight function $\tilde E$}
% \end{subfigure}
% \begin{subfigure}[b]{0.49\textwidth}            
                
%                 \includegraphics[width=\textwidth, keepaspectratio=true]{GM_Hre_N_gam05.eps}
%                 \subcaption{Weight function $\tilde H$}
% \end{subfigure}

%                 \caption{{\bf OLD} weight functions computed by IDFT for the double mass-spring chain for various numbers of sampling points $N$, $\beta =0.5$, $\gamma= 0.5$, $\tilde k_0=40$.}
%                 \label{FH_Nconv}
% \end{figure}
% 
\begin{figure}[H]
\centering%
\begin{subfigure}[b]{0.49\textwidth}

                \includegraphics[width=\textwidth, keepaspectratio=true]{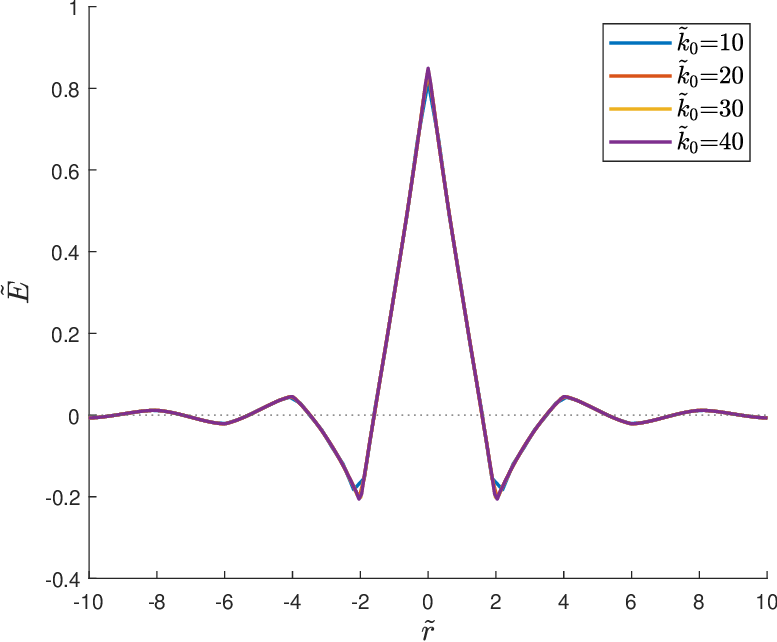}
                \subcaption{Weight function $\tilde E$}
\end{subfigure}
\begin{subfigure}[b]{0.49\textwidth}            
                
                \includegraphics[width=\textwidth, keepaspectratio=true]{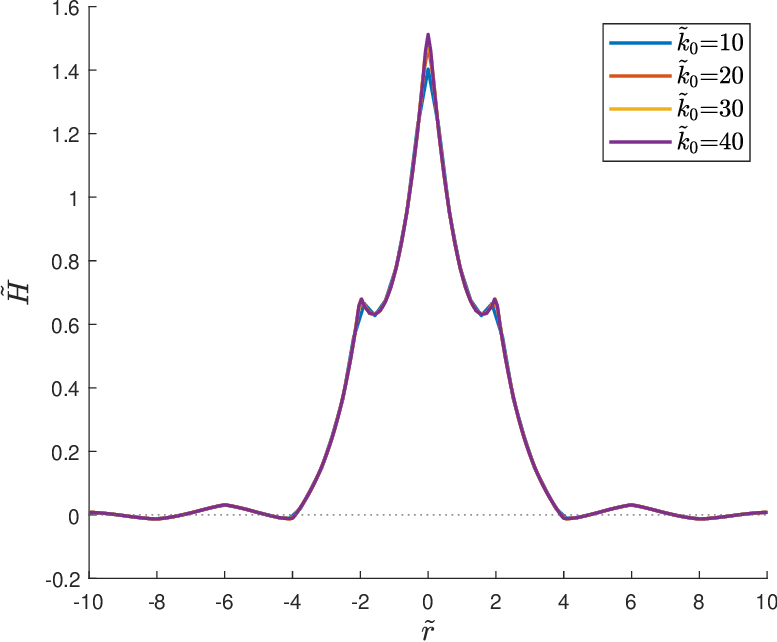}
                \subcaption{Weight function $\tilde H$}
\end{subfigure}

                \caption{Weight functions computed by IDFT for the double-mass-spring chain for various lengths of the sampling interval $k_0$; model parameters $\beta =0.5$, $\lambda=1$, $\gamma= 0.5$ and numerical parameter $N=400$}
                \label{FH_k0conv}
\end{figure}

One can note that the shape of the weight functions seems to be stabilized for $N=200$, since the result 
is visually indistinguishable from the result
for $N=400$. Subsequently, the effect of the sampling interval length is studied, with the number of sampling points set to $N=400$. In Figure \ref{FH_k0conv}, the resulting graphs are displayed. It is apparent that, except for the blue line corresponding to $\tilde k_0=10$, all the other curves are almost identical.        
% \begin{figure}[H]
% \centering%
% \begin{subfigure}[b]{0.49\textwidth}

%                 \includegraphics[width=\textwidth, keepaspectratio=true]{GM_Ere_k0_gam05.eps}
%                 \subcaption{Weight function $\tilde E$}
% \end{subfigure}
% \begin{subfigure}[b]{0.49\textwidth}            
                
%                 \includegraphics[width=\textwidth, keepaspectratio=true]{GM_Hre_k0_gam05.eps}
%                 \subcaption{Weight function $\tilde H$}
% \end{subfigure}

%                 \caption{{\bf OLD} Weight functions computed by IDFT for the double mass-spring chain for various lengths of the sampling interval $k_0$, $\beta =0.5$, $\gamma= 0.5$, $N=801$}
%                 \label{FH_k0conv}
% \end{figure}
% 

\begin{figure}[H]
\centering%
\begin{subfigure}[b]{0.49\textwidth}

                \includegraphics[width=\textwidth, keepaspectratio=true]{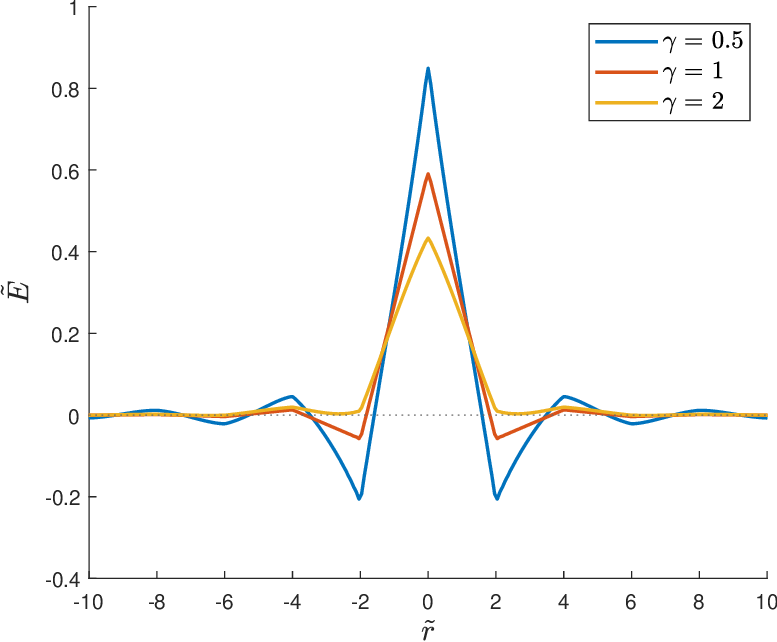}
                 \subcaption{Weight function $\tilde E$}
\end{subfigure}
\begin{subfigure}[b]{0.49\textwidth}            
               
                \includegraphics[width=\textwidth, keepaspectratio=true]{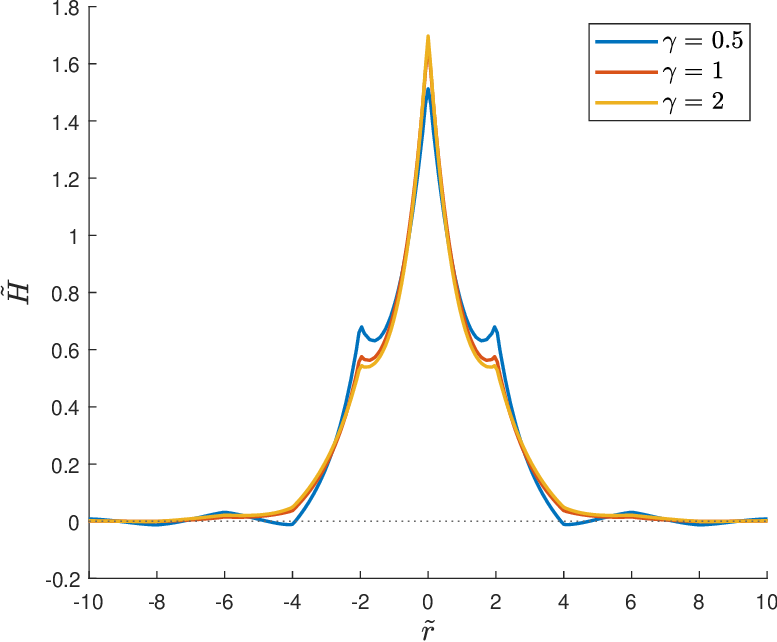}
                 \subcaption{Weight function $\tilde H$}
\end{subfigure}

                \caption{Weight functions computed by IDFT for the double-mass-spring chain; model parameters $\beta =0.5$, $\lambda=1$ and numerical parameters $\tilde k_0=40$, $N=400$}
                \label{FH_IFTall}
\end{figure}

Overall, the numerically approximated results obtained with parameters $\tilde k_0=40$, and $N=400$ are considered as sufficiently accurate and are adopted for further processing. Figure \ref{FH_IFTall} shows the resulting approximations of the dimensionless weight functions evaluated for various coefficients $\gamma$. 
%Note that  the same values of parameters $\tilde k_0=40$, and $N=801$ were used for all considered parameters $\gamma$. 
% \begin{figure}[H]
% \centering%
% \begin{subfigure}[b]{0.49\textwidth}

%                 \includegraphics[width=\textwidth, keepaspectratio=true]{GM_Eall.eps}
%                  \subcaption{Weight function $\tilde E$}
% \end{subfigure}
% \begin{subfigure}[b]{0.49\textwidth}            
               
%                 \includegraphics[width=\textwidth, keepaspectratio=true]{GM_Hall.eps}
%                  \subcaption{Weight function $\tilde H$}
% \end{subfigure}

%                 \caption{ {\bf OLD} weight functions computed by IDFT for the double mass-spring chain, $\beta =0.5$, $\tilde k_0=40$, $N=801$}
%                 \label{FH_IFTall}
% \end{figure}
% 

The accuracy of the obtained pointwise approximations of weight functions $\tilde E$ and $\tilde H$ can be assessed by evaluating their Fourier images and comparing them to the analytical expressions in \eqref{ellPar} and \eqref{Hfun2a}--\eqref{Hfun2b}. 
By the IDFT one obtains approximate values of the weight function at non-negative positions $\tilde r_j = j\,h$ where $j=0,1,2,\dots N$ and $h=2\pi N/\tilde{k}_0 /(2N+1)$  
% \textbf{(I think $h$ should be defined as $h=\frac{2 \pi}{\tilde{k}_0}\frac{N}{2N+1}$)} 
is the spacing in the dimensionless physical domain; these values can then be mirrored to the
region of negative coordinates, since the weight functions are even.
% These points are located symmetrically with respect to the origin and their number is always odd. 
%Due to the symmetry, the values of the approximated functions for the negative coordinates $\tilde r$ are identical to those for positive $\tilde r$.
To approximate the dimensionless weight function, linear interpolation is used between neighboring positions $\tilde{r}_j$. The piecewise linear approximation can be expressed as
% \beq
% \tilde E(\tilde r) \approx \sum_{ j=1}^N \tilde E_j \phi_j(\tilde r)
% \eeq
\beq
\tilde E(\tilde r) \approx \sum_{ j=0}^N \tilde E_j \phi_j(\tilde r) + \sum_{ j=1}^N \tilde E_j \phi_j(-\tilde r)
\eeq
where $\tilde E_j$ denotes the approximate value of the dimensionless weight function $\tilde{E}$ at position $\tilde r_j$, and $\phi_j(\tilde r)$ stands for the hat function defined as 
%  \[  \phi_j(\tilde r) = \begin{cases} 
%           0 & \tilde r \in [-\infty,\tilde r_{j-1}] \\
%           \frac{\tilde r-\tilde r_{j-1}}{h} & \tilde r \in [\tilde r_{j-1},\tilde r_{j}] \\
%           \frac{\tilde r_{j+1}-\tilde r}{h} & \tilde r \in [\tilde r_{j},\tilde r_{j+1}] \\
%           0 & \tilde r \in [\tilde r_{j+1},\infty] 
%        \end{cases}
% \]
\beq
   \phi_j(\tilde r) = \begin{cases} 
          0 & \tilde r \in (-\infty,\tilde r_{j}-h) \\
          \frac{\tilde r-(\tilde r_{j}-h)}{h} & \tilde r \in [\tilde r_{j}-h,\tilde r_{j}] \\
          \frac{\tilde r_{j}+h-\tilde r}{h} & \tilde r \in [\tilde r_{j},\tilde r_{j}+h] \\
          0 & \tilde r \in (\tilde r_{j}+h,\infty) 
       \end{cases}
\eeq
%  \[ \begin{split}
%  \phi^+_j(\tilde r) = \begin{cases} 
%           0 & \tilde r \in [-\infty,\tilde r_{j}-h] \\
%           \frac{\tilde r-(\tilde r_{j}-h)}{h} & \tilde r \in [\tilde r_{j}-h,\tilde r_{j}] \\
%           \frac{\tilde r_{j}+h-\tilde r}{h} & \tilde r \in [\tilde r_{j},\tilde r_{j}+h] \\
%           0 & \tilde r \in [\tilde r_{j}+h,\infty] 
%        \end{cases}
%         \end{split}
%         \ \ \ \ \ \ \ \ 
%         \begin{split}
%  \phi^-_j(\tilde r) = \begin{cases} 
%           0 & \tilde r \in [-\infty,\tilde r_{j}-h] \\
%           \frac{\tilde r-(\tilde r_{j}-h)}{h} & \tilde r \in [\tilde r_{j}-h,\tilde r_{j}] \\
%           \frac{\tilde r_{j}+h-\tilde r}{h} & \tilde r \in [\tilde r_{j},\tilde r_{j}+h] \\
%           0 & \tilde r \in [\tilde r_{j}+h,\infty] 
%        \end{cases}
%         \end{split}
% \]
% with $h$ being the distance between the evaluation positions $\tilde r_{j}$ in the space domain. 
The Fourier transform  of the hat function $\phi_j(\tilde r)$ can be evaluated analytically and is given by
\beq
\Phi_j(\tilde k) = \int^\infty_{-\infty} \phi_j(\tilde r) \exp{(-i \tilde k\tilde r)} \mathrm{d}\tilde r = h \left(\dfrac{\sin{\dfrac{\tilde k h}{2}}}{\dfrac{\tilde k h}{2}}\right)^2 \exp{(-i \tilde k\tilde r_j)}
=\frac{4}{\tilde k^2h}\sin^2\frac{\tilde kh}{2}\exp{(-i \tilde k\tilde r_j)}
\eeq
By virtue of the similarity theorem $\mathcal{F}(\phi_j(-\tilde r)) = \Phi_j(-\tilde k)$. 
Therefore, the Fourier transform of the approximated weight function is expressed as
\bea\nonumber
\tilde \cE(\tilde k) &\approx& \sum_{ j=0}^N \tilde E_j \Phi_j(\tilde k)+\sum_{ j=1}^N \tilde E_j \Phi_j(-\tilde k)= \frac{4}{\tilde k^2h}\sin^2\frac{\tilde kh}{2}\left(\sum_{ j=0}^N \tilde E_j \exp{(-i \tilde k\tilde r_j)}+\sum_{ j=1}^N \tilde E_j \exp{(i \tilde k\tilde r_j)}\right) =\\  &=& 
\frac{4}{\tilde k^2h}\left(\tilde E_{0}+2\sum_{ j=1}^{N} \tilde E_j \cos{\tilde k\tilde r_j} \right)\sin^2\frac{\tilde kh}{2}
 \label{Efapprox}
\eea
% \beq \label{Efapprox}
% \tilde \cE(\tilde k) \approx \sum_{ j=1}^N \tilde E_j \tilde\phi_j(\tilde k)= h \left(\dfrac{\sin{\dfrac{\tilde k h}{2}}}{\dfrac{\tilde k h}{2}}\right)^2\sum_{ j=1}^N \tilde E_j \exp{(-i \tilde k\tilde r_j)} = 
% h \left(\dfrac{\sin{\dfrac{\tilde k h}{2}}}{\dfrac{\tilde k h}{2}}\right)^2\left(2\sum_{ j=1}^{\frac{N-1}{2}} \tilde E_j \cos{(\tilde k\tilde r_j)} +\tilde E_{\frac{N+1}{2}}\right)
% \eeq
In the same spirit, also the approximation of function $\tilde \cH(\tilde k)$ can be constructed. 
In Figure \ref{EfHf}, the obtained approximations of the Fourier images are visualized by star markers and compared to the exact functions.
% In Figure \ref{EfHf} the exact Fourier images $\tilde \cE$, and $\tilde \cA$, and their approximations are compared for all the investigated $\gamma$ coefficients. 
It is confirmed that the agreement is excellent.   
% \begin{figure}[H]
% \centering%       
% \begin{subfigure}[b]{0.49\textwidth}

%                 \includegraphics[width=\textwidth, keepaspectratio=true]{GM_EfappAll.eps}
%                 \subcaption{Function $\tilde\cE$}
% \end{subfigure}
% \begin{subfigure}[b]{0.49\textwidth}            
                
\begin{figure}[H]
\centering%       
\begin{subfigure}[b]{0.49\textwidth}

                \includegraphics[width=\textwidth, keepaspectratio=true]{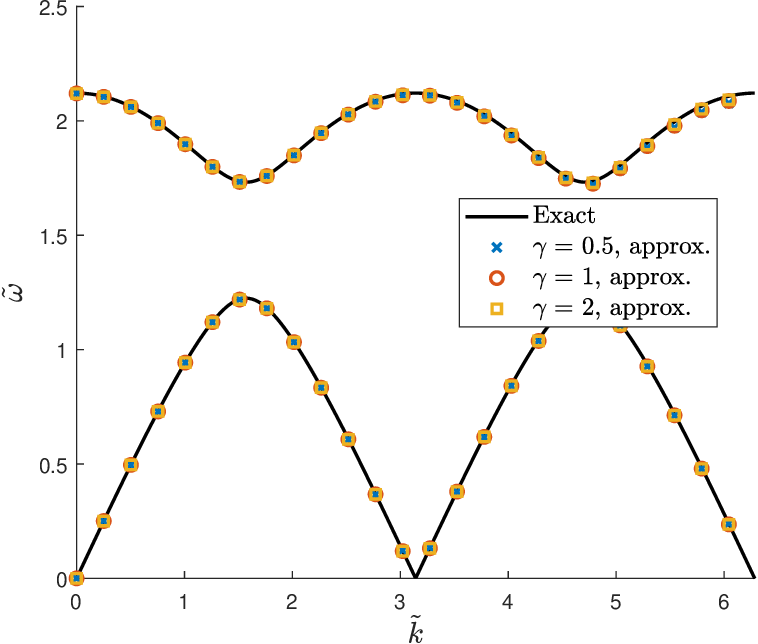}
                \subcaption{Dispersion diagram}\label{chainappA}
\end{subfigure}
\begin{subfigure}[b]{0.49\textwidth}            
                
                \includegraphics[width=\textwidth, keepaspectratio=true]{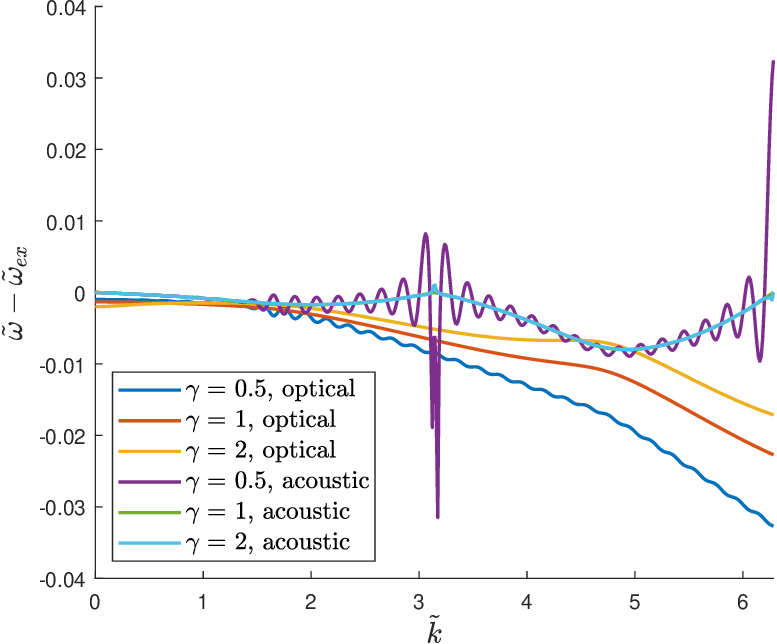} 
                \subcaption{Absolute error}\label{chainappB}
\end{subfigure}
                \caption{Comparison of the exact dispersion diagram of the double-mass-spring chain and its enriched continuum approximation for model parameters $\beta=0.5$ and $\lambda=1$ and numerical parameters $\tilde k_0=40$, $N=400$}
                \label{chainapp}
\end{figure}

%                 \includegraphics[width=\textwidth, keepaspectratio=true]{GM_HfappAll.eps}  
%                 \subcaption{Function $\tilde\cH$}
% \end{subfigure}
%                 \caption{Comparison of the Fourier images of the weight functions and their approximations, $\beta=0.5$}
%                 \label{EfHfapp}
% \end{figure} 
Finally, one can assess the difference between the dispersion diagram of the double-mass-spring chain and the dispersion diagram of the enriched continuum model evaluated with the approximated weight functions that are plotted in Figure~\ref{FH_IFTall}. The dispersion relation of the investigated continuum model is obtained by solving 
equation \eqref{mm4} for the unknown $\omega$. In the dimensionless form and after substituting $\tilde \cA =\gamma \tilde \cE$, the solution reads
\bea
\tilde\omega_{1,2}^2 &=&  
 \dfrac{\tilde\cH+(\tilde\cE(1+\gamma)+\tilde\cH)\tilde k^2}{2} \pm  \dfrac{\sqrt{\left( \tilde\cH+(\tilde\cE(1+\gamma)+\tilde\cH)\tilde k^2\right)^2 - 4\left(\gamma\tilde\cE(\tilde\cE+\tilde\cH)\tilde k^4+\tilde\cE\tilde\cH \tilde k^2\right)}}{2}
 \label{eq:77}
\eea
In Figure \ref{chainapp}, the results obtained from (\ref{eq:77}) with the Fourier images $\tilde\cE$ and $\tilde\cH$ replaced by their approximations  \eqref{Efapprox} are compared to the exact dispersion relation of the double-mass-spring chain given in  \eqref{eq57}. The dispersion diagrams are displayed in Figure \ref{chainappA}, and the difference between the exact and approximated results is visualized for the acoustic as well as optical branch in Figure \ref{chainappB}. 
% Note that in both graphs squared circular frequencies are visualized, since the approximated values of $\tilde \omega^2$ might be negative for wave numbers close to multiples of $\pi$, which would result in complex values of $\tilde \omega$.
% \begin{figure}[H]
% \centering%       
% \begin{subfigure}[b]{0.49\textwidth}

%                 \includegraphics[width=\textwidth, keepaspectratio=true]{GM_chainAppGen.eps}
%                 \subcaption{Dispersion diagram}\label{chainappA}
% \end{subfigure}
% \begin{subfigure}[b]{0.49\textwidth}            
                
%                 \includegraphics[width=\textwidth, keepaspectratio=true]{GM_chainErr.eps} 
%                 \subcaption{Absolute Error}\label{chainappB}
% \end{subfigure}
%                 \caption{{\bf OLD} Comparison of the exact dispersion diagram of the double mass-spring chain and its enriched continuum approximation, $\beta=0.5$}
%                 \label{chainapp}
% \end{figure} 
% 

Figure \ref{chainappB} indicates that the error is greater for larger wave numbers. Moreover, the deviations are more severe for smaller parameters $\gamma$. From Figure \ref{Efall} it is apparent that for decreasing parameter $\gamma$ the amplitudes of the weight function get larger. Hence, a larger sampling window and a higher number of terms are required for the IDFT to achieve similar accuracy as for the function with larger values of $\gamma$.  

\subsubsection{Choice of characteristic length}

\begin{figure}[H]
\centering%
\begin{subfigure}[b]{0.49\textwidth}    
               
                \includegraphics[width=\textwidth, keepaspectratio=true]{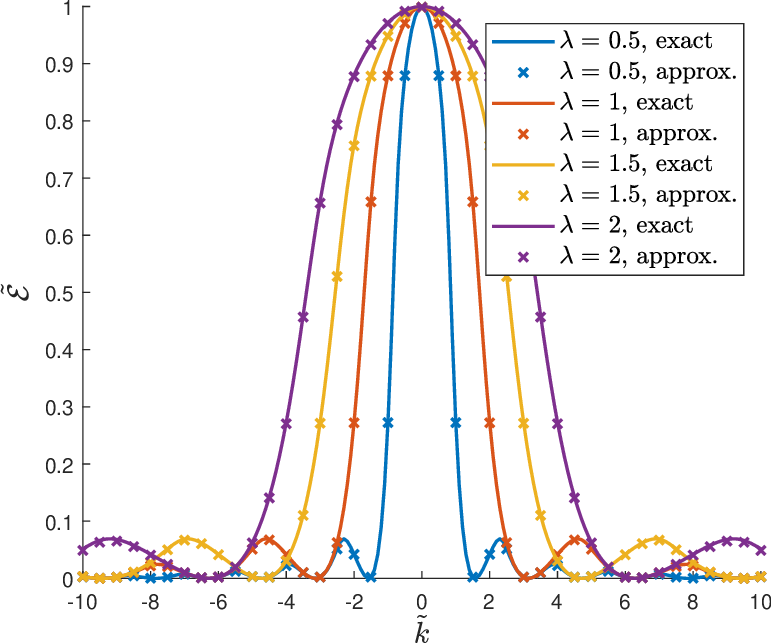}
                \subcaption{Function $\tilde\cE$} \label{EfallLam}
\end{subfigure}
\begin{subfigure}[b]{0.49\textwidth}     
                     \includegraphics[width=\textwidth, keepaspectratio=true]{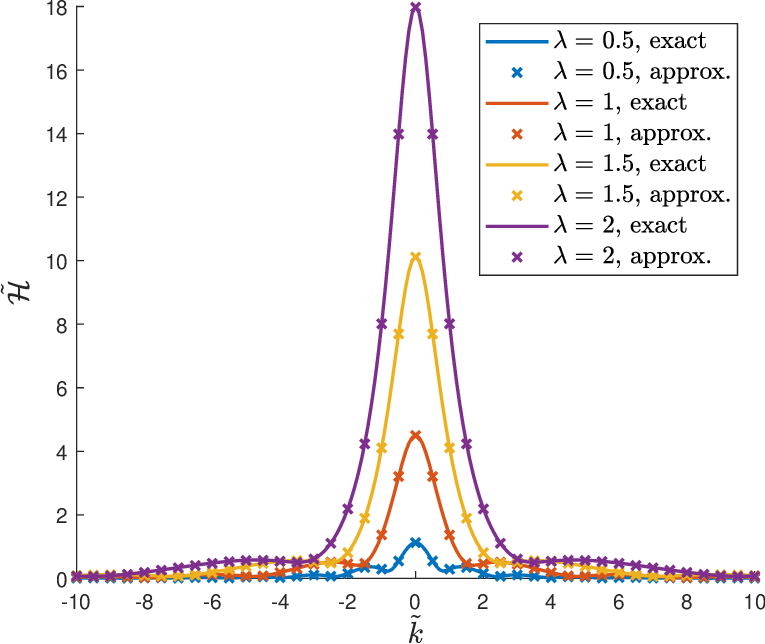}
                     \subcaption{Function $\tilde \cH$} \label{HfallLam} 
 \end{subfigure}               
                \caption{Fourier images of weight functions and their approximations evaluated for a double-mass-spring chain  using fixed parameters $\beta =0.5$ and $\gamma=1$ and various values of $\lambda$}
                \label{EfHfLam}
\end{figure}

So far, we have been working with a fixed value of $\lambda=1$, meaning that the characteristic length of the continuum model is chosen equal to the particle spacing in the discrete chain. It is now interesting to look at the effect of parameter $\lambda$ on the shape of the evaluated weight functions. 
Figure \ref{EfHfLam} shows the dimensionless Fourier images
$\tilde\cE$ and $\tilde\cH$ obtained for a double-mass-spring
chain with $\beta=0.5$ using values of $\lambda$ between 0.5 and 2. Parameter $\gamma$ has a fixed value of 1, which means
that functions $\tilde\cE$ and $\tilde\cA$ are identical.
Function $\tilde\cE$ is by definition
normalized such that $\tilde\cE(0)=1$.
For increasing $\lambda$, the graph of $\tilde\cE$ is stretched horizontally, which corresponds to a change
of scale. On the other hand, function $\tilde\cH$ is not constrained by a normalizing condition, and its maximum
increases with increasing $\lambda$, simultaneously with
horizontal stretching and certain changes in shape.

\begin{figure}[H]
\centering%
\begin{subfigure}[b]{0.49\textwidth}

                \includegraphics[width=\textwidth, keepaspectratio=true]{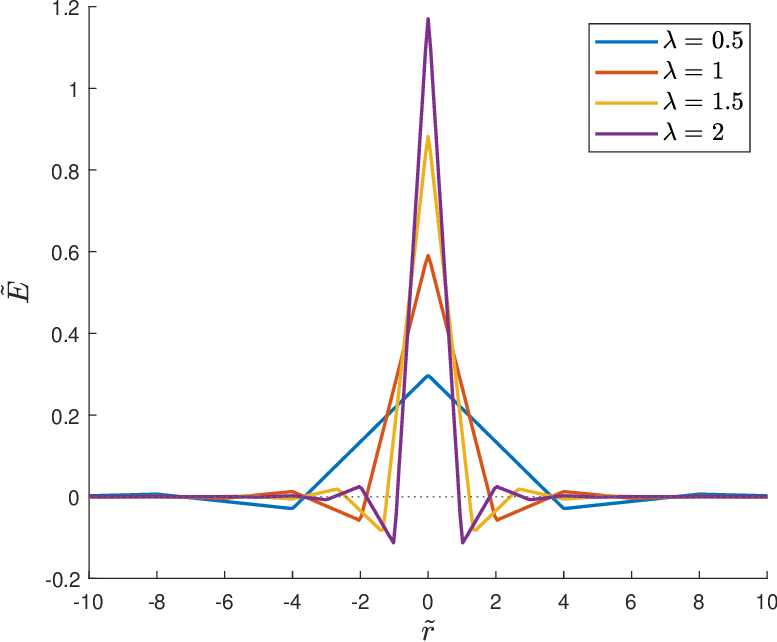}
                 \subcaption{Weight function $\tilde E$}
\end{subfigure}
\begin{subfigure}[b]{0.49\textwidth}            
               
                \includegraphics[width=\textwidth, keepaspectratio=true]{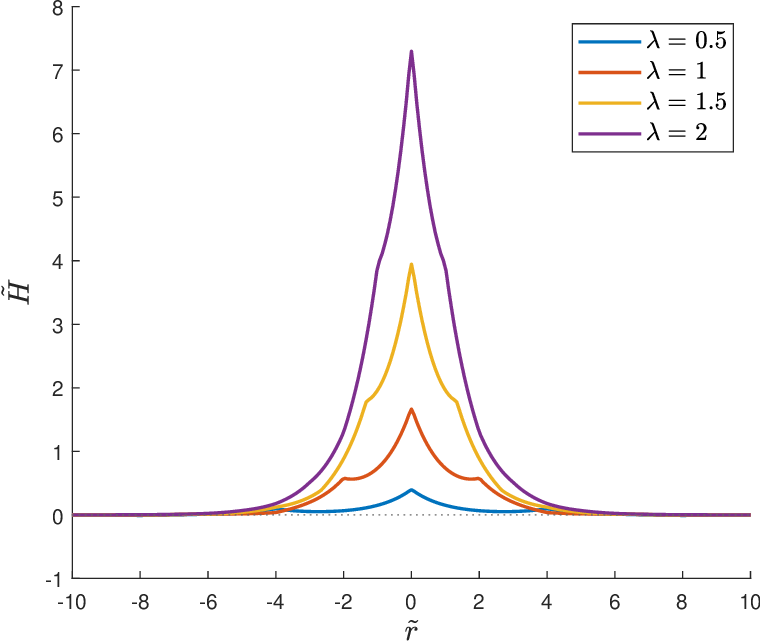}
                 \subcaption{Weight function $\tilde H$}
\end{subfigure}

                \caption{Weight functions computed by IDFT for the double-mass-spring chain  using fixed parameters $\beta =0.5$ and $\gamma=1$ and various values of $\lambda$}
                \label{FH_IFTallLam}
\end{figure}

\begin{figure}[H]
\centering%
\begin{subfigure}[b]{0.49\textwidth}
 
                \includegraphics[width=\textwidth, keepaspectratio=true]{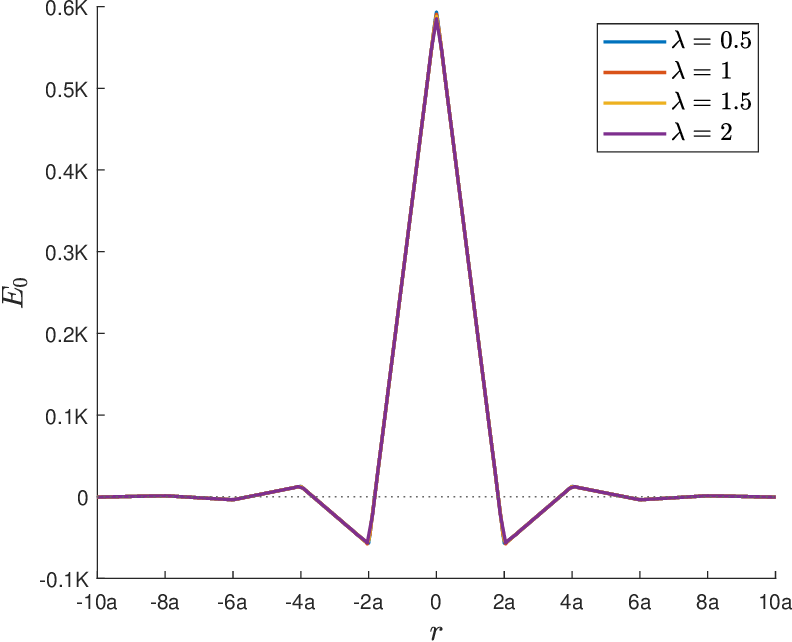}
                 \subcaption{Weight function $E_0$, $\gamma=1$}
\end{subfigure}
\begin{subfigure}[b]{0.49\textwidth}            
               
                \includegraphics[width=\textwidth, keepaspectratio=true]{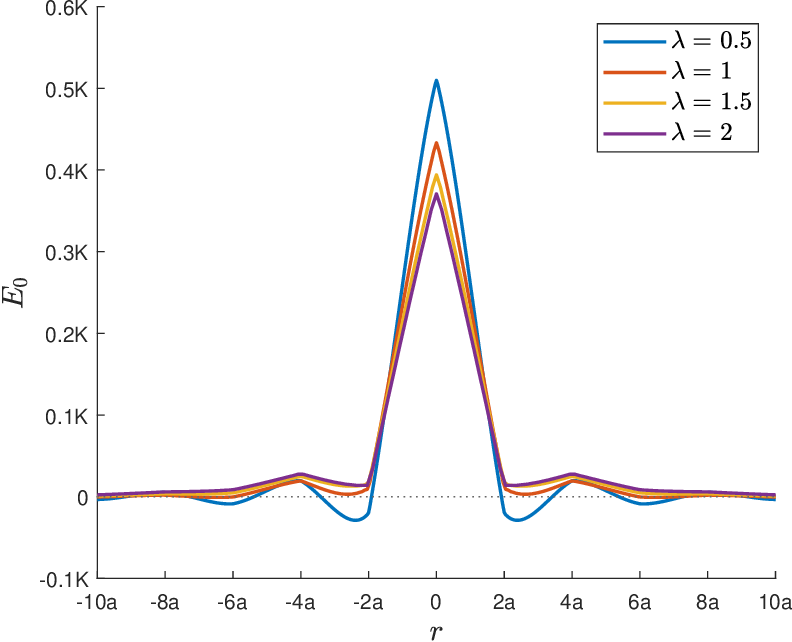}
                 \subcaption{Weight function $E_0$, $\gamma=2$}
\end{subfigure}
\\
\begin{subfigure}[b]{0.49\textwidth}
 
                \includegraphics[width=\textwidth, keepaspectratio=true]{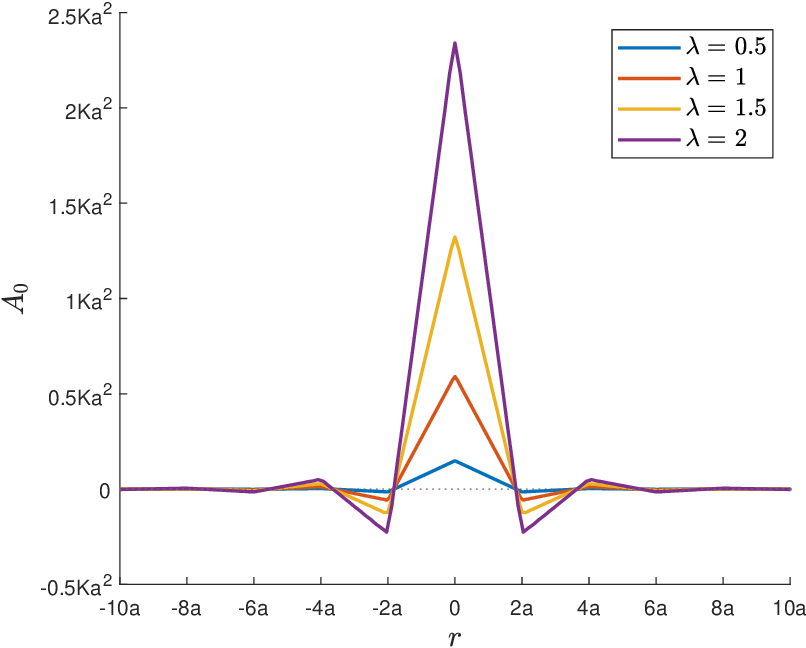}
                 \subcaption{Weight function $A_0$, $\gamma=1$}
\end{subfigure}
\begin{subfigure}[b]{0.49\textwidth}            
               
                \includegraphics[width=\textwidth, keepaspectratio=true]{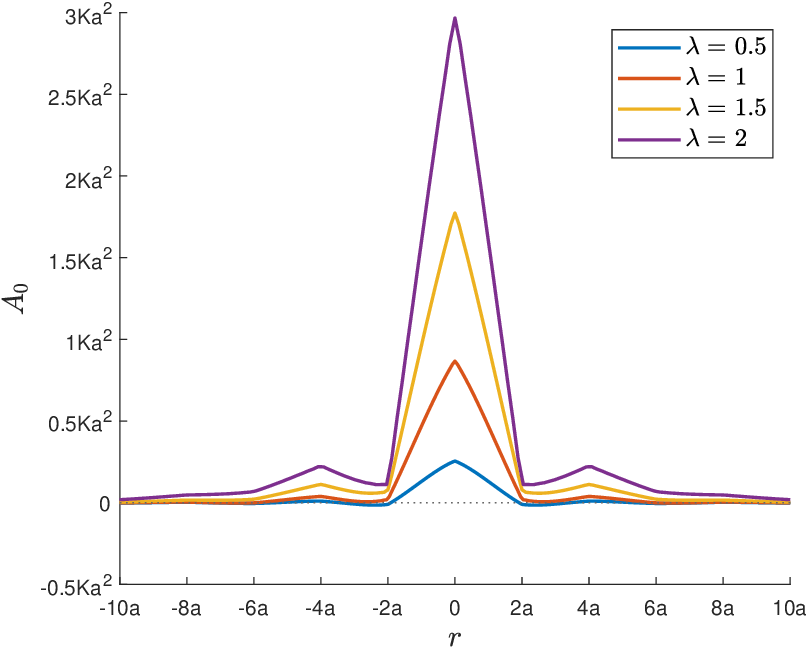}
                 \subcaption{Weight function $A_0$, $\gamma=2$}
\end{subfigure}
\\
\begin{subfigure}[b]{0.49\textwidth}
 
                \includegraphics[width=\textwidth, keepaspectratio=true]{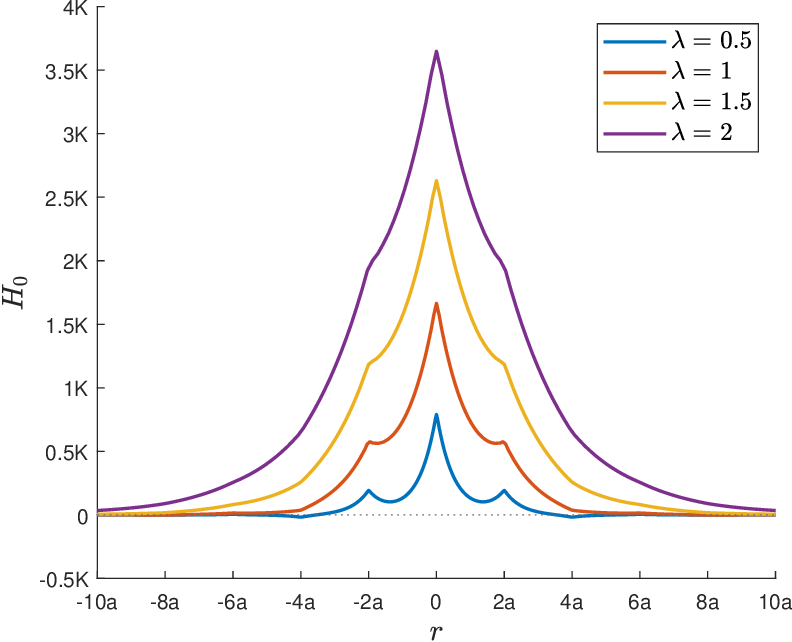}
                 \subcaption{Weight function $H_0$, $\gamma=1$}
\end{subfigure}
\begin{subfigure}[b]{0.49\textwidth}            
               
                \includegraphics[width=\textwidth, keepaspectratio=true]{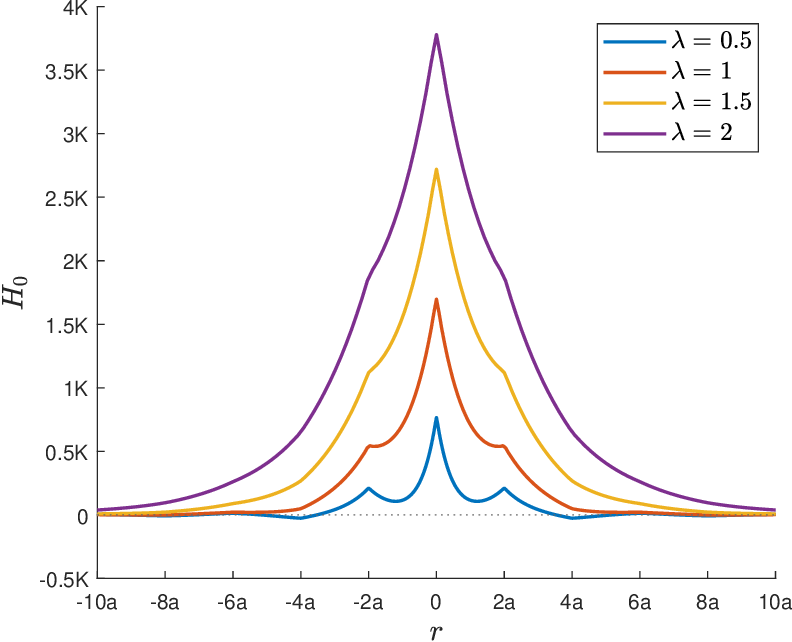}
                 \subcaption{Weight function $H_0$, $\gamma=2$}
\end{subfigure}

                \caption{Physical (not dimensionless) weight functions evaluated for the double-mass-spring chain using various values of $\lambda$, two  representative values of $\gamma$, and fixed $\beta =0.5$}
                \label{FH_IFTallLam2}
\end{figure}

Dimensionless weight functions $\tilde E$ and $\tilde H$ obtained by inverse Fourier transformation are shown in Figure \ref{FH_IFTallLam}. For larger values of $\lambda$, 
function $\tilde E$ is concentrated in a more narrow interval
and its peak value is higher, while function $\tilde H$ spreads
over the same interval and its peak value is also higher.
However, it is important to consider that the meaning of the
dimensionless coordinate $\tilde r$ depends on $\lambda$.
The value of $\tilde r=1$ corresponds to distance $r=l=\lambda a$
in the physical space. Moreover, when the dimensionless 
weight functions $\tilde E$ and $\tilde H$ 
are transformed into the physical ones, 
their values are multiplied by $\bar E/l$, as follows 
from (\ref{eq:scaleE}) and (\ref{eq:scaleH}). Interestingly, 
the resulting weight function $E_0$  turns out to be
independent of the choice of  scaling factor $\lambda$,
as shown in Figure \ref{FH_IFTallLam2}a. 
This happens only in the particular case when $\gamma=1$.
Weight function $E_0$ constructed for instance for $\gamma=2$ 
is already slightly affected by $\lambda$, as shown in Figure \ref{FH_IFTallLam2}b. For the other weight functions,
$H_0$ and $A_0$, the effect of $\lambda$ is always felt,
even for $\gamma=1$; see Figure \ref{FH_IFTallLam2}c-f.

It is also interesting to observe that the weight functions,
in particular the nonlocal elastic modulus $E_0$ and the nonlocal micromorphic modulus $A_0$, strongly ``feel'' 
the interparticle spacing, $a$. The most important contribution
to the corresponding nonlocal integrals comes from
the neighborhood of radius $2a$. For the nonlocal coupling modulus,
$H_0$, the effective interaction distance increases with
increasing $\lambda$, i.e., with increasing characteristic 
length $l$, but even this function has clearly marked kinks
at distance $r=2a$. This means that these functions reflect
the discrete structure of the underlying mass-spring model,
which is periodic and consists of repeated units of size
$2a$.

When converting the nonlocal micromorphic model to a dimensionless format, the characteristic length was defined
as $l=\sqrt{\eta/\rho}$, i.e., as a parameter related to the
inertia effects. Alternatively, one could define 
a characteristic length related to stiffness, which would be
\beq\label{eq:78} 
l_s = \sqrt{\frac{\bar A}{\bar E}} = \sqrt{\frac{\int_{-\infty}^\infty A_0(r)\,{\rm d}r}{\int_{-\infty}^\infty E_0(r)\,{\rm d}r}}
\eeq 
Exact matches of the same dispersion diagram of the spring-mass chain
with alternating masses can be obtained with different combinations of length parameters $l$ and $l_s$, but if one of them is selected, the other is determined uniquely.
We have developed a procedure based on the selection of $l=\sqrt{\eta/\rho}$
as the primary independent parameter, 
%(subsequently transformed into $\lambda=l/a$). 
along with equivalent modulus $\bar E$ and characteristic
time $\tau=\sqrt{\eta/\bar{E}}$. After conversion to the
dimensionless form, we are able to determine dimensionless
weight functions; their identification is unique only 
if additional assumptions are made---function $\cA$
is assumed to be a multiple of function $\cE$ with
a fixed factor $\gamma$, and the ratio between the
characteristic length $\ell$ of the continuum model and
characteristic length $a$ of the discrete model is set
to a given value $\lambda$. The obtained dimensionless functions can be
transformed into their physical counterparts.
To complete the identification process, 
it is necessary to express the basic parameters of the
continuum model in terms of the four basic parameters of the
discrete model from Fig.~\ref{chain_scheme}, namely $a$, $m$, $M$, and $K$.
The relations that have already been established, i.e.,
$l=\lambda a$ and $\tau=\lambda\sqrt{(M+m)/(2K)}$,
can be rewritten in terms of the original parameters as
\bea \label{eq:79}
\frac{\eta}{\rho} &=& \lambda^2 a^2 \\
\label{eq:80}
\frac{\eta}{\bar E} &=& \lambda^2\frac{M+m}{2K}
\eea 
Note that $\lambda$ is not a true parameter of the discrete
model---along with $\gamma$ it is one of two parameters that
can be chosen more or less arbitrarily (within certain limits).
Therefore, the parameters of the continuum and discrete models
must satisfy only one constraint,
\beq \label{eq:81}
\frac{\bar E}{\rho} = \frac{2Ka^2}{M+m}
\eeq 

If the only objective was to capture the given dispersion diagram, we could select one of parameters $\bar E$ and $\rho$
arbitrarily and then evaluate the other one from the above
condition. Physically, this means that we would only need to
make sure that the long-wave limit of the elastic wave speed is the same for both models. However, it makes good sense to
add a simple condition that will endow $\rho$ with the physical
meaning of macroscopic density, and then $\bar E$ will
automatically get the meaning of macroscopic elastic modulus
that links, under homogeneous deformation, strain to stress.
The added condition can be based on the matching of kinetic
energy that corresponds to the motion of the whole bar
or chain as a rigid body at constant speed $v$. 
For the micromorphic continuum model, this motion
is characterized by $u(x,t)=u_0+vt$ and $\chi(x,t)=0$,
and the corresponding kinetic energy density (per unit volume)
is $\rho v^2/2$. For the discrete model with alternating masses $m$ and $M$ at distance $a$, the kinetic energy
per unit length would be $(M+m)v^2/(4a)$. To be able to
compare these expressions, we would need to define the area
of the continuous bar that represents the mass-spring chain.
However, this would be just an additional factor for conversion
between 3D and 1D, which does not need to be introduced if
we consider masses $M$ and $m$ as taken per unit area
(of the equivalent continuous bar) and also stiffness
$K$ as the actual spring stiffness divided by the area.
The fractions on the right-hand sides of (\ref{eq:80}) and (\ref{eq:81}) then remain the same while the kinetic energy $(M+m)v^2/(4a)$
is taken per unit volume and thus can be directly
set equal to $\rho v^2/2$. The resulting condition
\beq 
\rho = \frac{M+m}{2a}
\eeq 
naturally reflects the fact that the equivalent density of the continuous bar is obtained by uniform smearing
of the concentrated masses along the bar.
When this is substituted into (\ref{eq:81}), we obtain
\beq 
\bar E=Ka
\eeq 
as the expression for equivalent macroscopic modulus
of the continuum model. In this way, parameters
$\bar E$ and $\rho$, which would be preserved by the
standard continuum theory without any micromorphic or nonlocal
enhancements, can be directly deduced from the properties
of the mass-spring chain. Of course, the standard continuum
model is non-dispersive and provides only a linear 
acoustic branch of the dispersion diagram.

The non-standard parameter $\eta$ related to micromorphic
inertia can be expressed according to (\ref{eq:79}) as
\beq \label{e90}
\eta = \lambda^2 a^2\rho = \lambda^2 a\frac{M+m}{2}
\eeq 
However, $\lambda$ is a chosen parameter, independent of the
given characteristics of the mass-spring chain, and so 
$\eta$ is in fact arbitrary and $\lambda$ is just a 
conveniently transformed dimensionless value that 
can be used for comparison of the characteristic length
induced by the choice of $\eta$ with the spacing
between masses in the discrete model.

Once $\bar E$ and $\rho$ have been set to values that reflect the
properties of the mass-spring chain and dimensionless
parameters $\lambda$ (as a transformed version of the micromorphic
inertia parameter $\eta$) and $\gamma$ have been selected (within reasonable bounds),
the procedure described above leads to certain uniquely
defined nonlocal weight functions $E_0$, $A_0$, and $H_0$. 
It is then meaningful to characterize the relative ``strength''
of individual nonlocal effects by the ratios between the
integrals of the weight functions over the whole real axis.
In fact, the ratio
\beq\label{eq:78x} 
 \frac{\int_{-\infty}^\infty A_0(r)\,{\rm d}r}{\int_{-\infty}^\infty E_0(r)\,{\rm d}r} = \frac{\bar A}{\bar E} = l_s^2
\eeq 
is the square of the alternative (stiffness-related) characteristic length $l_s$, which was introduced in (\ref{eq:78}). 
Functions $E_0$ and $H_0$ have the same physical dimension, and so the fraction
\beq 
\frac{\int_{-\infty}^\infty H_0(r)\,{\rm d}r}{\int_{-\infty}^\infty E_0(r)\,{\rm d}r} = \frac{\bar H}{\bar E} = \kappa
\eeq 
is dimensionless and measures the ratio between the
coupling micromorphic stiffness and the standard stiffness.

Making use of  our previous results and assumptions, we can express
\bea 
\bar H &=& \int_{-\infty}^\infty H_0(r)\,{\rm d}r = \frac{\bar E}{l}\int_{-\infty}^\infty \tilde H_0(\tilde r)\,l\,{\rm d}\tilde r = \bar{E} \tilde\cH(0)
\\
\bar A&=&\int_{-\infty}^\infty A_0(r)\,{\rm d}r =\bar{E} l\int_{-\infty}^\infty \tilde A_0(\tilde r)\,l\,{\rm d}\tilde r = \bar{E}l^2 \tilde\cA(0) = \bar{E}l^2\gamma
\eea 
from which
\bea \label{eq:89}
l_s^2 &=& \frac{\bar A}{\bar E} = l^2\gamma \\
\kappa &=& \frac{\bar H}{\bar E} = \tilde\cH(0) = \tilde\omega_2^2(0)=\tau^2
\omega_2^2(0) = \frac{\eta\omega_2^2(0)}{\bar E}
\label{eq:90}
\eea 
Relation (\ref{eq:89}) means that, by choosing parameter $\gamma$, we set
the ratio between the squares of characteristic lengths $l_s$ and $l$,
which are related to micromorphic stiffness and micromorphic inertia,
respectively. From this point of view, it makes sense that $\gamma$
needs to be taken from a certain interval which always contains
value 1. For $\gamma=1$, both characteristic lengths coincide.
Relation (\ref{eq:90}) links the ratio $\kappa$ to parameters $\eta$
and $\bar E$ and to the circular frequency at the starting point
of the optical branch. For the considered discrete model, this frequency
is $\omega_2(0)=\sqrt{2K(M+m)/(mM)}$, and we get
\beq 
\kappa = \frac{2K(M+m)}{mM}\lambda^2\frac{M+m}{2K} = \lambda^2 \frac{(M+m)^2}{mM}
\eeq 
The fraction that involves $M$ and $m$ depends on the discrete model,
but parameter $\lambda$ is determined by our choice of the 
characteristic length related to micromorphic inertia. When more
weight is given to micromorphic inertia effects, then the area
under the weight function that couples the micromorphic strain
to the standard strain must also be increased, in proportion to the
square of the characteristic length.

\subsection{Remarks on the complexity of the model}

The approach used so far has been based on the assumption
that weight function $\tilde A$ is a multiple of weight function
$\tilde E$. It has been shown that, with a suitable choice
of the proportionality factor, this constraint
leads to a unique solution for all three weight functions involved. It is now logical to ask whether the class of
considered formulations could not be restricted in another
way, e.g., by reducing the general nonlocality 
imposed on all three terms of the free energy \eqref{free_en}
to
a formulation for which one of the terms remains local. 
This is formally equivalent to replacing one of the general weight functions by a multiple of the Dirac delta distribution, and the Fourier transform of such a weight function is a constant function for every wave number $\tilde k$. 

Based on Figure \ref{ellipse} it is possible to explain why neither $\tilde A(\tilde r)$ nor $\tilde E(\tilde r)$ can be chosen as a multiple of the Dirac distribution. Each individual ellipse depicts the set of all admissible solutions for a fixed wave number. A constant Fourier image $\tilde\cA$ (or $\tilde\cE$) would correspond to finding one fixed vertical (or horizontal) line that intersects the ellipses constructed for all wave numbers $\tilde k$, which is not possible.
The reason is that as the wave number increases, the corresponding
ellipse
shrinks and its center approaches the origin in the 
$(\tilde\cE,\tilde\cA)$ plane. The projection
of the ellipse
on the vertical axis is a finite interval $[\tilde\cA_{min},\tilde\cA_{max}]$, and from the description of the ellipse
by equation (\ref{ellGen}) it follows that 
\bea
\tilde\cA_{max} &=& \dfrac{(\tilde\omega_1^2+\tilde\omega_2^2)\tilde k+\vert\tilde\omega_2^2-\tilde\omega_1^2\vert\sqrt{\tilde k^2+1}}{2\tilde k^3} 
\eea 
%\bea
%\tilde\cE_{min} &=& \dfrac{\tilde\omega_2^2+\tilde\omega_1^2-\vert\tilde\omega_2^2-\tilde\omega_1^2\vert\sqrt{\tilde k^2+1}}{2\tilde k^2} \\
%\tilde\cE_{max} &=& \dfrac{\tilde\omega_2^2+\tilde\omega_1^2+\vert\tilde\omega_2^2-\tilde\omega_1^2\vert\sqrt{\tilde k^2+1}}{2\tilde k^2} 
%\eea 
This upper limit depends on the wave number $\tilde k$ not only
explicitly, but also through the circular frequencies $\tilde\omega_{1,2}$. Substituting the specific expressions (\ref{eq57})
that describe the dimensionless dispersion diagram of the double-mass-spring model, we obtain
\beq\label{eq:Amax}
\tilde\cA_{max}(\tilde k) =\frac{\lambda^2(1+\beta)^2}{2\beta \tilde k^3} \left(\tilde k+\sqrt{(\tilde k^2+1)\left(1-\frac{4\beta}{(1+\beta)^2}\sin^2\frac{\tilde k}{\lambda}\right)}\right) \le 
\frac{\lambda^2(1+\beta)^2}{2\beta} \cdot\frac{\tilde k+\sqrt{\tilde k^2+1}}{ \tilde k^3}
\eeq
where $\lambda$ and $\beta$ are given positive parameters (notice that $4\beta/(1+\beta)^2\le 1$). Based on the inequality
in (\ref{eq:Amax}), it is easy to prove that $\lim_{\tilde k\to\infty}\tilde\cA_{max}(\tilde k)=0$. Therefore, for any fixed
 $\tilde\cA^*>0$, there exist certain wave numbers $\tilde k$
 for which $\tilde\cA_{max}(\tilde k)<\tilde\cA^*$ and thus
 the horizontal straight line
at $\tilde\cA=\tilde\cA^*$ does not intersect the corresponding
ellipses. Consequently,
 the dispersion relation cannot be satisfied 
for all wave numbers if the weight function $\tilde A$ is
a multiple of the Dirac distribution. 

In a very similar way, it is possible
to show that the upper limit 
\beq \label{eq:Emax}
%\tilde\cE_{min}(\tilde k) &=&\frac{\lambda^2(1+\beta)^2}{2\beta \tilde k^2} \left(1-\sqrt{(\tilde k^2+1)\left(1-\frac{4\beta}{(1+\beta)^2}\sin^2\frac{\tilde k}{\lambda}\right)}\right)
%\\
\tilde\cE_{max}(\tilde k) =\frac{\lambda^2(1+\beta)^2}{2\beta \tilde k^2} \left(1+\sqrt{(\tilde k^2+1)\left(1-\frac{4\beta}{(1+\beta)^2}\sin^2\frac{\tilde k}{\lambda}\right)}\right) \le 
\frac{\lambda^2(1+\beta)^2}{2\beta} \cdot\frac{1+\sqrt{\tilde k^2+1}}{ \tilde k^2}
\eeq
of the interval 
$[\tilde\cE_{min},\tilde\cE_{max}]$ that represents the
projection of the ellipse on the horizontal axis also tends to zero as the wave number tends to infinity, and so it is not possible to choose a fixed positive value $\tilde\cE^*$ of Fourier image $\tilde\cE$ and obtain a real solution
$\tilde\cA$ of equation (\ref{constEq}) for all wave numbers.
In fact, the only candidate would be $\tilde\cE^*=1$
because the adopted conversion to the dimensionless formulation
always yields $\tilde\cE(0)=1$, but this does not invalidate
the reasoning.

%To prove such a statement, it is sufficient to find two wave numbers for which the projection of the first ellipse on the vertical (or horizontal) axis has no intersection with the projection of the second ellipse. 
%Such an example needs to be found for both axes $\tilde\cA$ and $\tilde\cE$. 
 %From the description of the ellipse
%by equation (\ref{ellGen}) it is clear that the projection
%on the horizontal axis is the interval $[\tilde\cE_{min},\tilde\cE_{max}]$ where
%\bea
%\tilde\cE_{min} &=& \dfrac{\tilde\omega_2^2+\tilde\omega_1^2-\vert\tilde\omega_2^2-\tilde\omega_1^2\vert\sqrt{\tilde k^2+1}}{2\tilde k^2} \\
%\tilde\cE_{max} &=& \dfrac{\tilde\omega_2^2+\tilde\omega_1^2+\vert\tilde\omega_2^2-\tilde\omega_1^2\vert\sqrt{\tilde k^2+1}}{2\tilde k^2} 
%\eea 
%These bounding values depend on the wave number $\tilde k$ not only
%explicitly, but also through the circular frequencies $\tilde\omega_{1,2}$. Substituting the specific expressions (\ref{eq57})
%that describe the dimensionless dispersion diagram of the double-mass-spring model, we obtain
%\bea 
%\tilde\cE_{min}(\tilde k) &=&\frac{\lambda^2(1+\beta)^2}{2\beta \tilde k^2} \left(1-\sqrt{(\tilde k^2+1)\left(1-\frac{4\beta}{(1+\beta)^2}\sin^2\frac{\tilde k}{\lambda}\right)}\right)
%\\
%\tilde\cE_{max}(\tilde k) &=&\frac{\lambda^2(1+\beta)^2}{2\beta \tilde k^2} \left(1+\sqrt{(\tilde k^2+1)\left(1-\frac{4\beta}{(1+\beta)^2}\sin^2\frac{\tilde k}{\lambda}\right)}\right) \le 
%\frac{\lambda^2(1+\beta)^2}{2\beta} \cdot\frac{1+\sqrt{\tilde k^2+1}}{ \tilde k^2}
%\eea 

So far, it has been demonstrated that the terms with
weight functions $E_0$ and $A_0$ must remain truly nonlocal (i.e., regular weight functions need to be used).  
The last case to be explored is what happens if function  $\tilde H(\tilde r)$ is chosen as a multiple of the Dirac distribution, leading to a constant Fourier image $\tilde\cH(\tilde k)$. In view of (\ref{eq:90}), the constant value would correspond
to the previously introduced parameter 
$\kappa =\bar H/\bar E$. 
In our general framework, $\tilde\cH$ is linked to the
other Fourier images $\tilde\cE$ and $\tilde\cA$ by
relation (\ref{mm9x}), and so the assumption that $\tilde\cH(\tilde k)=\kappa$ leads to 
\beq\label{mm9z}
\frac{\tilde\omega_1^2+\tilde\omega_2^2-(\tilde\cA+\tilde\cE)\,\tilde k^2}{1+\tilde k^2}=\kappa
\eeq
Using this constraint, $\tilde\cA$ can be expressed as
\beq\label{cAdirx}
 \tilde\cA = \dfrac{\tilde\omega_1^2+\tilde\omega_2^2-(1+\tilde k^2)\kappa}{\tilde k^2}-\tilde\cE
\eeq
and eliminated from \eqref{constEq}, which leads to a quadratic equation for $\tilde\cE$ with a discriminant  given by
\beq 
D = \tilde k^4\left(\left(\tilde\omega_2^2-\tilde\omega_1^2\right)^2-4\tilde k^2\kappa^2\right)
\eeq 
To obtain a real Fourier image $\tilde\cE$, we would need
$D\ge 0$ for all $\tilde k\ge 0$. For $\kappa>0$, this would be possible
only if $\tilde\omega_2^2-\tilde\omega_1^2=O(\tilde k)$ 
as $\tilde k\to\infty$. Some dispersion diagrams may satisfy
such a condition, but if both branches are bounded,
the condition is violated and the only possibility would
be to set $\kappa$ to zero, which leads to a model with
no coupling between the displacement and the micromorphic strain. For instance, for the double-mass-spring chain, 
the dimensionless dispersion relation (\ref{eq57}) implies that
\beq 
\tilde\omega_2^2-\tilde\omega_1^2 = \lambda^2\frac{1+\beta}{\beta}\sqrt{(1+\beta)^2-4\beta\sin^2\frac{\tilde k}{\lambda}}
\le \lambda^2\frac{(1+\beta)^2}{\beta}
\eeq 
and the discriminant becomes negative for sufficiently large
wave numbers (of course, unless $\kappa=0$).

In summary, in this subsection it has been shown that
nonlocality is needed in all three terms of the free energy function
in order to reproduce the dispersion diagram of the double-mass-spring chain.

\subsection{Mass-spring chain with equal masses}
\label{sec:equalmasschain}
In this section, we investigate how the weight functions look when the masses in the double-mass-spring chain are equal, i.e., when $\beta=m/M=1$. 
For such a simple mass-spring chain, the dispersion diagram 
has only one (acoustic) branch. 
However, substituting $\beta=1$ into equation
\eqref{eq57} that describes the double-mass chain,
we obtain
\beq
 \tilde\omega_{1,2}^2 = 2\lambda^2 \left(1\pm  \cos{\frac{\tilde k }{\lambda}} \right) =\begin{cases}
4\lambda^2  \cos^2{\dfrac{ \tilde k }{2\lambda}} \\[2mm]
4\lambda^2  \sin^2{\dfrac{ \tilde k }{2\lambda}}
\end{cases}
\eeq
The corresponding dispersion diagram is visualized by the dash-dotted line in Figure \ref{doubleMassDispRef}. Note that two branches still exist, since the two masses are treated separately.   

The branches obtained for $\beta=1$ intersect at those wave numbers $\tilde k$ for which 
$\cos(\tilde k/\lambda)=0$. This has important consequences. If $\tilde\omega_1=\tilde\omega_2$
for some $\tilde k$, the ellipse described by (\ref{ellGen})
reduces to a single point with coordinates
$\tilde\cE=\tilde\cA=\tilde\omega_1^2/\tilde k^2$,
and it is intersected only by one line passing through the origin---the one with the unit slope. Therefore, $\gamma=1$ is the only possible
choice for this model. Furthermore, when $\tilde k$
passes through one of the points at which $\cos(\tilde k/\lambda)=0$, the expression under the square root in (\ref{Efsol0})
becomes zero and it is possible, for the subsequent interval until the next $\cos(\tilde k/\lambda)=0$, to revert the sign before the square root without losing continuity.  
As a result, the expression for the Fourier image $\tilde \cE$ based on formula \eqref{ellPar} can be written as
\beq\label{ellParx}
\tilde\cE =
\frac{2\lambda^2}{\tilde k^2}\left(1-\cos\frac{\tilde k}{\lambda}\right)
\eeq
without the need to replace the cosine term with its absolute value. 
The inverse Fourier transform of such a function can be evaluated analytically, leading to 
\beq
\tilde E(\tilde r) = \lambda \,\max\left(1-\lambda|\tilde r|,0\right) = \lambda \,\langle 1-\lambda|\tilde r|\rangle
\eeq
where the Macauley brackets $\langle\ldots\rangle$ denote the positive part. Since $\gamma=1$, we have $\tilde\cA=\tilde\cE$
and $\tilde A=\tilde E$. After transformation into the physical space based on scaling relations (\ref{eq:scaleE})--(\ref{eq:scaleA}), the weight functions are given by
\bea \label{eq:E0r}
E_0(r) &=& \frac{\bar{E}}{l}\, \tilde{E}\left(\frac{r}{l}\right) =
\frac{\bar{E}}{l}\lambda \,\left\langle 1-\frac{\lambda|r|}{l}\right\rangle = \frac{\bar{E}}{a}\,\left\langle 1-\frac{|r|}{a}\right\rangle
\\
A_0(r) &=&  \bar{E}l \,\tilde{A}\left(\frac{r}{l}\right) =
\bar{E}l\lambda \,\left\langle 1-\frac{\lambda|r|}{l}\right\rangle = l^2\frac{\bar{E}}{a}\,\left\langle 1-\frac{|r|}{a}\right\rangle
\eea 
Here we have exploited the relation $\lambda=l/a$ to express
the final results in terms of the particle spacing $a$ and
characteristic length $l$. It is remarkable that weight
function $E_0$ depends only on the physical parameters, 
$\bar{E}$ and $a$, and is independent of the choice of $l$. 
The second weight function, $A_0$, is simply $E_0$ multiplied by the square of the selected characteristic length, which can be arbitrarily adjusted.

The Fourier image $\tilde \cH$ of the third weight function is in the present case given by 
\beq
\tilde\cH=
\dfrac{4 \lambda^2}{1+\tilde k^2}\cos{\frac{\tilde k }{\lambda}}
\eeq
and the corresponding dimensionless weight function obtained analytically by the inverse Fourier transform has the form 
\beq
\tilde H(\tilde r) = 
\lambda^2{\left({\mathrm{e}}^{-\left|1/\lambda +\tilde r\right|} +{\mathrm{e}}^{-\left|1/\lambda -\tilde r\right|} \right)}
\eeq
from which
\beq 
H_0(r) =  \frac{\bar{E}}{l} \tilde H\left(\frac{r}{l}\right)
=\frac{\bar{E}}{l}\lambda^2{\left({\mathrm{e}}^{-\left|1/\lambda +r/l\right|} +{\mathrm{e}}^{-\left|1/\lambda -r/l\right|} \right)}=\frac{\bar{E}l}{a^2}\left({\mathrm{e}}^{-\left|r+a\right|/l} +{\mathrm{e}}^{-\left|r-a\right|/l} \right)
\eeq 
The model with the derived weight functions $E_0$, $A_0$, and $H_0$ should always give the same dispersion diagram, independently of the choice of parameter $l$. 
As $l\to 0^+$, functions $A_0$ and $H_0$ tend to zero functions and the model reduces to a simpler nonlocal
strain model, with no micromorphic terms. 
This is consistent with the results described in \cite{jirasek2004nonlocal}, where the dispersion 
diagram of a simple mass-spring chain was shown to be
exactly reproduced by a nonlocal strain model with
weight function (\ref{eq:E0r}).

Recall that,
according to (\ref{e90}), the micromorphic inertia $\eta$ is
proportional to $\lambda^2 a$ and therefore to $l^2$
(for fixed physical parameters $a$, $m$, and $M$ that characterize the mass-spring chain). In the limit,
when $l=0$, the micromorphic inertia vanishes 
and the micromorphic strain field does not affect the
kinetic or potential energy. The optical branch of
the dispersion diagram then totally vanishes. 
Since $\eta$ was used in the definitions of the characteristic time and characteristic length, this case needs to be analyzed separately, starting again from the dispersion equation \eqref{mm4}.
For $\eta=0$, this equation simplifies to
\beq\label{eq:110}
- \rho(\cA k^2+\cH)\omega^2+\cA(\cE+\cH)k^4+\cE\cH k^2 =0
\eeq
which is a single equation for three unknown functions $\cE$, $\cA$, and $\cH$. Let us now substitute the dispersion relation for a single mass-spring chain, given by
\beq
\omega^2 = \dfrac{4K}{m} \sin^2{\frac{a  k }{2}}
\eeq
and exploit (\ref{eq:110}) to express 
\beq
\cE = \dfrac{\rho \omega^2 (\cH +\cA k^2) - \cA \cH k^4 }{\cH k^2 + \cA k^4} = 
\dfrac{4 \rho K}{mk^2} \sin^2{\dfrac{ak}{2}} 
-
\dfrac{ \cA \cH k^2 }{\cH + \cA k^2} 
\eeq
In the special case of vanishing $\cA$, which corresponds to the standard integral continuum (non-micromorphic), the previous equation reduces to 
\beq
\cE =  \dfrac{ 4 \rho K}{mk^2}\sin^2{\dfrac{ak}{2}} =\dfrac{ 2 \rho K}{mk^2}\left(1-\cos{ak}\right) 
\eeq
which is the same function as was derived in 
\eqref{ellParx} for the case of nonvanishing microinertia, just not normalized.

\subsection{Mass-spring chain with alternating masses and second-nearest-neighbor interactions}
\begin{figure}[H]
\centering%
\includegraphics[width=0.45\textwidth, keepaspectratio=true]{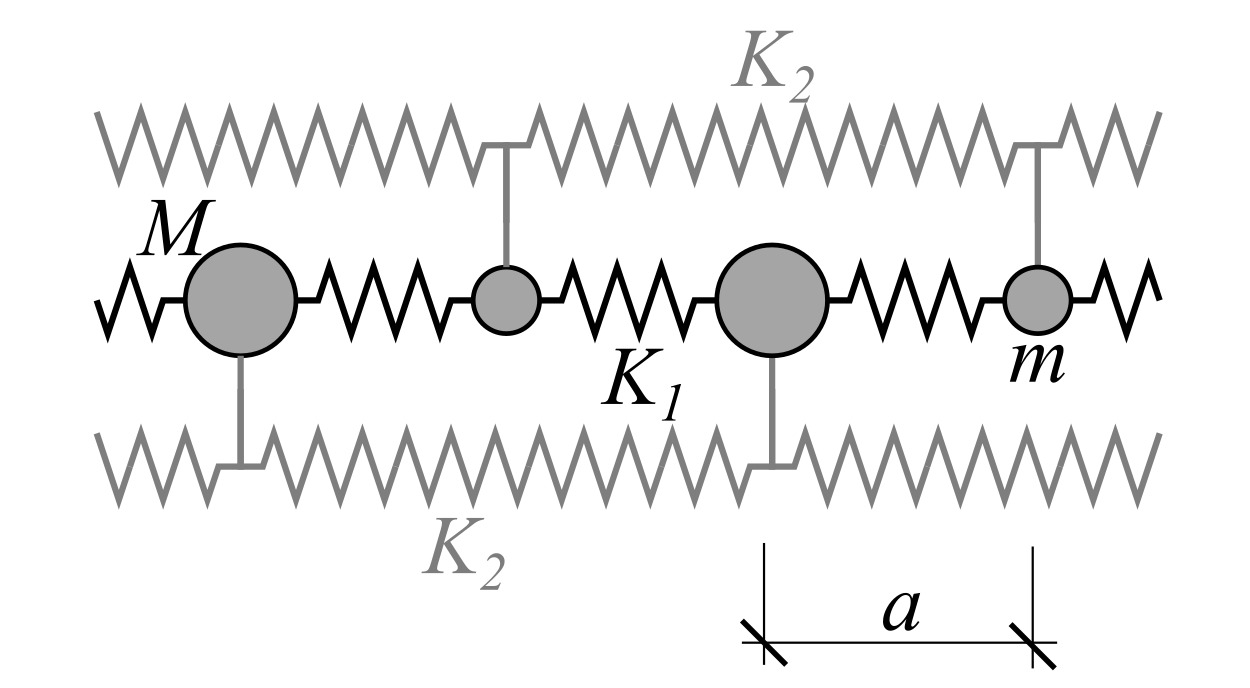}
                \caption{Mass-spring chain with two alternating masses                
                }
                \label{chain_scheme2}
\end{figure} 
A slightly more complex dispersion diagram can be obtained by adding second-nearest-neighbor interactions to the double-mass-spring chain; see Figure \ref{chain_scheme2}. By a specific combination of the mass and stiffness parameters, the initial curvature of the dispersion diagram can be changed to positive, while without the long-range interaction, it is always negative.
Let us denote the displacement of the larger mass $M$ at position $x=2na$ by $u_{2n}$ and the displacement of the smaller mass $m$ at the neighbouring position $x=(2n+1)a$ by $v_{2n+1}$, and let $K_1$ be the stiffness of the springs connecting the nearest neighbors and $K_2$ the stiffness of the springs between the second-nearest neighbors. 
The equations of motion then read 
\bea
M \ddot{u}_{2n} &=& K_1(v_{2n+1} - 2u_{2n} + v_{2n-1}) + K_2(u_{2n+2} - 2u_{2n} + u_{2n-2})
\\
m \ddot{v}_{2n+1} &=& K_1(u_{2n+2} - 2v_{2n+1} + u_{2n})+ K_2(v_{2n+3} - 2v_{2n+1} + v_{2n-1})
\eea
Substituting the harmonic ansatz
\begin{eqnarray}
u_{2n}(t) &=& U\,{\rm e}^{i(2nka-\omega t)}
\\
v_{2n+1}(t) &=& V\,{\rm e}^{i((2n+1)ka-\omega t)}
\end{eqnarray}
we obtain the system of equations 
\begin{eqnarray}
\nonumber
-M\omega^2 U &=& K_1\left(V({\rm e}^{ika}+{\rm e}^{-ika}) - 2U\right) + K_2\left(U({\rm e}^{2ika}+{\rm e}^{-2ika}) - 2U\right)  
\\
\nonumber
-m\omega^2 V &=& K_1\left(U({\rm e}^{ika}+{\rm e}^{-ika}) - 2V\right)+ K_2\left(V({\rm e}^{2ika}+{\rm e}^{-2ika}) - 2V\right)  
\end{eqnarray}
which has a nontrivial solution only if 
\begin{eqnarray}\label{eq91}
{\rm det}\begin{pmatrix}
-M\omega^2 + 2K_1+ 4K_2 \sin^2{ka}  & -2K_1\cos{ka} 
\\
-2K_1\cos{ka}  & -m\omega^2 + 2K_1+ 4K_2 \sin^2{ka}
\end{pmatrix} = 0
\end{eqnarray}
Solving this characteristic equation, we get the dispersion relation
% \begin{equation} \label{dispRel}
% \omega_{1,2}^2 = \dfrac{K_1}{m} \left(1+\beta + 2\beta \xi \sin^2{ k a} \pm \sqrt{(1+\beta + 2\beta \xi \sin^2{ k a})^2  - 4\beta(2\xi+1)\sin^2{ k a} }\right)
% \end{equation}
\begin{equation} \label{dispRel2}
\omega_{1,2}^2 = \dfrac{K_1}{m} \left((1+\beta )(1+2\xi \sin^2{ k a}) \pm \sqrt{(1+\beta)^2 (1+2\xi \sin^2{ k a})^2  - 4\beta(4\xi^2\sin^2{ k a}+4\xi+1)\sin^2{ k a} }\right)
\end{equation}
in which $\xi=K_2/K_1$ is the stiffness ratio. 

Same as in the case of the original discrete model, 
transformation to the dimensionless form should be done such that
the slope of the acoustic branch at the origin equals 1.
%in order to determine the ratio $l/\tau$ of the , 
Therefore, one needs to evaluate 
\beq
\frac{{\rm d}\omega_1(0)}{{\rm d}k} = \sqrt{\frac{2\beta K_1(1+4\xi)}{(1+\beta)m}}\,a = \sqrt{\frac{2(K_1+4K_2)}{M+m}}\,a 
\eeq 
and then select the normalizing parameters $l$ and $\tau$ such that
their ratio meets the condition
\beq\label{eq:117}
\frac{l}{\tau} = \sqrt{\frac{2\beta K_1(1+4\xi)}{(1+\beta)m}}\,a = \sqrt{\frac{2(K_1+4K_2)}{M+m}}\,a 
\eeq 
which guarantees that  ${\rm d}\tilde\omega(0)/{\rm d}\tilde k=1$.
For instance, one can set the characteristic length $l$ to a given multiple of the distance between the masses, i.e.,\ $l=\lambda a$, 
and determine the characteristic time $\tau$ from (\ref{eq:117}).
The resulting dimensionless form of \eqref{dispRel2} reads
\begin{equation} \label{eq95}
\tilde\omega_{1,2}^2 = \lambda^2 \frac{1+\beta}{2\beta (1+4\xi)} \left((1+\beta )\left(1+2\xi \sin^2{ \frac{\tilde k}{\lambda}}\right) \pm \sqrt{(1+\beta)^2 \left(1+2\xi \sin^2{\frac{\tilde k}{\lambda}}\right)^2  - 4\beta\left(4\xi^2\sin^2{ \frac{\tilde k}{\lambda}}+4\xi+1\right)\sin^2{\frac{\tilde k}{\lambda}} }\right)
\end{equation}

\begin{figure}[H]
\centering%
\begin{subfigure}[b]{0.49\textwidth}

                \includegraphics[width=\textwidth, keepaspectratio=true]{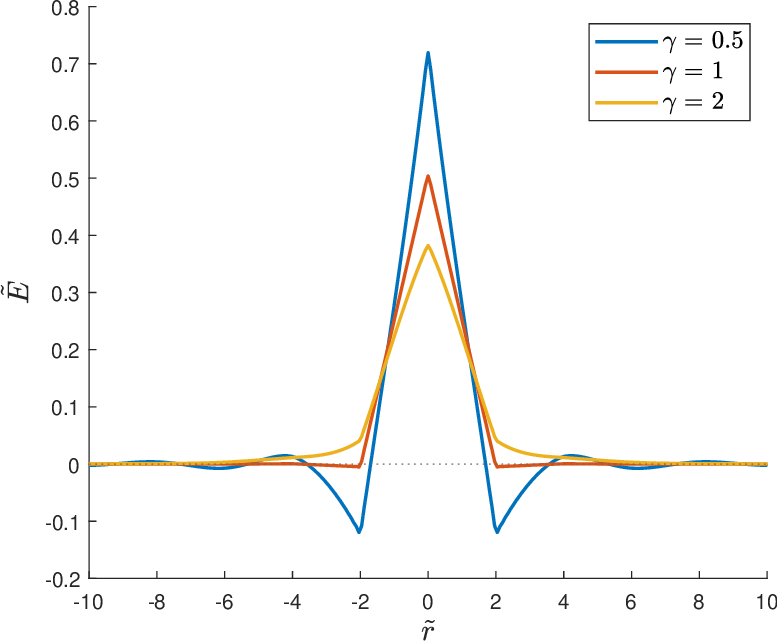}
                 \subcaption{Weight function $\tilde E$}
\end{subfigure}
\begin{subfigure}[b]{0.49\textwidth}            
               
                \includegraphics[width=\textwidth, keepaspectratio=true]{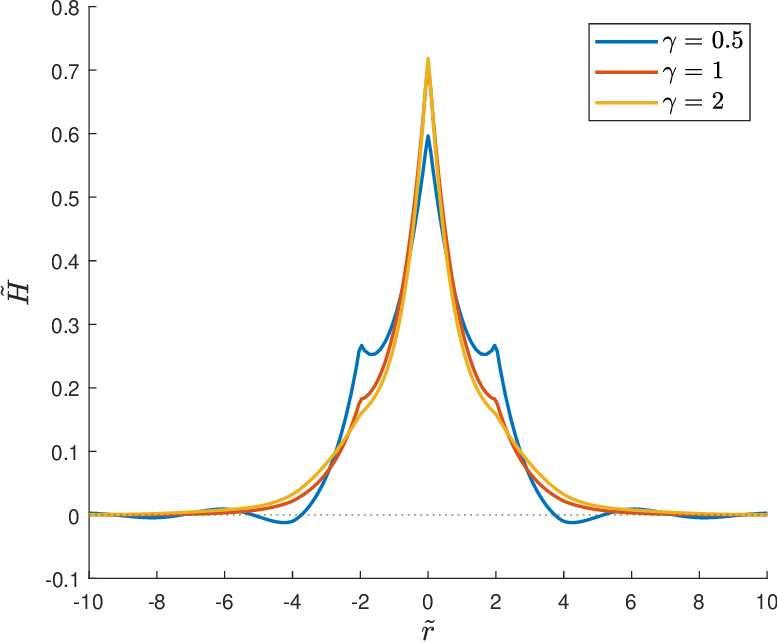}
                 \subcaption{Weight function $\tilde H$}
\end{subfigure}

                \caption{Weight functions computed by IDFT for the double-mass-spring chain with second-neighbour interactions, using model parameters $\beta =0.5$, $\xi=0.4$ and $\lambda=1$ and numerical parameters $\tilde k_0=40$ and $N=400$}
                \label{FH_IFTall_2}
\end{figure}

\begin{figure}[H]
\centering%
\begin{subfigure}[b]{0.49\textwidth}
 
                \includegraphics[width=\textwidth, keepaspectratio=true]{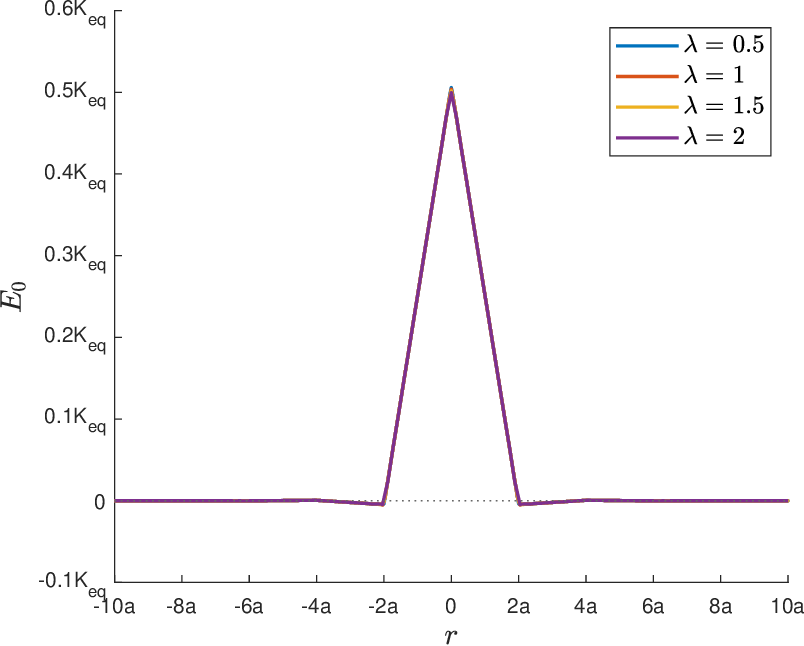}
                 \subcaption{Weight function $E_0$, $\gamma=1$}
\end{subfigure}
\begin{subfigure}[b]{0.49\textwidth}            
               
                \includegraphics[width=\textwidth, keepaspectratio=true]{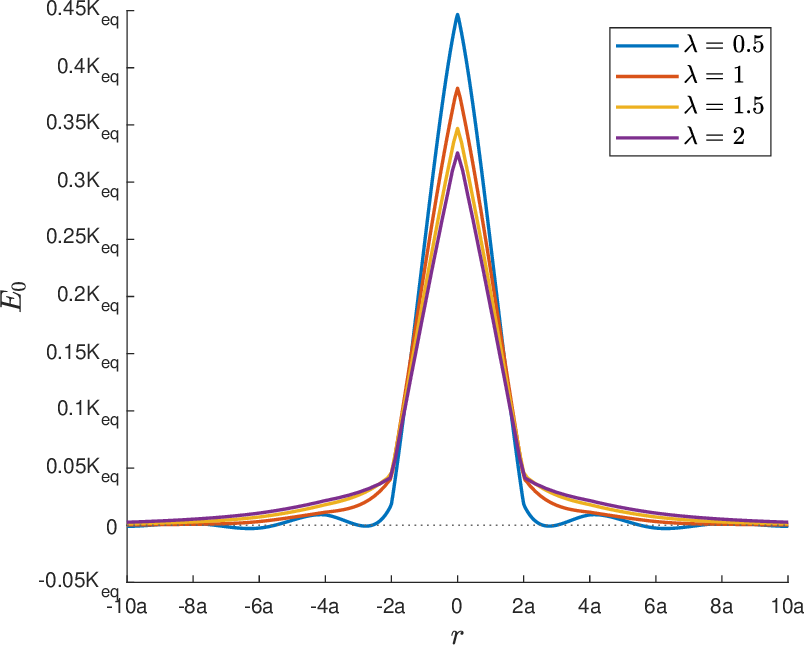}
                 \subcaption{Weight function $E_0$, $\gamma=2$}
\end{subfigure}
\\
\begin{subfigure}[b]{0.49\textwidth}
 
                \includegraphics[width=\textwidth, keepaspectratio=true]{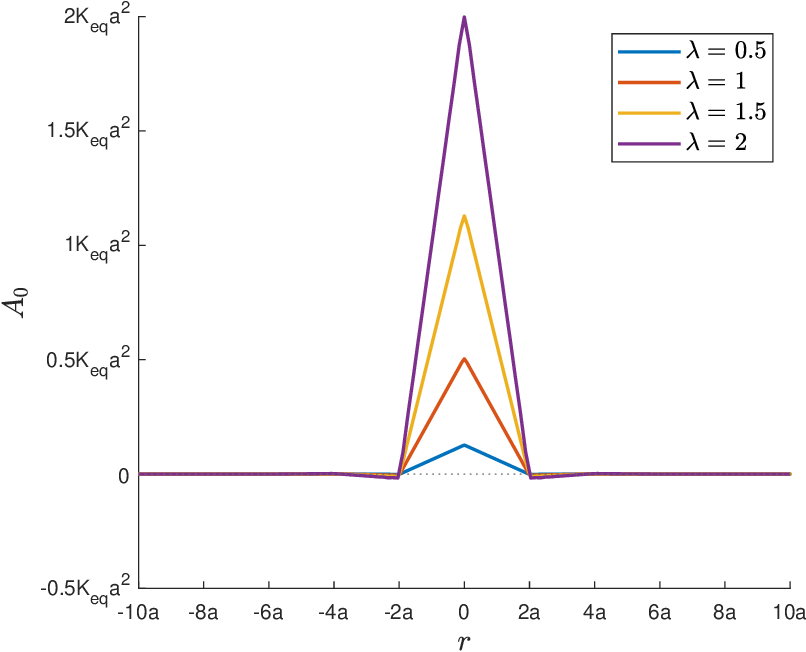}
                 \subcaption{Weight function $A_0$, $\gamma=1$}
\end{subfigure}
\begin{subfigure}[b]{0.49\textwidth}            
               
                \includegraphics[width=\textwidth, keepaspectratio=true]{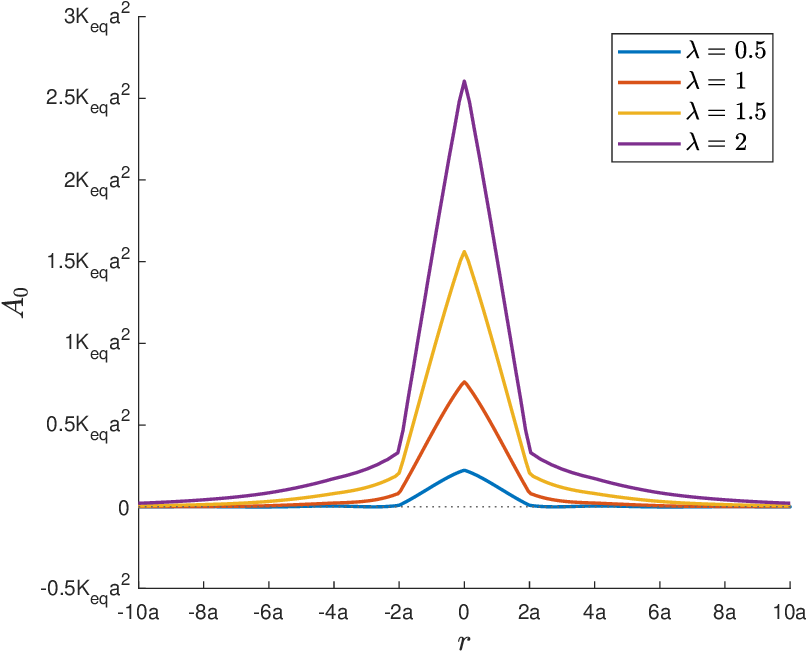}
                 \subcaption{Weight function $A_0$, $\gamma=2$}
\end{subfigure}
\\
\begin{subfigure}[b]{0.49\textwidth}
 
                \includegraphics[width=\textwidth, keepaspectratio=true]{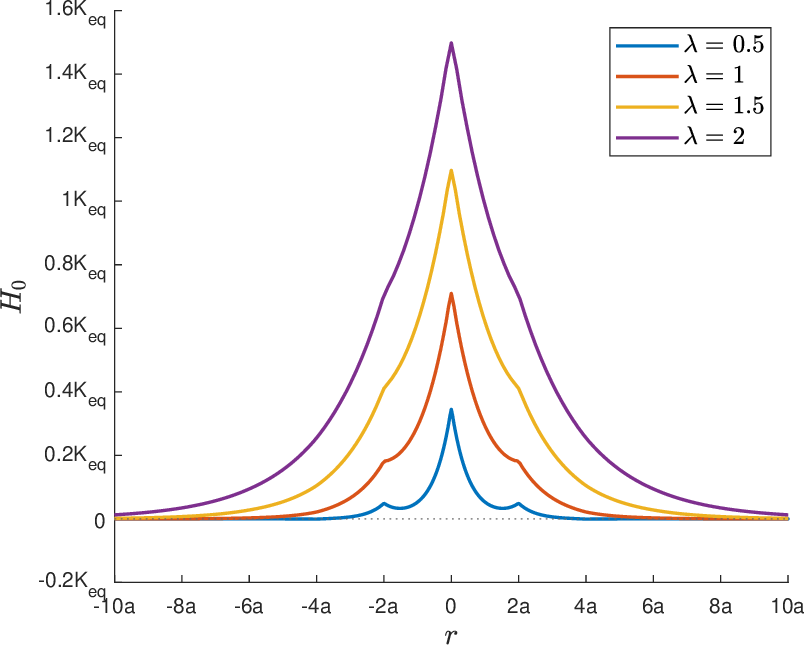}
                 \subcaption{Weight function $H_0$, $\gamma=1$}
\end{subfigure}
\begin{subfigure}[b]{0.49\textwidth}            
               
                \includegraphics[width=\textwidth, keepaspectratio=true]{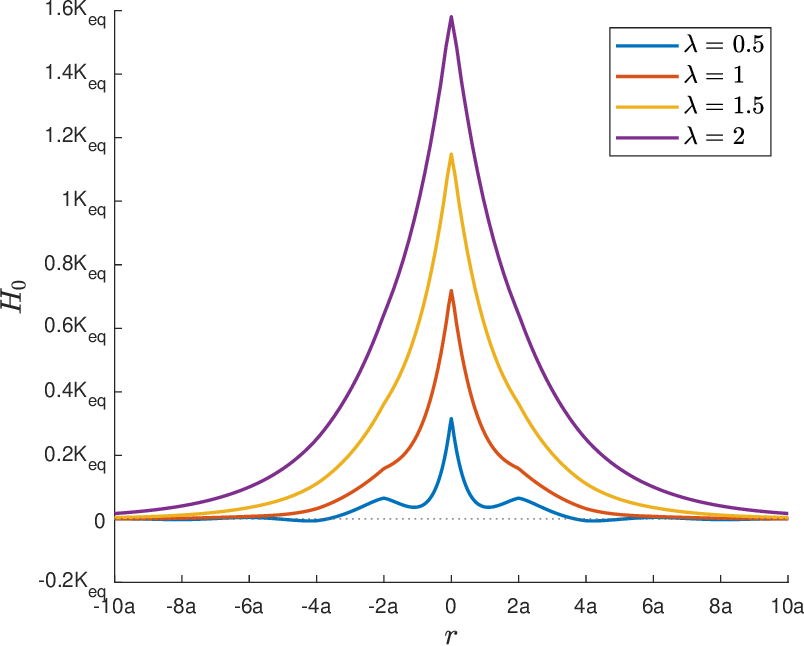}
                 \subcaption{Weight function $H_0$, $\gamma=2$}
\end{subfigure}

                \caption{Weight functions (in physical space) evaluated for the double-mass-spring chain with second-nearest-neighbor interactions using various values of parameter $\lambda$ and two different values of parameter $\gamma$, with fixed $K_{eq}=K_1+4K_2$, $\beta =0.5$ and $\xi=0.4$}
                \label{FH_IFTallLam23}
\end{figure}

\begin{figure}[H]
\centering%       

                \includegraphics[width=0.55\textwidth, keepaspectratio=true]{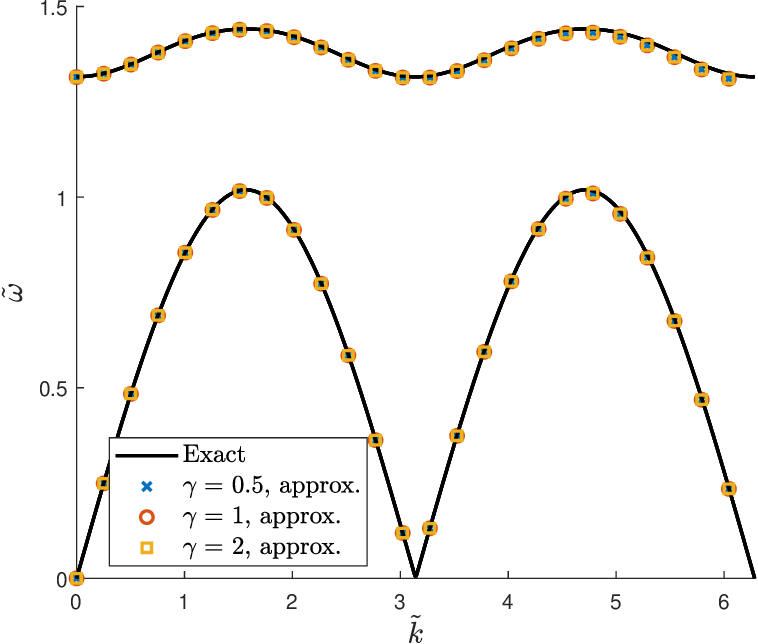}

                \caption{Comparison of the original dispersion diagram of the double-mass-spring chain with second-neighbor interactions and its replica based on the enriched continuum formulation with parameters $\beta = 0.5$, $\xi=0.4$ and $\lambda=1$ }
                \label{chainapp_2}
\end{figure}

Adopting the same procedure as described in Section~\ref{sec:doublemasschain} for the chain with only nearest-neighbor interactions, we can construct 
an enriched continuum model that exactly reproduces the
dispersion relation. 
In Figure \ref{FH_IFTall_2}, the numerically evaluated dimensionless weight functions obtained for parameters $\xi=K_2/K_1=0.4$ and $\beta=m/M=0.5$ are displayed. The IDFT of the Fourier images of weight functions has been performed with numerical parameters $\tilde k_0=40$ and $N=400$. To ensure that function $\tilde \cE$ is real for every wave number $\tilde k$, parameter $\gamma$ must be selected from  $[0.4984, 2.6042]$. 
Interestingly, for $\gamma=1$, the resulting weight function $E_0$ scaled into the physical space does not depend on parameter $\lambda$
that sets the characteristic length for conversion into the dimensionless format. 
The length scale is dictated directly by the particle spacing $a$; see Figure \ref{FH_IFTallLam23}a. Weight function $A_0$
has the same support and shape for different values of $\lambda$, but its magnitude is affected by the choice of $\lambda$; see Figure \ref{FH_IFTallLam23}c.
For other choices of $\gamma$, slightly different weight functions are obtained but 
their overall look remains similar.
 The agreement between the original dispersion diagram of the discrete model and its replica based on the continuum model is excellent
for all parameter choices, as shown in Figure \ref{chainapp_2}.
% In Figure \ref{chainapp_2} the exact and approximated dispersion diagrams are plotted for the parameters $\xi=1.5$ and $\beta=m/M=0.5$. The parameter $\gamma$ must be selected from interval $[0.4998 \ 2.2766]$ in order to ensure that the function $\tilde \cE$ is real for every wave number $\tilde k$. 
% % The IDFT of the Fourier images of the weight functions was performed with parameters $\tilde k_0=40$, $N=800$. In Figure \ref{chainappA2} the comparison of the exact dispersion diagram and its approximations is plotted for various parameters $\gamma$, and in Figure \ref{chainappB2} the absolute error of both optical and acoustic branches is visualized. Note that the results for all the considered $\gamma$ parameters are almost identical, hence in Figure \ref{chainapp_2} all the lines for single branch coincide. 
% The IDFT of the Fourier images of the weight functions was performed with parameters $\tilde k_0=40$, $N=800$. 

% \begin{figure}[H]
% \centering%       
% \begin{subfigure}[b]{0.49\textwidth}

%                 \includegraphics[width=\textwidth, keepaspectratio=true]{GM_chainAppGen_2.eps}
%                 \subcaption{Dispersion diagram}\label{chainappA2}
% \end{subfigure}
% \begin{subfigure}[b]{0.49\textwidth}                  \includegraphics[width=\textwidth, keepaspectratio=true]{GM_chainErr_2.eps} 
%                 \subcaption{Absolute Error}\label{chainappB2}
% \end{subfigure}
%                 \caption{Comparison of the exact dispersion diagram of the double mass-spring chain with second neighbour interactions and its enriched continuum approximation, $\beta = 0.8$, $\xi=1$ }
%                 \label{chainapp_2}
% \end{figure} 

\subsection{Reconstruction of the dispersion diagram with band gap obtained for local micromorphic model with vanishing micromorphic modulus}

In \cite{jirasek2022integral} it was shown that, in the case of the 1D local micromorphic model, a band gap in the dispersion diagram can only be obtained when the micromorphic modulus (in \cite{jirasek2022integral} denoted by $A$) is set to zero. 
The dispersion relation then reads
\bea\label{e119}
\omega_{1,2}^2 =  \frac{k^2(\bar E+\bar H)\eta+\bar H\rho}{2\rho \eta} \pm \frac{\sqrt{\left(k^2(\bar E+\bar H)\eta-\bar H\rho\right)^2 + 4k^2\bar H^2\rho\eta}}{2\rho\eta}  
\eea
Normalizing the frequency and the wave number as described in section \ref{dimFor} and introducing parameter $\kappa=\bar H/ \bar E$, we can rewrite equation (\ref{e119}) as
\bea \label{disp_mm}
\tilde \omega_{1,2}^2 =  \frac{\tilde k^2(1+\kappa)+\kappa}{2} \pm \frac{\sqrt{\left(\tilde k^2(1+\kappa)-\kappa\right)^2 + 4\tilde k^2\kappa^2}}{2} 
\eea
from which 
\bea
\tilde\omega_1^2+\tilde\omega_2^2 &=& \tilde k^2(1+\kappa)+\kappa \label{omsum2} \\
\tilde\omega_1^2\tilde\omega_2^2 &=&\tilde k^2 \kappa \label{ommult2}
\eea
In this section, we show how the present nonlocal micromorphic model approximates such dispersion relations. 

\begin{figure}[H]
\centering%
\includegraphics[width=0.55\textwidth, keepaspectratio=true]{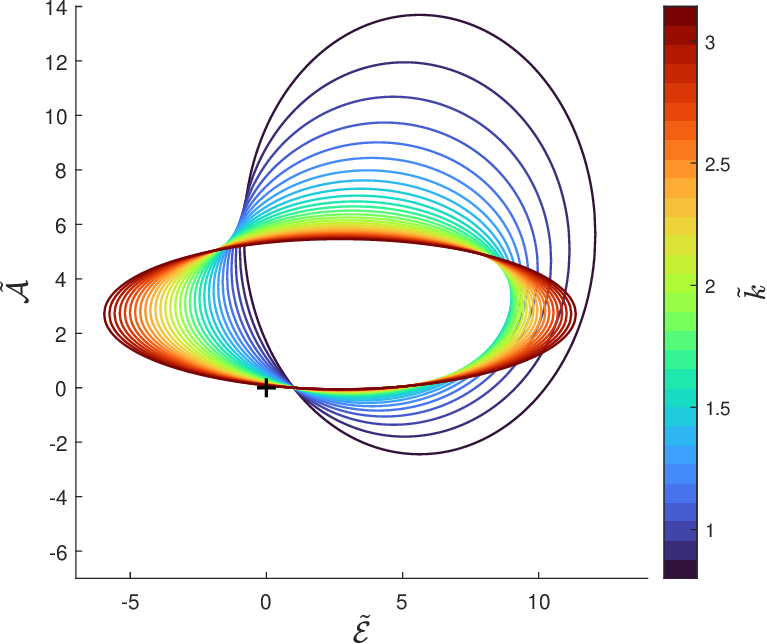}
                \caption{Solutions $\tilde \cA$ and $\tilde \cE$ visualized in the $\tilde\cE-\tilde\cA$ space for various wave numbers, $\kappa=4$}
                \label{ellipse_mm}
\end{figure}

Since the local micromorphic formulation is a special case of the model introduced in this paper, one possible choice is to set $\tilde E(\tilde r)= \sqrt{{\eta}/{\rho}}\, \delta(\tilde r)$, $\tilde A(\tilde r)= 0$, and $\tilde H(\tilde r)=\kappa  \sqrt{{\eta}/{\rho}}\, \delta(\tilde r) $, where $\delta(\tilde r)$ denotes the Dirac delta distribution. The corresponding Fourier images are $\tilde \cE(\tilde k)= 1$, $\tilde \cA(\tilde k)= 0$, and $\tilde \cH(\tilde k)= \kappa $, and one can indeed check that equation~\eqref{constEq} is satisfied when relations  \eqref{omsum2}--\eqref{ommult2} are substituted. However, there exist infinitely many other combinations of weight functions for which the dispersion diagram is exactly reproduced. This can be demonstrated by plotting the set of all admissible solutions described by  \eqref{ellGen} with the squares of dimensionless circular frequencies substituted from \eqref{disp_mm}.  
% The equation of the ellipse depicting the set of possible solutions for this particular case is obtained by plugging the previous relation into Equation\ \eqref{ellGen}. 
The corresponding graphs are displayed in Figure \ref{ellipse_mm}.

\begin{figure}[H]
\centering%       
\begin{subfigure}[b]{0.49\textwidth}

                \includegraphics[width=\textwidth, keepaspectratio=true]{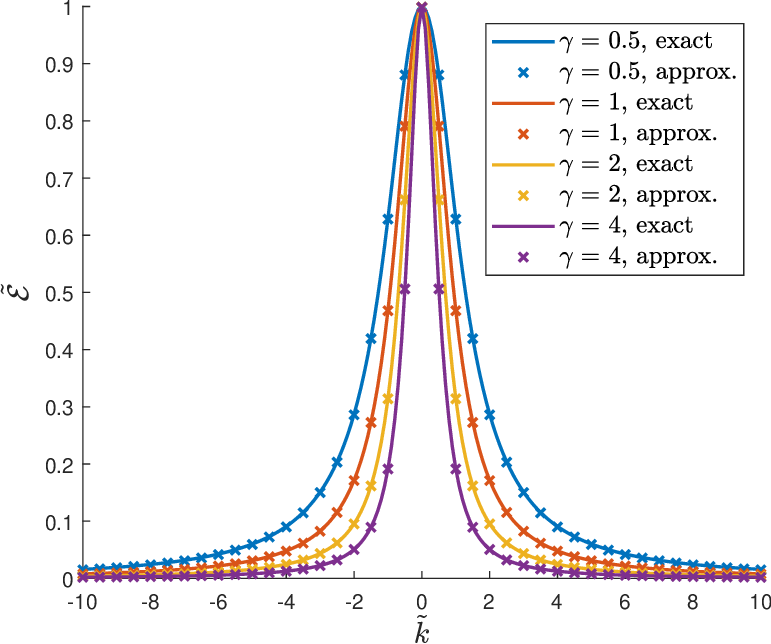}
                \subcaption{Function $\tilde\cE$}
\end{subfigure}
\begin{subfigure}[b]{0.49\textwidth}            
                
                \includegraphics[width=\textwidth, keepaspectratio=true]{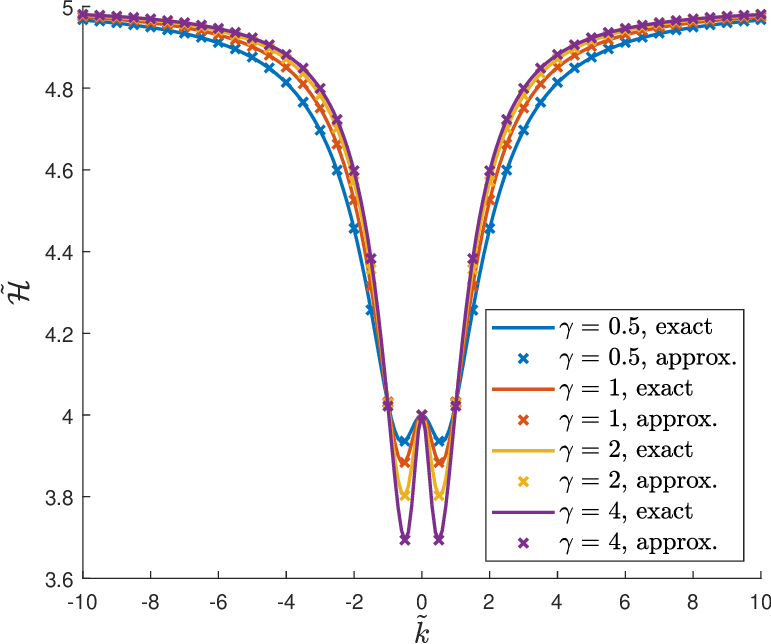}  
                \subcaption{Function $\tilde\cH$}
\end{subfigure}
                \caption{Comparison of the Fourier images of dimensionless weight functions and their approximation for $\kappa=4$ and for various values of $\gamma$} 
                \label{EfHfapp_mm}
\end{figure}

One can now clearly see that all the ellipses pass through the point where $\tilde\cE(\tilde k)=1$ and $\tilde\cA(\tilde k)=0$, which corresponds to the local micromorphic model described above. 
Other possible models can be constructed by adopting the procedure described in section \ref{sol}. Substituting \eqref{omsum2} and \eqref{ommult2} into \eqref{Efsol} and \eqref{Hfun}, we express the dimensionless Fourier images  as
\bea \label{E_mm}\label{e123}
\tilde\cE &=& \dfrac{\left(\tilde k^2(1+\kappa)+\kappa\right)(\gamma \tilde k^2+1)  - \sqrt{{\left(\tilde k^2(1+\kappa)+\kappa\right)}^2{(\gamma \tilde k^2+1)}^2 - 4(\gamma^2 \tilde k^2+ 1)(1+\tilde k^2)\tilde k^2 \kappa}}{2(\gamma^2 \tilde k^4+ \tilde k^2)} 
\\
\tilde\cH &=& \kappa +   \frac{\tilde k^2} {1+\tilde k^2}-(1+\gamma)\frac{\tilde k^2 }{1+\tilde k^2} \tilde\cE
\label{e124}
\eea
Parameter $\gamma = \tilde A / \tilde E$ can have an arbitrary positive value, since the expression in \eqref{E_mm} is real for any $\gamma>0$. 
In Figure \ref{EfHfapp_mm}, the graphs of Fourier images $\tilde\cE$ and $\tilde\cH$ are plotted for $\kappa=4$ and for values of $\gamma$ ranging from 0.5 to 4. 
% \begin{figure}[H]
% \centering%       
% \begin{subfigure}[b]{0.49\textwidth}

%                 \includegraphics[width=\textwidth, keepaspectratio=true]{GM_EfappAll_mm2.eps}
%                 \subcaption{Function $\tilde\cE$}
% \end{subfigure}
% \begin{subfigure}[b]{0.49\textwidth}            
                
%                 \includegraphics[width=\textwidth, keepaspectratio=true]{GM_HfappAll_mm2.eps}  
%                 \subcaption{Function $\tilde\cH$}
% \end{subfigure}
%                 \caption{{\bf JUST FOR ILLUSTRATION, line for $\gamma=0.01$ added}, Comparison of the Fourier images of the weight functions and their approximations, $\kappa=4$}
 
% \end{figure} 

\begin{figure}[H]
\centering%
\begin{subfigure}[b]{0.49\textwidth}

                \includegraphics[width=\textwidth, keepaspectratio=true]{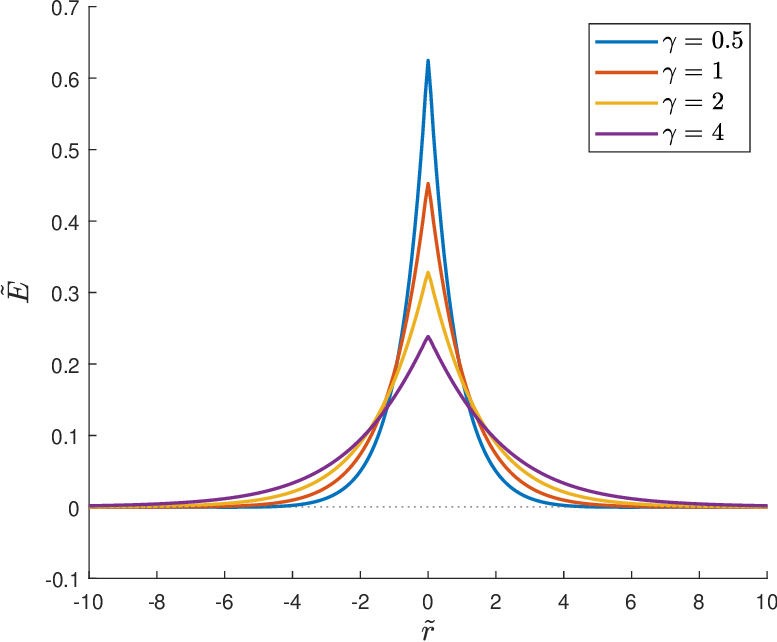}
                 \subcaption{Weight function $\tilde E$}
\end{subfigure}
\begin{subfigure}[b]{0.49\textwidth}            
               
                \includegraphics[width=\textwidth, keepaspectratio=true]{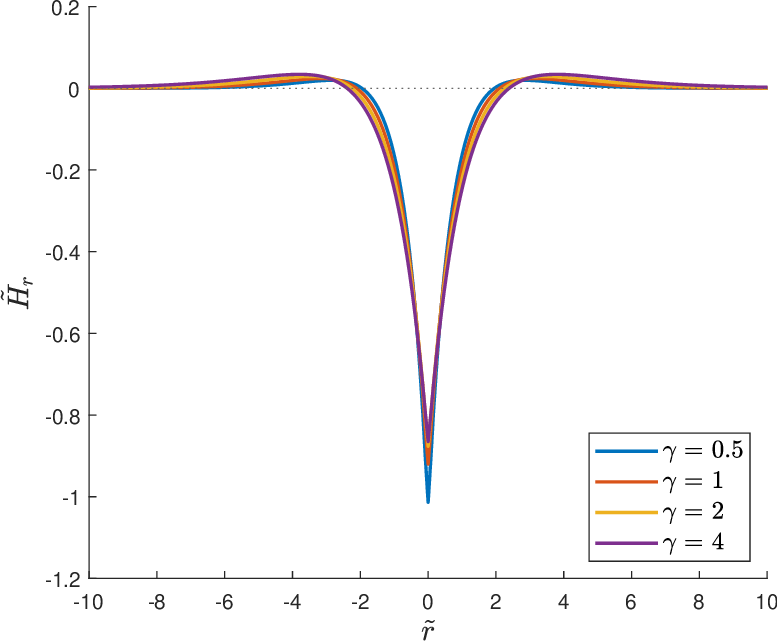}
                 \subcaption{Regular part $\tilde H_r$ of weight function $\tilde H$ }
\end{subfigure}

                \caption{Weight functions computed by IDFT for the approximation of the dispersion diagram obtained from the local micromorphic model with vanishing micromorphic modulus, using various values of $\gamma$ and  fixed values of model parameter $\kappa=4$ and numerical parameters $\tilde k_0=40$ and $N=400$
                % {\bf REPLACE the graph in part (b) by the graph that is now displayed in the next figure (to be deleted)
                % }
                }
                \label{FH_IFTall_mm}
\end{figure}
% \begin{figure}[H]
% \centering%
% \begin{subfigure}[b]{0.55\textwidth}            
               
%                 \includegraphics[width=\textwidth, keepaspectratio=true]{GM_Hall_mm2.eps}
%                  \subcaption{Weight function $\tilde H$}
% \end{subfigure}

%                 \caption{{\bf JUST FOR ILLUSTRATION}, weight function $\tilde H$ without the DIrac contribution, $\kappa=4$, $\tilde k_0=40$, $N=800$}
% \end{figure}

It is interesting to note that, in the limit of $\tilde k\to\infty$, the value of $\tilde\cE$ tends to zero but 
$\tilde\cH$ approaches (from below) a positive limit $\kappa+1$. This is caused by the first two terms in (\ref{e124}). Their inverse Fourier transform can be evaluated analytically
and contains a multiple of the Dirac distribution:
\beq
\mathcal{F}^{-1} \left(\kappa +   \frac{\tilde k^2} {1+\tilde k^2} \right) =(1+\kappa)\,\delta(\tilde r) - \half {\rm e}^{-|\tilde r|} 
\eeq
Therefore, the resulting weight function 
\beq 
\tilde H = \mathcal{F}^{-1}\left(\tilde\cH\right) = \tilde H_s + \tilde H_r
\eeq 
consists of a singular part
\beq 
 \tilde H_s = (1+\kappa)\,\delta(\tilde r)
\eeq 
and a regular part
\beq 
 \tilde H_r = - \half {\rm e}^{-|\tilde r|}  -(1+\gamma)\,\mathcal{F}^{-1} \left(\frac{\tilde k^2 }{1+\tilde k^2} \tilde\cE \right)
\eeq 
The regular part, evaluated numerically,
is plotted in Figure \ref{FH_IFTall_mm}b, along with
the full
weight function $\tilde E$ plotted in Figure \ref{FH_IFTall_mm}a.

\begin{figure}[H]
\centering%       

                \includegraphics[width=0.55\textwidth, keepaspectratio=true]{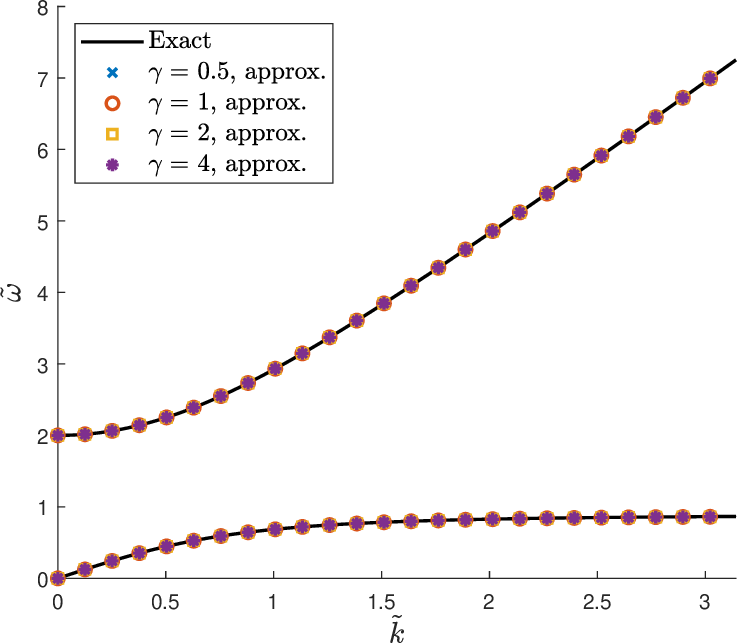}

                \caption{Comparison of the original dispersion diagram obtained from the local micromorphic model with vanishing micromorphic modulus
                and its replica obtained using the integral micromorphic model, with fixed $\kappa=4$ and various values of $\gamma$}
                \label{chainapp_mm}
\end{figure}

Finally,
Figure \ref{chainapp_mm} shows an excellent agreement
between the dispersion diagram of the original local micromorphic model with $A=0$, given by \eqref{disp_mm}, and the dispersion diagram obtained using the present integral micromorphic model with various values of $\gamma$ and with the corresponding weight functions $\tilde E$, $\tilde H$ and $\tilde A$ obtained by inverse Fourier transform of the functions in (\ref{e123})--(\ref{e124}) and by setting $\tilde A=\gamma\tilde E$. 
% In Figure \ref{chainapp_mm} the dispersion diagram of the micromorphic model \eqref{disp_mm} is compared to its approximation based on the present integral micromorphic model, and the discrepancy between both diagrams is also visualized separately for acoustic and optical branch. Also for this test case the results for various $\gamma$ coefficients are almost identical.
% \begin{figure}[H]
% \centering%       
% \begin{subfigure}[b]{0.49\textwidth}

%                 \includegraphics[width=\textwidth, keepaspectratio=true]{GM_chainAppGen_mm.eps}
%                 \subcaption{Dispersion diagram}\label{chainappAmm}
% \end{subfigure}
% \begin{subfigure}[b]{0.49\textwidth}            
                
%                 \includegraphics[width=\textwidth, keepaspectratio=true]{GM_chainErr_mm.eps} 
%                 \subcaption{Absolute Error}\label{chainappBmm}
% \end{subfigure}
%                 \caption{Comparison of the exact dispersion diagram obtained from the local micromorphic model with vanishing micromorphic modulus
%                 and its enriched continuum approximation, $\kappa=4$}
%                 \label{chainapp_mm}
% \end{figure} 

\section{Energetic Aspects}

\subsection{General considerations}

Since the proposed generalized continuum model has been consistently derived from expressions for the kinetic and potential energy, it is interesting to look at the relative contribution of individual terms in those expressions.

%Let us start with the potential energy, which is evaluated according to Eq.~\eqref{free_en}, with the outer integrals taken over one wavelength. The result is then divided by the wave length, to get the energy density per unit length. Firstly, we substitute the harmonic ansatz for the displacement and micromorphic variable fields. As the potential energy functional is quadratic in both fields, it is no longer possible to use the complex exponential notation, but one must adopt the trigonometric one. Moreover, from Eq.~\eqref{eqHom} one obtains 

For a harmonic wave characterized by a certain wave number $k$ and the corresponding circular frequency $\omega$ (linked to $k$ by the dispersion equation \eqref{mm4}), the matrix in \eqref{eqHom} is singular and, assuming that $\cH_1=\cH_2=\cH$,
we can express
the ratio of the complex amplitudes 
\beq
\frac{\hat{\chi}}{\hat{u}} = \dfrac{\rho\omega^2-\left(\cE+\cH\right)k^2}{\cH ik } = \dfrac{\left(\cE+\cH\right)k^2-\rho\omega^2}{\cH k } \, i = X\,i
\eeq
where
\beq \label{eqX}
X = \dfrac{\left(\cE+\cH\right)k^2-\rho\omega^2}{\cH k } = \dfrac{\cH k}{\cH+\cA k^2-\eta\omega^2}
\eeq 
is real.
This indicates that the phase shift between the micromorphic variable and the displacement field is always either $\pi/2$, or $-\pi/2$ (depending on the sign of $X$).
Consequently, if the real representation of the displacement field is given by
\beq 
u(x,t) = \hat{u}\cos{(kx-\omega t)}
\eeq 
where $\hat{u}$ is real, then the micromorphic strain field 
is described by
\beq 
\chi(x,t) =  -X\hat{u}\sin{(kx-\omega t)}
\eeq 
Based on this, it is possible to evaluate the corresponding kinetic and potential energy associated with the harmonic elastic wave. 

In an infinite body, the energy would also be infinite, but it makes sense to determine its (uniform) spatial average, which remains finite. 
For a harmonic wave, it is sufficient to average over one wave length. 
According to \eqref{eq3}, the spatially averaged kinetic energy is expressed as
\bea\nonumber
\bar{T} &=&\frac{k}{2\pi} \int_0^{\frac{2\pi}{k}} \left(\frac{1}{2}\rho{\dot{u}(x,t)}^2 + \frac{1}{2}\eta{\dot{\chi}(x,t)}^2 \right)\; \dx = \\
&=&\frac{k}{4\pi} \rho \omega^2 \hat{u}^2 \int_0^{\frac{2\pi}{k}} \sin^2{(kx-\omega t)}  \; \dx +\frac{k}{4\pi}\eta \omega^2X^2\hat{u} ^2\int_0^{\frac{2\pi}{k}} \cos^2{(kx-\omega t)}  \; \dx = 
\frac{1}{4} \left( \rho +\eta X^2  \right) \omega^2 \hat{u}^2
\label{eq133}
\eea
The result is independent of time, even though we have applied averaging in space only. The reason is that the considered wave travels through the infinite space without changing its form and the distribution of velocities at two time instants is only shifted in space. 

Evaluation of the potential energy is more tedious because it requires double integration. 
The local strain is given by
\beq 
u'(x,t) = -k\hat{u}\sin{(kx-\omega t)}
\eeq
and thus it is useful to prepare the weighted spatial average
\bea \nonumber
\intL E_0(x-\xi)\sin(k\xi-\omega t) \;{\rm d}\xi &=&
\intL E_0(r)\sin(kx-kr-\omega t) \;{\rm d}r = \\
&=& \intL E_0(r)\cos kr \;{\rm d}r \,\sin(kx-\omega t) - \intL E_0(r)\sin kr\;{\rm d}r\,\cos(kx-\omega t) = \nonumber\\
&=& \cE(k)\sin(kx-\omega t)
\eea 
Averaging integrals that involve weights $A_0(r)$ and $H_0(r)$ are 
evaluated similarly, and the same procedure can be used if the sine function is replaced by the cosine. 
The potential energy spatially averaged over one wave length consists of the following three contributions:
\bea \nonumber
\bar{\Psi}_E &=&  \frac{k}{2\pi}\int_0^{\frac{2\pi}{k}}\frac{1}{2}k^2\hat{u}^2\left(\intL E_0(x-\xi)\sin{(k\xi-\omega t)} \;{\rm d}\xi\right) \sin{(kx-\omega t)}\dx =
\frac{1}{4\pi}k^3\hat{u}^2\cE(k)\int_0^{\frac{2\pi}{k}}\sin^2{(kx-\omega t)}\dx = \\
&=& \frac{1}{4}k^2\hat{u}^2\cE(k)
\\ 
\bar{\Psi}_A &=&  \frac{k}{2\pi}\int_0^{\frac{2\pi}{k}}\frac{1}{2}k^2X^2\hat{u}^2\left(\intL A_0(x-\xi) \cos{(k\xi-\omega t)} \;{\rm d}\xi\right) \cos{(kx-\omega t)}\dx =\frac{1}{4}k^2X^2\hat{u}^2\cA(k)
\\ 
\bar{\Psi}_H &=&  \frac{k}{2\pi}\int_0^{\frac{2\pi}{k}}\frac{1}{2}(k\hat{u}-X\hat{u})^2\left(\intL H_0(x-\xi)\sin{(k\xi-\omega t)} \;{\rm d}\xi\right) \sin{(kx-\omega t)}\dx =
\frac{1}{4}(k-X)^2\hat{u}^2\cH(k)
\eea 

It is worth noting that the spatially averaged total potential energy
\beq \label{e145}
\bar{\Psi} = \bar{\Psi}_E +  \bar{\Psi}_A +  \bar{\Psi}_H =
\frac{\hat{u}^2}{4} \left(k^2\cE(k) + k^2X^2\cA(k) + (k-X)^2\cH(k)\right)
\eeq 
is always equal to the spatially averaged kinetic energy given by (\ref{eq133}). To prove that, it is sufficient to 
verify the identity
\beq 
k^2\cE + k^2X^2\cA + \left(k-X\right)^2\cH = \left( \rho +\eta X^2  \right) \omega^2
\eeq 
in which, for conciseness, the dependence of the Fourier images $\cE$, $\cA$ and $\cH$ on the wave number $k$ is not marked explicitly.
If $X$ is replaced by the expression in (\ref{eqX}),
the identity can be shown to be equivalent to the dispersion equation (\ref{mm4}).

Relations (\ref{eq133}) and (\ref{e145}) indicate that
the spatial averages of the kinetic and potential energy are proportional to the square of the amplitude, which can be arbitrary. However,
the relative contribution of individual terms (standard or enriching ones) is amplitude-independent and can provide further insight into
the role of the enrichment. If the spatial average of kinetic energy, $\bar{T}$, is considered as the sum of the standard term $\bar{T}_\rho=\rho\omega^2\hat{u}^2/4$ and the micromorphic term $\bar{T}_\eta=\eta X^2\omega^2\hat{u}^2/4$, the
relative contribution of the enrichment can be expressed as
\beq 
\frac{\bar{T}_\eta}{\bar{T}_\rho+\bar{T}_\eta} = \frac{\eta X^2}{\rho+\eta X^2} = \frac{l^2X^2}{1+l^2X^2} = \frac{\tilde{X}^2}{1+\tilde{X}^2}
\eeq 
where $\tilde{X} = lX$
is the dimensionless counterpart of $X$, i.e., of the ratio between the amplitudes of the micromorphic strain and of the displacement.
Based on (\ref{eqX}) and (\ref{mm9}) converted to the dimensionless form, it is possible to show that
\beq \label{e148}
\tilde{X} = \tilde{k}\,\dfrac{\tilde\omega_1^2+\tilde\omega_2^2-(1+1/\tilde{k}^2)\tilde\omega^2+(1-\gamma\tilde{k}^2)\tilde\cE}{\tilde\omega_1^2+\tilde\omega_2^2-(1+\gamma)\tilde{k}^2 \tilde\cE}
\eeq 
where $\tilde\cE$ can also be written in terms of $\tilde{k}$, $\tilde\omega_1$, $\tilde\omega_2$ and $\gamma$; see (\ref{Efsol}).
When (\ref{e148}) is evaluated for a specific branch of the 
dispersion diagram (acoustic or optical), $\tilde\omega$ should
be understood as the frequency on that branch, but the sum $\tilde\omega_1^2+\tilde\omega_2^2$ contains contributions of both branches
and remains the same. 

\subsection{Example: double-mass-spring chain with nearest-neighbor interactions}

% {\bf We should explain which model was used to construct the graphs. Probably it was the double-mass-spring chain with nearest-neighbor interactions. Is it true? Michal, please check. Please also update the reference to our paper [34].}

\begin{figure}[H]
\centering%       
\begin{subfigure}[b]{0.49\textwidth}
    
                \includegraphics[width=\textwidth, keepaspectratio=true]{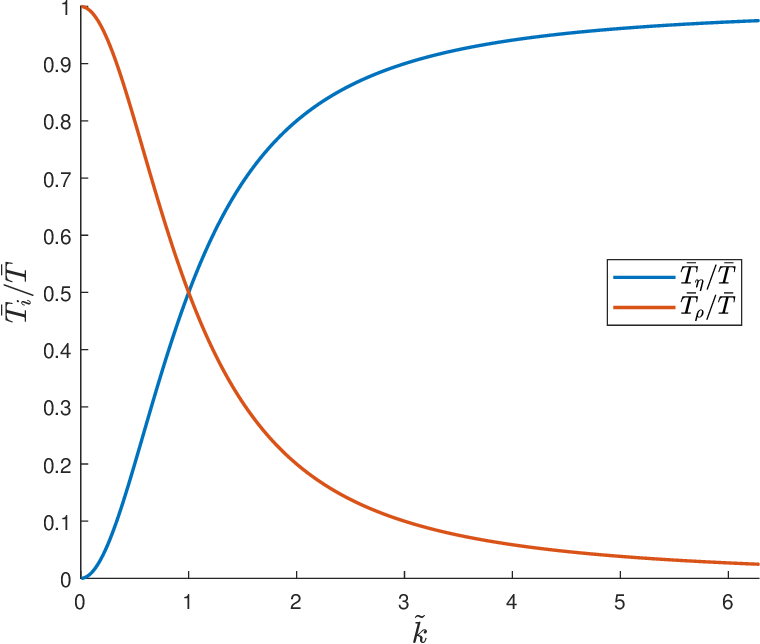}
                \subcaption{Acoustic branch}
\end{subfigure}
\begin{subfigure}[b]{0.49\textwidth}                \includegraphics[width=\textwidth, keepaspectratio=true]{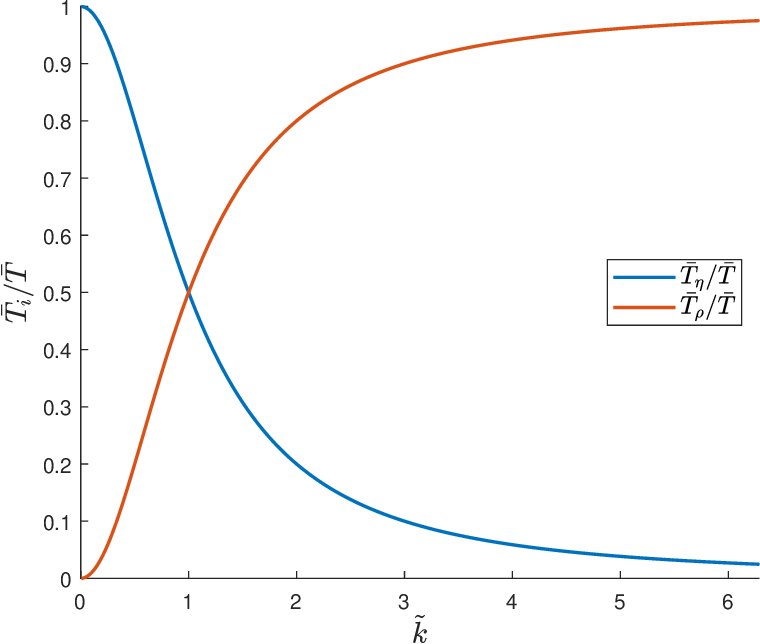}  
                \subcaption{Optical branch}
\end{subfigure}
                \caption{Relative kinetic energy contributions for parameters $\beta=0.5$, $\gamma=1$ and $\lambda=1$}
                \label{ener_kin1}
\end{figure}

\begin{figure}[H]
\centering%       
\begin{subfigure}[b]{0.49\textwidth}
    
                \includegraphics[width=\textwidth, keepaspectratio=true]{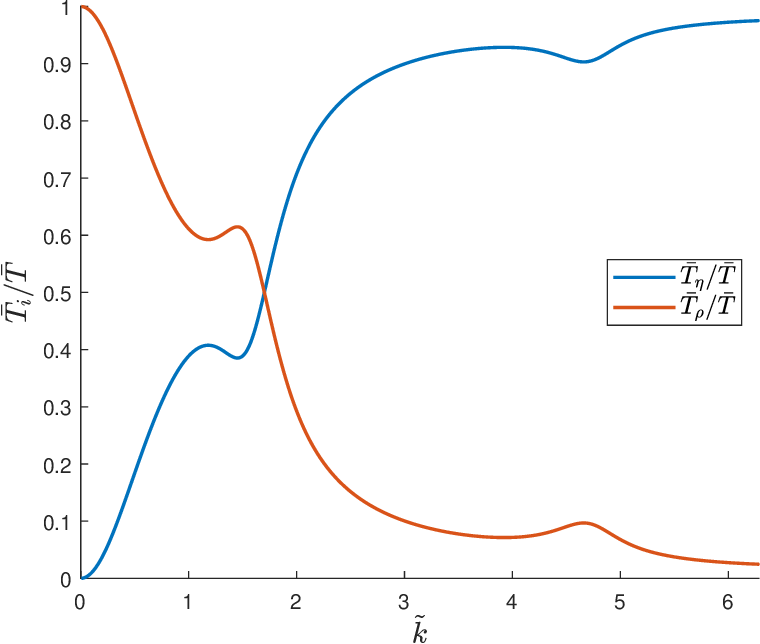}
                \subcaption{Acoustic branch}
\end{subfigure}
\begin{subfigure}[b]{0.49\textwidth}                \includegraphics[width=\textwidth, keepaspectratio=true]{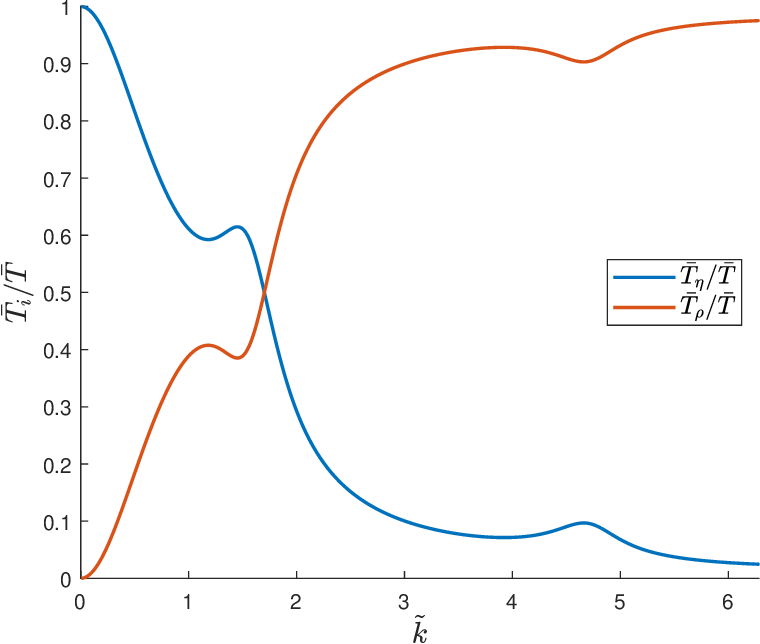}  
                \subcaption{Optical branch}
\end{subfigure}
                \caption{Relative kinetic energy contributions for parameters $\beta=0.5$, $\gamma=2$ and $\lambda=1$}
                \label{ener_kin2}
\end{figure}

For illustration, Figures \ref{ener_kin1}--\ref{ener_kin2} show the relative
contributions of the standard and micromorphic kinetic energy terms
for a range of wave numbers, separately for the acoustic and optical branches.The graphs reveal  that, for a given set of parameters,
the curves are the same for the acoustic and optical branch, only their meaning is flipped. This is a consequence 
of the fact that the amplitude ratios $X_1$ and $X_2$ evaluated for the acoustic and optical branches, respectively, 
satisfy the condition $X_1X_2=-\rho/\eta$, which then leads to
$\tilde X_1^2=1/\tilde X_2^2$ and, consequently, to
\beq 
\frac{\bar T_{\eta 1}}{\bar T_{\rho 1}+\bar T_{\eta 1}} =
\frac{\tilde X_1^2}{1+\tilde X_1^2}  =  \frac{1/\tilde X_2^2}{1+1/\tilde X_2^2} = \frac{1}{1+\tilde X_2^2} = \frac{\bar T_{\rho 2}}{\bar T_{\rho 2}+\bar T_{\eta 2}}
\eeq 
As seen in Figures \ref{ener_kin1}--\ref{ener_kin2},
for the acoustic branch, the kinetic energy of long waves (with small wave numbers) is dominated by the standard  term (red curve)
and the kinetic energy of short waves by the micromorphic term (blue curve),
while for the optical branch it is the contrary.


\begin{figure}[H]
\centering%       
\begin{subfigure}[b]{0.49\textwidth}
    
                \includegraphics[width=\textwidth, keepaspectratio=true]{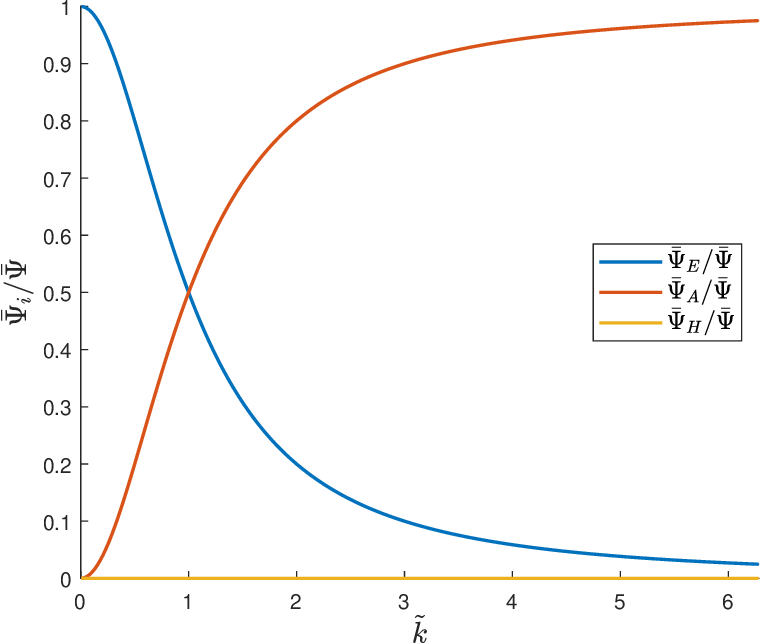}
                \subcaption{Acoustic branch}
\end{subfigure}
\begin{subfigure}[b]{0.49\textwidth}                \includegraphics[width=\textwidth, keepaspectratio=true]{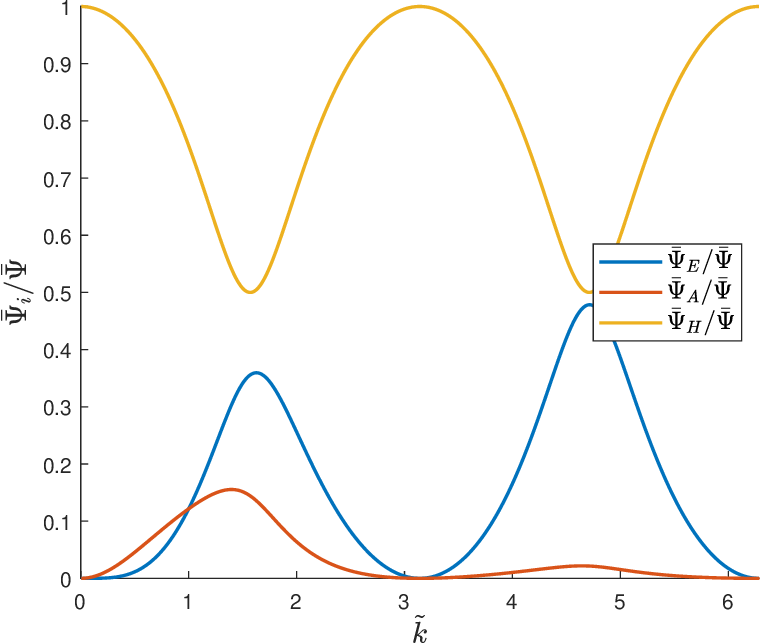}  
                \subcaption{Optical branch}
\end{subfigure}
                \caption{Relative contributions to potential energy for parameters $\beta=0.5$, $\gamma=1$ and $\lambda=1$  }
                \label{ener_pot1}
\end{figure}

\begin{figure}[H]
\centering%       
\begin{subfigure}[b]{0.49\textwidth}
    
                \includegraphics[width=\textwidth, keepaspectratio=true]{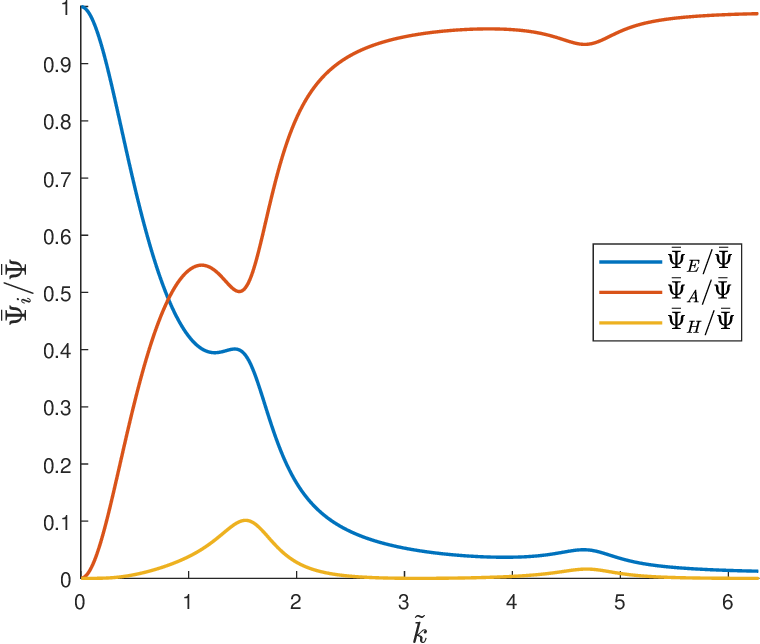}
                \subcaption{Acoustic branch}
\end{subfigure}
\begin{subfigure}[b]{0.49\textwidth}                \includegraphics[width=\textwidth, keepaspectratio=true]{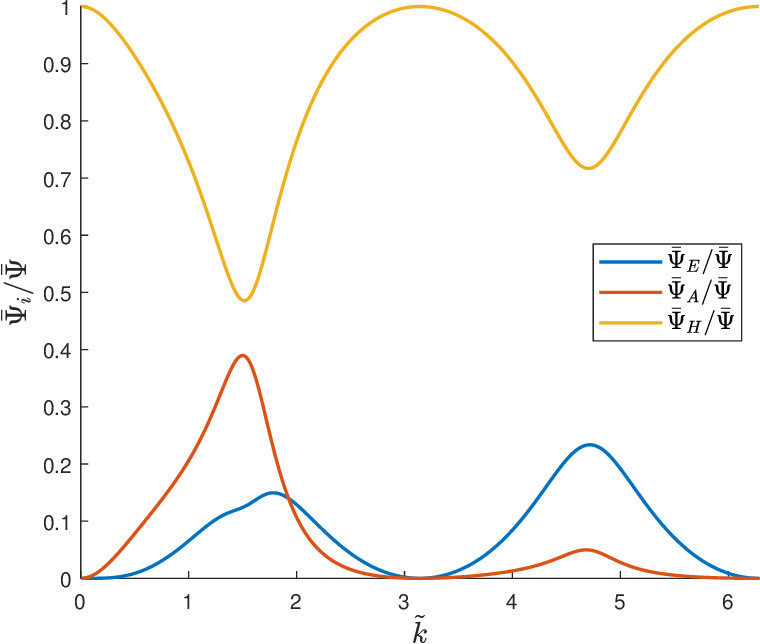}  
                \subcaption{Optical branch}
\end{subfigure}
                \caption{Relative contributions to potential energy for parameters  $\beta=0.5$, $\gamma=2$ and $\lambda=1$  }
                \label{ener_pot2}
\end{figure}

Similarly, we can evaluate the relative importance of three terms in the nonlocal expression for potential energy. 
The ratios
\bea 
\frac{\bar\Psi_E}{\bar\Psi} &=& \frac{\bar\Psi_E}{\bar{T}}=\frac{k^2\cE}{(\rho+\eta X^2)\omega^2} = \frac{\tilde{k}^2\tilde{\cE}}{(1+\tilde{X}^2)\tilde\omega^2} 
\\
\frac{\bar\Psi_A}{\bar\Psi} &=&  \frac{\bar\Psi_A}{\bar{T}}=\frac{k^2X^2\cA}{(\rho+\eta X^2)\omega^2} = \frac{\gamma\tilde{k}^2\tilde{X}^2\tilde{\cE}}{(1+\tilde{X}^2)\tilde\omega^2} 
\eea 
characterize the relative contribution of the nonlocal strain term and the nonlocal micromorphic strain gradient term; the relative contribution of the nonlocal coupling term is the complement to 1, as shown in Figures \ref{ener_pot1}--\ref{ener_pot2}.
For the acoustic branch (parts (a) of these figures), 
the potential energy stored in the coupling term (orange curve)
is very small and, for $\gamma=1$, it even vanishes.
On the other hand, for the optical branch it is the 
dominant one. Furthermore, for the acoustic branch,
the strain energy (blue curve) dominates for small wave numbers
and the energy stored in the gradients of micromorphic
strain (red curve) dominates for large wave numbers.
For the optical branch, both of these energy terms 
are negligible for small wave numbers and their
relative contribution is the largest near $\tilde k=\pi/2$.


\section{Summary and Conclusions}

We designed a one-dimensional micromorphic integral continuum model that can exactly reproduce the dispersion behavior of an arbitrary material with a dispersion diagram consisting of two branches, including cases when a band gap separates the branches. The proposed model enhances the classical micromorphic model by nonlocal effects introduced into all three terms in the free energy density function. Furthermore, we developed a procedure that directly constructs the corresponding three weight functions based on the given dispersion relation describing two given branches of the dispersion diagram. 

It was shown that for each dimensionless wave number $\tilde k$, the values of the dimensionless Fourier images of weight functions $\tilde \cE(\tilde k)$ and $\tilde \cA(\tilde k)$ must satisfy a quadratic equation that graphically corresponds to an ellipse in the $\tilde \cE-\tilde \cA$ plane. The value of the Fourier image of the third weight function $\tilde \cH(\tilde k)$ is subsequently computed from $\tilde \cE(\tilde k)$ and $\tilde \cA(\tilde k)$. Therefore, there is additional flexibility in the determination of the three weight functions, e.g., two of them can be chosen to be proportional. From the Fourier images, the original functions in the space domain are obtained by inverse Fourier transform, which needs to be performed numerically and constitutes the only source of potential error.     

In the results section, the developed identification procedure was applied to simple discrete models, which can simulate, e.g., atomic interactions. A mass-spring chain with two alternating masses is perhaps one of the simplest models for which a band gap in the dispersion curve is observed and hence is chosen as the basic toy problem. To keep the two functions $\tilde \cE$ and $\tilde \cA$ proportional and simultaneously ensure that both are real, the proportionality constant must be chosen within certain limits. This feature is nicely visualized when the ellipses depicting the set of possible solutions are plotted. It was also shown that, in this basic case, the nonlocality needs to be present in all three terms in the free energy density function, meaning that none of the weight functions can be chosen as a multiple of the Dirac delta distribution.   
The obtained weight functions are real, finite valued, and of reasonable shapes.
The deviations between the dispersion diagram acquired with the proposed enriched continuum model using the approximated weight functions and the analytical one increase for larger wave numbers. The error is exclusively due to the numerical evaluation of the inverse Fourier transform, and it can be made arbitrarily small by increasing the number of sampling points and the length of the sampling window. 
In the special case of a mass-spring chain with equal masses, analytical expressions for the weight functions are obtained when the two functions $\tilde E$ and $\tilde A$ are enforced to be identical. 

A slightly more complex dispersion diagram is obtained when second-neighbour interactions are added to the basic double-mass-spring chain model. Also for this test case, the weight functions are reasonable and well-behaved. 

The last example was focused on the dispersion behavior of the local micromorphic model with vanishing micromorphic stiffness, which is the only case in which the local model exhibits a band gap. Using the proposed methodology,
one obvious solution can be obtained when the weight function $\tilde A$ vanishes and the other two are set to multiples of the Dirac distribution, since then the nonlocal model degenerates to the local one to be reproduced. Nevertheless, additional solutions with all nonvanishing weight functions exist, and they are acquired by applying the developed procedure. The resulting functions $\tilde E$ and $\tilde A$ are reasonable and finite valued, but the function $\tilde H$ contains a contribution of the Dirac distribution.  

Finally, individual contributions to the energy associated with a harmonic wave were evaluated. It was shown that the spatial averages of kinetic energy and of potential energy are the same. Furthermore, the relative contribution of individual terms in the energy functional was analyzed 
for the acoustic and optical branches, and their importance
for long and short waves was discussed. For waves that
correspond to the acoustic branch, the major contribution to the kinetic energy of long waves comes from the standard term 
and to the potential energy of long waves from the strain-related nonlocal term. As the wave length gets comparable
to the internal length of the model, the contribution
of the micromorphic inertia to the kinetic energy and of the nonlocal term involving the gradient of micromorphic strain to the potential energy increases.
For waves that
correspond to the optical branch, the major contribution to the kinetic energy of long waves comes from the micromorphic inertia
and to the potential energy of long waves from the nonlocal term that penalizes the difference between standard and micromorphic strain.  As the wave length gets comparable
to the internal length of the model, the contribution
of the standard inertia to the kinetic energy and of the nonlocal terms involving the standard strain and the gradient of micromorphic strain to the potential energy increases.

\clearpage
\section*{Acknowledgements}
Financial support received in 2023 
from the Czech Science Foundation (project No.\ 19-26143X)
is gratefully acknowledged.
In 2024, this work was co-funded by the European Union under the ROBOPROX project (reg. no. CZ.02.01.01/00/22\_008/0004590) and by MŠMT - ERC CZ project No.\  LL2310.

%\clearpage

 \bibliographystyle{elsarticle-num-names.bst} 
 \bibliography{Reference.bib}

% else use the following coding to input the bibitems directly in the


% % \bibitem[Author(year)]{label}


% \end{thebibliography}
\end{document}